\documentclass[aps,
               pra,
               showpacs,
               amssymb,
               nofootinbib,
               superscriptaddress,
               twocolumn,
               longbibliography]{revtex4-2}
\usepackage[ansinew]{inputenc}
\usepackage{bbold}
\usepackage{bbm}
\usepackage{bm}
\usepackage{amsbsy}
\usepackage{amsthm}
\usepackage{amssymb}
\usepackage{amsfonts}
\usepackage{amsmath, mathtools}
\usepackage{dsfont}
\usepackage{graphicx}
\usepackage{epsfig}
\usepackage{epstopdf}
\usepackage{dsfont}
\usepackage{mathrsfs}
\usepackage{multibib}
\usepackage[dvipsnames]{xcolor}
\usepackage[colorlinks]{hyperref}
\makeatletter
\newcommand\org@hypertarget{}
\let\org@hypertarget\hypertarget
\renewcommand\hypertarget[2]{%
  \Hy@raisedlink{\org@hypertarget{#1}{}}#2%
  }
\makeatother
\usepackage[figure,table]{hypcap}

\usepackage{MnSymbol}
\usepackage{enumerate}
\usepackage{enumitem} %needed for physical circuits
\usepackage{float}
\usepackage{comment}
\usepackage{braket}
\usepackage{subfigure}
\usepackage{orcidlink}
\usepackage{tcolorbox}
\makeatletter
\newcommand{\providetcbcountername}[1]{%
  \@ifundefined{c@tcb@cnt@#1}{%
    --undefined--%
  }{%
    tcb@cnt@#1%
  }
}

\newcommand{\settcbcounter}[2]{%
  \@ifundefined{c@tcb@cnt@#1}{%
    \GenericError{Error}{counter name #1 is no tcb counter }{}{}%
  }{%
    \setcounter{tcb@cnt@#1}{#2}%
   }%
}%
\tcbuselibrary{theorems}
\newtcbtheorem{Definitions}{Definition}%
{colback=mycolor!5,colframe=mycolor,fonttitle=\bfseries,
left=4pt, right=4pt, top=5pt, bottom=5pt}{defi}
\newtcbtheorem{Theorems}{Theorem}%
{colback=mycolor!5,colframe=mycolor,fonttitle=\bfseries,
left=4pt, right=4pt, top=5pt, bottom=5pt}{theorem}
\definecolor{mycolor}{rgb}{0.122, 0.435, 0.698}
\usepackage{subfigure}
\usepackage{makecell}

\hypersetup{
	bookmarksnumbered,
	pdfstartview={FitH},
	citecolor={darkgreen},
	linkcolor={darkred},
	urlcolor={darkblue},
	pdfpagemode={UseOutlines}}
\definecolor{darkgreen}{RGB}{50,190,50}
\definecolor{somegreen}{RGB}{25,150,25}
\definecolor{darkblue}{RGB}{0,0,190}
\definecolor{darkred}{RGB}{238,0,0}
\definecolor{mycolor}{rgb}{0.122, 0.435, 0.698}
\usepackage{soul}
\usepackage{booktabs}
\usepackage{adjustbox}

\definecolor{prepcolor}{HTML}{E3F2FD}
\definecolor{flagcolor}{HTML}{FFF8E1}
\definecolor{mergecolor}{HTML}{E8F5E9}
\definecolor{cczcolor}{HTML}{FFF3E0}
\definecolor{hcolor}{HTML}{F3E5F5}
\definecolor{meascolor}{HTML}{ECEFF1}
\definecolor{swapcolor}{HTML}{FCE4EC}

\newcommand{\nl}{\ensuremath{\hspace*{-0.5pt}}}

\newcommand{\subtiny}[3]{\ensuremath{_{\hspace{#1 pt}\protect\raisebox{#2 pt}{\tiny{$ #3$}}}}}
\newcommand{\suptiny}[3]{\ensuremath{^{\hspace{#1 pt}\protect\raisebox{#2 pt}{\tiny{$ #3$}}}}}
\newcommand{\expval}[1]{\ensuremath{\left\langle\right.\hspace*{-1pt} #1 \hspace*{-1pt}\left.\right\rangle}}

\renewcommand{\thesection}{\Roman{section}}
\renewcommand{\thesubsection}{\Roman{section}.\Alph{subsection}}
\renewcommand{\thesubsubsection}{\Roman{section}.\Alph{subsection}.\arabic{subsubsection}}
\makeatletter
\renewcommand{\p@subsection}{}
\renewcommand{\p@subsubsection}{}
\makeatother

\newcommand{\Zbar}[1]{\widebar{Z}_{#1}}
\newcommand{\Xbar}[1]{\widebar{X}_{#1}}

\usepackage{eczoo}

\usepackage{tikz}
\usetikzlibrary{quantikz2}
\tikzset{
    qubit/.style={draw, circle, minimum width=1.3em, inner sep=1pt}
}

\newcommand{\widebar}[1]{\mkern 1.5mu\overline{\mkern-1.5mu#1\mkern-1.5mu}\mkern 1.5mu}
\usepackage{orcidlink}

\begin{document}

\title{Genuine Multipartite Entanglement between Logical Qubits via Cross-Code Lattice Surgery}
\author{Alex Steiner\,\orcidlink{0000-0001-5443-7871}}
\thanks{These authors contributed equally to this work. Contact: \href{mailto:alex.steiner@uibk.ac.at}{alex.steiner@uibk.ac.at}, \href{mailto:tomasz.andrzejewski@tuwien.ac.at}{tomasz.andrzejewski@tuwien.ac.at}}
\affiliation{Universit{\"a}t Innsbruck, Institut f{\"u}r Experimentalphysik, Innsbruck, Austria}

\author{Tomasz Andrzejewski\,\orcidlink{0009-0007-6546-8578}}
\thanks{These authors contributed equally to this work. Contact: \href{mailto:alex.steiner@uibk.ac.at}{alex.steiner@uibk.ac.at}, \href{mailto:tomasz.andrzejewski@tuwien.ac.at}{tomasz.andrzejewski@tuwien.ac.at}}
\affiliation{Technische Universit{\"a}t Wien, Atominstitut \& Vienna Center for Quantum Science and Technology (VCQ), Stadionallee 2, 1020 Vienna, Austria}

\author{Phila Rembold\,\orcidlink{0000-0003-1405-730X}}
\affiliation{Technische Universit{\"a}t Wien, Atominstitut \& Vienna Center for Quantum Science and Technology (VCQ), Stadionallee 2, 1020 Vienna, Austria}

\author{Hendrik Poulsen Nautrup\,\orcidlink{0000-0001-7815-7006}}
\affiliation{Institute for Theoretical Physics, University of Innsbruck, Technikerstr. 21a, A-6020 Innsbruck, Austria}

\author{Christian D. Marciniak\,\orcidlink{0000-0001-8401-3981}}
\altaffiliation{Present address: neQxt GmbH, 80686 M{\"u}nchen, Germany}
\affiliation{Universit{\"a}t Innsbruck, Institut f{\"u}r Experimentalphysik, Innsbruck, Austria}

\author{Robert Freund\,\orcidlink{0009-0007-8401-2322}}
\affiliation{Universit{\"a}t Innsbruck, Institut f{\"u}r Experimentalphysik, Innsbruck, Austria}

\author{Ivan Pogorelov\,\orcidlink{0009-0001-5103-9410}}
\affiliation{Universit{\"a}t Innsbruck, Institut f{\"u}r Experimentalphysik, Innsbruck, Austria}

\author{Thomas Monz\,\orcidlink{0000-0001-7410-4804}}
\affiliation{Universit{\"a}t Innsbruck, Institut f{\"u}r Experimentalphysik, Innsbruck, Austria}
\affiliation{Alpine Quantum Technologies GmbH, Innsbruck, Austria}
\author{Philipp Schindler\,\orcidlink{0000-0002-9461-9650}}
\email{philipp.schindler@uibk.ac.at.}
\affiliation{Universit{\"a}t Innsbruck, Institut f{\"u}r Experimentalphysik, Innsbruck, Austria}
\author{Marcel Meyer\,\orcidlink{0000-0002-7832-5927}}
\affiliation{Universit{\"a}t Innsbruck, Institut f{\"u}r Experimentalphysik, Innsbruck, Austria}
\author{Nicolai Friis\,\orcidlink{0000-0003-1950-8640}}
\email{nicolai.friis@tuwien.ac.at}
\affiliation{Technische Universit{\"a}t Wien, Atominstitut \& Vienna Center for Quantum Science and Technology (VCQ), Stadionallee 2, 1020 Vienna, Austria}

\date{\today}

\begin{abstract} 
Universal quantum computers are expected to generate arbitrary complex quantum states of logical qubits encoded in many physical qubits. This capability hinges on a fault-tolerantly implemented universal gate set, which no single quantum error-correction code admits transversally but which becomes accessible by joining complementary codes via lattice surgery. Here we report on the experimental generation and certification of logical genuine multipartite entanglement in a trapped-ion quantum processor using a transversally implemented universal logical gate set. The gate set is accessed via lattice surgery across two different codes and comprises a Hadamard gate on a four-qubit surface code and a doubly controlled Pauli-$Z$ ($\widebar{\mathrm{CCZ}}$) gate on an eight-qubit 3D colour code. To showcase this lattice-surgery toolbox, we generate both stabiliser (Greenberger{\textendash}Horne{\textendash}Zeilinger) and non-stabiliser ($\ket{\widebar{\mathrm{CCZ}}}$) states of three logical qubits and verify their genuine multipartite entanglement{\textemdash}a form of correlation beyond statistical mixtures of bipartite entanglement across any bipartition. We further use these cross-code primitives to demonstrate arbitrary rotations of single logical qubits via a $\widebar{\mathrm{CCZ}}$-based resource gadget accessing the full universal gate set through lattice surgery. Together, these demonstrations showcase the core building blocks of an architecture for fault-tolerant quantum computation and its ability to generate complex logical quantum states.
\end{abstract}

\date{\today}

\maketitle

%%%%%%%%%%%%%%%%%%%%%%%%%%%%%%%%%%%%%%%%%%%%%%%%%%%%%%%%%%%%%%%%%%%%%%%%%%%%%%%%

\begin{figure}[!htbp]
    \centering
    %%%trim={<left> <lower> <right> <upper>},trim={0cm 4.05cm 0cm 0cm},clip
    \includegraphics[width=\linewidth,trim={0.45cm 4.5cm 0.45cm 0.8cm},clip]{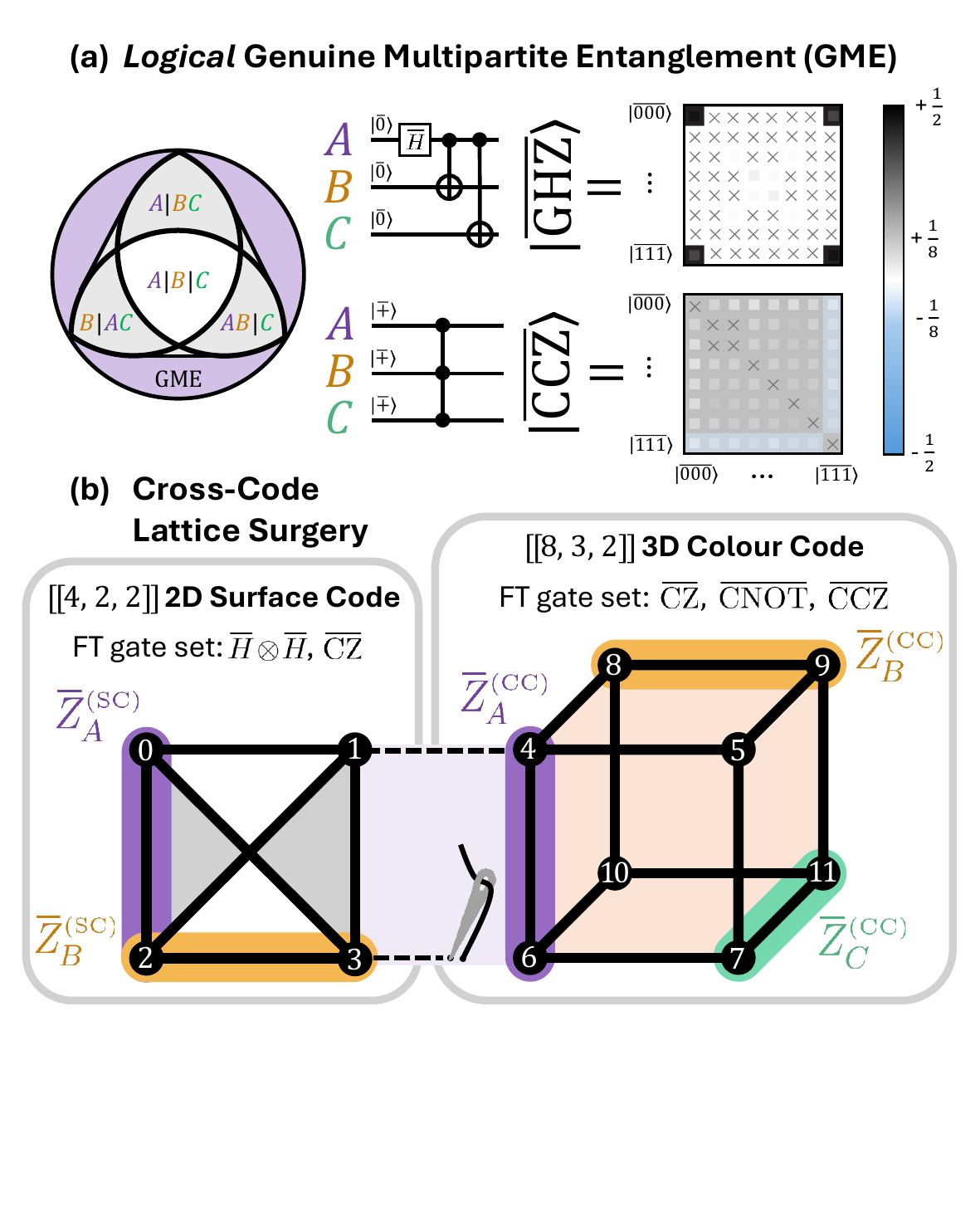}
    \caption{
    \textbf{Logical genuine multipartite entanglement and cross-code lattice surgery.}
    \textbf{(a)}~Logical genuine multipartite entanglement (GME): 
    States of three parties, $A$, $B$, and $C$, that are convex mixtures of states separable with respect to the bipartitions $A|BC$, $AB|C$, and $B|AC$ are biseparable, while states outside this convex set are GME. 
    Examples for GME states include the GHZ state (a stabiliser state) and the non-stabiliser state 
    $\ket{\widebar{\mathrm{CCZ}}}$. Circuits for the preparation of these states using logical $\widebar{\mathrm{CNOT}}$ and $\widebar{\mathrm{CCZ}}$ gates acting on logical states $\ket{\widebar{+}}=(\ket{\widebar{0}}+\ket{\widebar{1}})/\sqrt{2}$ and $\ket{\widebar{0}}$, as well as their density-matrix elements (with contours/fillings/$\times$ representing ideal/measured/not measured values) with respect to the logical computational basis are shown on the right-hand side. 
    \textbf{(b)}~QEC code blocks: 
    The $[\![4,2,2]\!]$ surface code~(SC) offers transversal and thus fault-tolerant (FT) $\widebar{H}\nl\otimes\nl\widebar{H}$ (up to a $\widebar{\mathrm{SWAP}}$) and $\widebar{\mathrm{CZ}}$ gates. The $[\![8,3,2]\!]$ 3D colour code~(CC) supports transversal $\widebar{\mathrm{CZ}}$ and $\widebar{\mathrm{CCZ}}$ gates, and a permutation-based FT $\widebar{\mathrm{CNOT}}$ gate. They are combined via a $\Zbar{\!A}\suptiny{0}{0}{\mathrm{(SC)}} \otimes \Zbar{\!A}\suptiny{0}{0}{\mathrm{(CC)}}$ parity measurement (dashed line) into a joint $[\![12,4,2]\!]$ merged code. Note that logical operators are equivalent up to multiplication with stabilisers.
    See Appendix~\ref{appendix:lattice surgery and codes} for more details.
    }
    \label{fig:conceptual_overview}
\end{figure}

%%%%%%%%%%%%%%%%%%%%%%%%%%%%%%%%%%%%%%%%%%%%%%%%%%%%%%%%%%%%%%%%%%%%%%%%%%%%%%%%
%%%%%%%%%%%%%%%%%%%%%%%%%%%%%%%%%%%%%%%%%%%%%%%%%%%%%%%%%%%%%%%%%%%%%%%%%%%%%%%%

\section{Introduction}
\label{sec:intro}
\vspace*{-1.5mm}

{\noindent}To outperform classical computers in selected tasks, quantum computers have to explore complex multipartite entangled states\;\cite{JozsaLinden2003}, whilst being protected from errors. Fault-tolerant quantum computing provides a framework to perform logical operations and prevent errors from spreading by encoding information in quantum error-correction (QEC) codes\;\cite{Preskill1997,NielsenChuang2010,LidarBrun2013}. Transversal gates provide fault tolerance by construction, but the Eastin{\textendash}Knill theorem\;\cite{EastinKnill2009} rules out a transversal universal gate set on any individual QEC code. 
Different strategies aim to overcome this restriction to combine fault tolerance with universality, including magic-state distillation\;\cite{BravyiKitaev2005,BeverlandEtAl2021} and cultivation\;\cite{GidneyEtAl2024}, code switching\;\cite{Bombin2016, ButtEtAl2024, PogorelovEtAl2025, WuZhongBrunLidar2026}, and lattice surgery\;\cite{HorsmanEtAl2012, ErhardEtAl2021, NautrupFrisBriegel2017, Litinski2019}. Here, we access a universal set of transversal logical gates by combining \emph{different} QEC codes via lattice surgery and use them to experimentally demonstrate the generation of genuine multipartite entanglement~(GME) between 3 logical qubits encoded in 12 physical qubits on a trapped-ion quantum processor. 

\vspace*{0.5mm}
Quantum states feature GME if they are entangled across all bipartitions, i.e., across all possible complementary subsystem pairs, but are not biseparable. That is, they cannot be expressed as statistical mixtures of states that are separable with respect to different bipartitions as shown in Fig.~\ref{fig:conceptual_overview}~(a). See~\cite{TothGuhne2005,FriisVitaglianoMalikHuber2019,BertlmannFriis2023} for reviews. As a fundamental resource for quantum communication~\cite{EppingKampermannMacchiavelloBruss2017, BaeumlAzuma2017, PivoluskaHuberMalik2018, RibeiroMurtaWehner2018, YamasakiPirkerMuraoDuerKraus2018} and quantum computation~\cite{RaussendorfBriegel2001, BriegelRaussendorf2001, JozsaLinden2003, Scott2004, BrussMacchiavello2011}, the detection of GME can be a key benchmark for multi-qubit quantum devices~\cite{FriisEtAl2018, CanteriEtAl2025} that serves as a ``\emph{first checkpoint for a fully operational QEC circuit}''~\cite{RodriguezBlancoEtAl2021}. We extend this benchmark to the logical level by experimentally certifying GME in two different joint states of three logical qubits: a Greenberger{\textendash}Horne{\textendash}Zeilinger (GHZ) state as the paradigmatic example of a stabiliser state, and the non-stabiliser state $\ket{\widebar{\mathrm{CCZ}}}$, which represents a non-Clifford resource for universal fault-tolerant quantum computation via magic state injection~\cite{veitch_resource_2014}. Not only do we certify 
non-stabiliserness, we also measure the logical state fidelity, which has, to our knowledge, not been reported before~\cite{LeeEtAl2025,daguerre_experimental_2025}.\\[-4mm]

We prepare these states by exploiting the fault-tolerant gate set provided by the $[\![4,2,2]\!]$ surface code and the $[\![8,3,2]\!]$ 3D colour code [see Fig.~\ref{fig:conceptual_overview}~(b)], which includes a Hadamard gate on one code and a doubly-controlled-$Z$ ($\widebar{\mathrm{CCZ}}$) gate~\cite{Aharonov2003} on the other. These gates generate the universal gate set $\{H,\mathrm{CCZ}\}$, are transversally implemented, and united in a joint logical circuit via cross-code lattice surgery. Ultimately, we combine these techniques to implement the continuous rotation of a single logical qubit using a $\widebar{\mathrm{CCZ}}$ resource gadget (adapted from~\cite{JerbiEtAl2023}) based on gate teleportation~\cite{GottesmanChuang1999, ZhouLeungChuang2000} via lattice surgery. 

Our demonstration thus goes beyond previous experiments reporting logical GME restricted to GHZ states within one code~\cite{hong_entangling_2024} or copies of the same code~\cite{BluvsteinEtAl2024}, and previous reports of lattice surgery between surface codes~\cite{ErhardEtAl2021, BesedinEtAl2025, WangEtAl2026} or between a surface and a Bacon{\textendash}Shor code~\cite{HetenyiWootton2024}, where the merged codes inherit only the transversal Clifford-gate set common to both blocks. Meanwhile, experiments on code switching~\cite{RyanAndersonEtAl2022, PogorelovEtAl2025} have accessed non-Clifford transversal gates by transferring a single logical qubit to a different code without generating logical entanglement~\cite{Bombin2016}. In contrast, merging via cross-code lattice surgery as carried out here combines both capabilities: generating (genuinely multipartite) entanglement \emph{between} logical qubits while bridging codes with complementary transversal gates, see Fig.~\ref{fig:conceptual_overview}~(b). Interactions between codes with complementary transversal gates have been proposed for the code pair we consider here~\cite{NelsonLandahlBaczewski2025}, and realised in neutral-atom systems~\cite{BluvsteinEtAl2024, BluvsteinEtAl2025} combining surface codes with the $[\![8,3,2]\!]$ colour code and the $[\![15,1,3]\!]$ Reed-Muller code. There, code blocks are coupled by $O(n)$ qubit-wise by transversal operations rather than by a joint boundary measurement. Whether transversal gate teleportation or surgery merging{\textemdash}here a single weight-four joint parity measurement, or an $O(d)$ boundary at distance~$d${\textemdash}is cheaper depends on the parallelisation capabilities of the specific architecture.\\[-5.5mm]

%%%%%%%%%%%%%%%%%%%%%%%%%%%%%%%%%%%%%%%%%%%%%%%%%%%%%%%%%%%%%%%%%%%%%%%%%%%%%%%%
%%%%%%%%%%%%%%%%%%%%%%%%%%%%%%%%%%%%%%%%%%%%%%%%%%%%%%%%%%%%%%%%%%%%%%%%%%%%%%%%

\section{Quantum Error Correction}
\label{sec:background}
\vspace*{-1.5mm}

{\noindent}\textbf{Stabiliser codes} are a powerful tool in QEC~\cite{GottesmanPhD1997}. An $n$-physical-qubit stabiliser code is defined by a stabiliser group $\mathcal{S}$ comprised of mutually commuting $n$-qubit Pauli operators (excluding $-\mathbb{1}$). The group is generated by $s<n$ independent stabilisers $S_{i}$ with $i=0,1,\ldots,s-1$. The joint $+1$ eigenstates of $\mathcal{S}$ span the code space, encoding $k=n-s$ logical qubits. 
Formally, we write $\mathcal{S}=\left\langle\right.\! S_{0}, S_{1},\ldots,S_{s-1}\!\left.\right\rangle$, and denote the corresponding code as $[\![n,k,d]\!]$, where $d$ is its distance, the minimum number of single-qubit Pauli errors whose combined effect is a non-trivial logical operator.\\[-3.0mm]

We consider two examples of stabiliser codes, the $[\![4,2,2]\!]$ and the $[\![8,3,2]\!]$ code, shown in Fig.~\ref{fig:conceptual_overview}~(b), with more details in Appendix~\ref{appendix:lattice surgery and codes}.
The $[\![4,2,2]\!]$ code~\eczoo{stab_4_2_2} is the smallest code of the hypercube family~\eczoo{hypercube_quantum} lying at the intersection of the surface- and colour-code families. We refer to it as the surface code~(SC)~\cite{Kitaev2003, DennisKitaevLandahlPreskill2002, FowlerMariantoniMartinisCleland2012, RaussendorfHarrington2007} because the logical gate we exploit is its fold-transversal Hadamard retained at every surface-code distance. With two stabilisers corresponding to weight-four $X$ and $Z$ operators on all four physical qubits $0$, $1$, $2$, and $3$, the non-trivial generators of the logical Pauli group of two encoded logical qubits $A$ and $B$ are $\Xbar{\!A}\suptiny{0}{0}{\mathrm{(SC)}} =X_0 X_1$ and $\Zbar{\!A}\suptiny{0}{0}{\mathrm{(SC)}}=Z_0 Z_2$ as well as $\Xbar{\!B}\suptiny{0}{0}{\mathrm{(SC)}} =X_1 X_3$ and $\Zbar{\!B}\suptiny{0}{0}{\mathrm{(SC)}} =Z_2 Z_3$. We use horizontal bars to better distinguish logical from physical gates (and states). Hadamard gates on all four qubits transversally implement the logical gate $\widebar{H}\otimes \widebar{H}$ up to a $\mathrm{SWAP}$ of~$A$ and~$B$.\\[-3.0mm]

\begin{figure*}[t!]
    \centering
    \includegraphics[width=\linewidth,trim={2cm 0cm 2cm 0cm},clip]{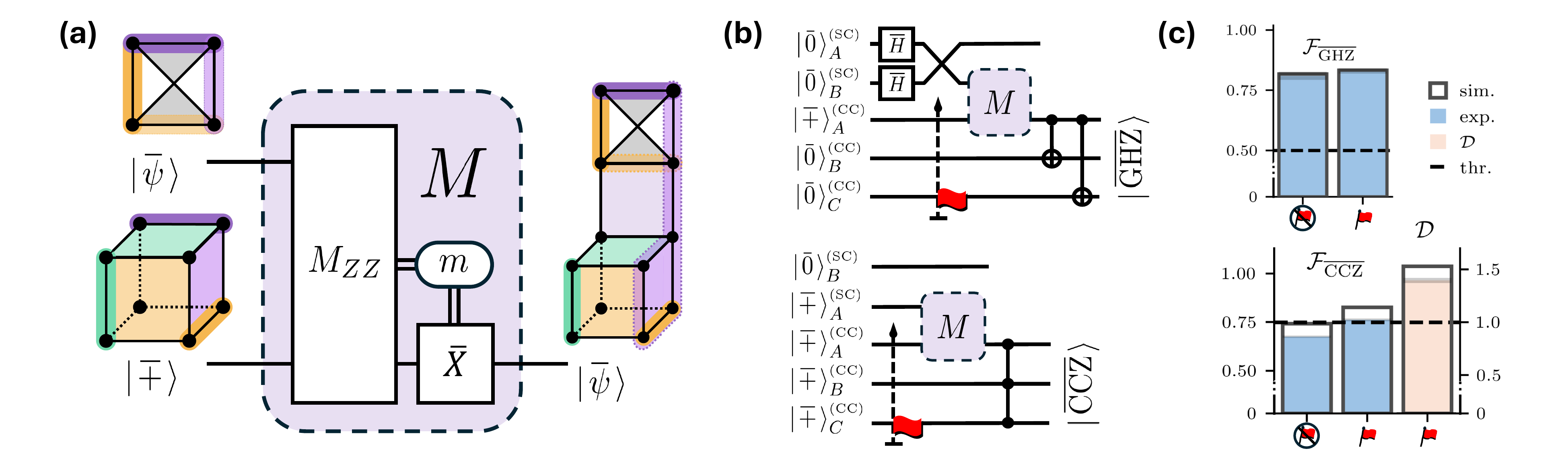} %,trim={1.15cm 19.9cm 1.15cm 0cm},clip]
    \vspace*{-5.0mm}
    \caption{
    \textbf{Lattice surgery and GME preparation.} 
    \textbf{(a)}~Smooth merge $M$ between two codes (e.g., the SC, top left, and the CC bottom left) with logical initial states $\ket{\widebar{\psi}}$ and $\ket{\widebar{+}}$, respectively, realised by measuring $M_{ZZ}$ in Eq.~(\ref{eq:MZZ}) with outcome $m$, and subsequent correction $\Xbar{}$ if $m=-1$, teleporting $\ket{\widebar{\psi}}$ to the merged code (MC, right). 
    \textbf{(b)}~Logical circuits employing transversal gates $\widebar{H}\otimes \widebar{H}$ (up to a $\widebar{\mathrm{SWAP}}$ of~$A$ and~$B$) on the SC, flags on the CC (illustrative flag poles connect the concerned logical qubits, see Appendix~\ref{appendix:flags}), transversal $\widebar{\mathrm{CCZ}}$ and fault-tolerant $\widebar{\mathrm{CNOT}}$ gates (the latter realised by relabelling of trapped-ion qubits) on the MC, and smooth merges $M$ are used to prepare the stabiliser and non-stabiliser GME states $\ket{\widebar{\mathrm{GHZ}}}$ and $\ket{\widebar{\mathrm{CCZ}}}$, respectively, see Sec.~\ref{sec:lattice surgery toolbox}. 
    \textbf{(c)}~Measured (blue) and simulated (grey outlines, depolarising-noise model, see Appendix~\ref{appendix:noise_model}) fidelities $\mathcal{F}\subtiny{0}{-1.5}{\widebar{\mathrm{GHZ}}}$ and $\mathcal{F}\subtiny{0}{-1.5}{\widebar{\mathrm{CCZ}}}$ with $\ket{\widebar{\mathrm{GHZ}}}$ and $\ket{\widebar{\mathrm{CCZ}}}$, respectively, and stabiliser norm $\mathcal{D}$ (peach) obtained from circuits without (left column), and with flags (right/middle column) are shown relative to the respective thresholds (dashed lines) of $0.5$ and $0.75$ for GME and $1$ for non-stabiliserness.
    }
    \label{fig:ghz_merged}
\end{figure*}

Meanwhile, the $[\![8,3,2]\!]$ 3D colour code~(CC)~\cite{KubicaYoshidaPastawski2015, CampbellHoward2017a, CampbellHoward2017b} is defined on eight physical qubits labelled $4,5,\ldots,11$ associated to the corners of a cube. Of the five stabiliser generators, four correspond to $Z$ operators on the four qubits of different cube faces, while the fifth comprises $X$ operators on all eight qubits. The logical operators defining the three encoded qubits $A$, $B$, and $C$ are realised by pairs of $Z$ operators on qubits associated with three orthogonal cube edges, $\Zbar{\!A}\suptiny{0}{0}{\mathrm{(CC)}} =Z_4 Z_6$, $\Zbar{\!B}\suptiny{0}{0}{\mathrm{(CC)}} =Z_8 Z_9$, and $\Zbar{\!C}\suptiny{0}{0}{\mathrm{(CC)}} =Z_7 Z_{11}$, complemented by logical $X$ operators corresponding to $X$ gates on the four physical qubits of cube faces that have one qubit in common with the respective $Z$-operator edges. The logical $\widebar{\mathrm{CZ}}$ and the non-Clifford gate $\widebar{\mathrm{CCZ}}$ can be implemented transversally on the $[\![8,3,2]\!]$ code~\cite{MenendezEtAl2023}, while logical $\widebar{\mathrm{CNOT}}$ gates can be realised fault-tolerantly by permutations on trapped-ion qubits due to their all-to-all connectivity which allows for straightforward relabelling. The physical operations are detailed in Table~\ref{tab:transversal appendix} of Appendix~\ref{appendix:lattice surgery and codes}.\\[-3.0mm]

We call the implementation of a logical operation fault-tolerant if a single physical fault does not lead to an unheralded error outside the detection or correction capability of the specific code: For the distance-two error-detecting codes described above, a single fault must not produce an undetected weight-two logical error. In our demonstration, such faults are either prevented by transversality or qubit permutations (relabelling), or detected by syndrome or flag information and removed by post-selection.\\[-2mm]

{\noindent}\textbf{Lattice surgery}. While no individual QEC code alone can support transversal implementations of a complete universal logical gate set~\cite{EastinKnill2009}, this restriction can be overcome via lattice surgery~\cite{HorsmanEtAl2012, ErhardEtAl2021, NautrupFrisBriegel2017}: A~merge realised by measuring a product of logical operators of both codes allows for the fault-tolerant implementation of logical entangling gates across these codes, while a split can be used to teleport logical information from one code to the other. We perform so-called smooth merges, illustrated in Fig.~\ref{fig:ghz_merged}~(a), corresponding to measurements of 
\begin{align}
M_{ZZ}&:=\,\Zbar{\!A}\suptiny{0}{0}{\mathrm{(SC)}}\otimes\Zbar{\!A}\suptiny{0}{0}{\mathrm{(CC)}} \,=\,Z_0 Z_2 Z_4 Z_6,
\label{eq:MZZ}
\end{align}
realised via $\mathrm{CNOT}$ gates between the qubits in question and an auxiliary qubit initialised in $\ket{0}$ and measured with respect to the computational basis at the end. The stabiliser generators of the resulting $[\![12,4,2]\!]$ merged code (MC) include the seven stabilisers of the initial codes, along with the new stabiliser $M_{ZZ}$, thus encoding four logical qubits. While this merge leaves the logical operators for the initial $B$ and $C$ qubits unchanged, the four logical operators for the two $A$ qubits are replaced by the pair
\begin{subequations}
\begin{align}
\Xbar{\!A}\suptiny{0}{0}{\mathrm{(MC)}} &=\,\Xbar{\!A}\suptiny{0}{0}{\mathrm{(SC)}}\otimes\Xbar{\!A}\suptiny{0}{0}{\mathrm{(CC)}}\,=\,
X_0 X_1 X_4 X_5 X_8 X_9,\\
\Zbar{\!A}\suptiny{0}{0}{\mathrm{(MC)}} &=\,\Zbar{\!A}\suptiny{0}{0}{\mathrm{(SC)}}\,=\,Z_0 Z_2.
\end{align}
\end{subequations}

Preparing initial logical states $\ket{\widebar{\psi}}_A=\alpha\ket{\widebar{0}}+\beta\ket{\widebar{1}}$ and 
$\ket{\widebar{+}}_A=\bigl(\ket{\widebar{0}}+\ket{\widebar{1}}\bigr)/\sqrt{2}$ of the SC and CC, respectively, 
and obtaining the outcome $m=+1$ in the $M_{ZZ}$ measurement results in an entangled state $\alpha\ket{\widebar{00}}+\beta\ket{\widebar{11}}$ of the two original $A$ qubits. The states $\ket{\widebar{00}}$ and $\ket{\widebar{11}}$ represent the computational-basis states of one logical qubit of the MC, thus completing the merge. A consecutive measurement of $\Xbar{\!A}\suptiny{0}{0}{\mathrm{(SC)}} =X_0 X_1$ with outcome~$m_s=+1$ projects into the state $\ket{\widebar{+}}\ket{\widebar{\psi}}$ allowing us to split the codes again. The result is a teleportation of the logical state $\ket{\widebar{\psi}}$ from SC to CC. Measurement outcomes $-1$ with suitable local corrections lead to the same result. Thus, lattice surgery provides access to the transversal logical gates of both codes by transferring the information between them. In particular, a complete universal logical gate set formed by the Hadamard and $\widebar{\mathrm{CCZ}}$ gates~\cite{Aharonov2003} becomes accessible.\\[-4.5mm]

%%%%%%%%%%%%%%%%%%%%%%%%%%%%%%%%%%%%%%%%%%%%%%%%%%%%%%%%%%%%%%%%%%%%%%%%%%%%%%%%
%%%%%%%%%%%%%%%%%%%%%%%%%%%%%%%%%%%%%%%%%%%%%%%%%%%%%%%%%%%%%%%%%%%%%%%%%%%%%%%%

\section{Experimental Setup}
\label{sec:experimental_setup}
\vspace*{-1.5mm}

{\noindent}We now describe the trapped-ion quantum processor used in our experiments, the encoding circuits that prepare logical states for the two code blocks, and the flag-qubit subroutine.\\[-2mm] 

%%%%%%%%%%%%%%%%%%%%%%%%%%%%%%%%%%%%%%%%%%%%%%%%%%%%%%%%%%%%%%%%%%%%%%%%%%%%%%%%

\begin{figure}[!t]
    \centering
    %%%trim={<left> <lower> <right> <upper>}
    \includegraphics[width=0.49\textwidth,trim={0cm 0.35cm 0cm 0cm},clip]{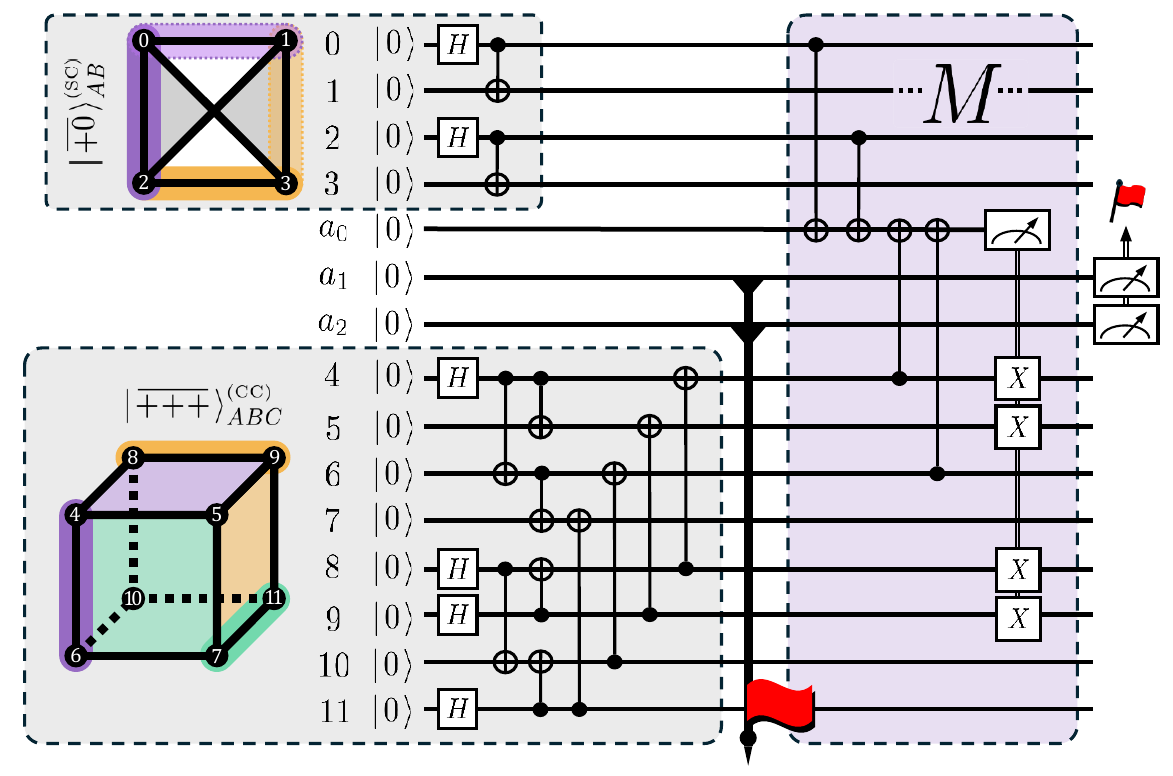}  %
    \vspace*{-4.0mm}
    \caption{\textbf{Physical circuit.} Encoding circuits (grey) for the preparation of logical initial states $\ket{\widebar{+0}}\suptiny{0}{0}{\mathrm{(SC)}}_{AB}$ and $\ket{\widebar{+\!+\!+}}\suptiny{0}{0}{\mathrm{(CC)}}_{ABC}$ of the SC and CC on physical qubits $0,1,2,3$ and $4,\ldots,11$, respectively, along with circuit for the smooth merge $M$ (lavender) using auxiliary qubit $a_0$. Two additional auxiliary qubits $a_1$ and $a_2$ are coupled to the CC (illustrative flag pole represents weight-four measurements on physical qubits ${4,6,9,11}$, see Appendix~\ref{appendix:flag_FT}) and subsequently measured as flags. 
    }
    \label{fig:circuits_diagrams_alt}
\end{figure}

%%%%%%%%%%%%%%%%%%%%%%%%%%%%%%%%%%%%%%%%%%%%%%%%%%%%%%%%%%%%%%%%%%%%%%%%%%%%%%%%

{\noindent}\textbf{Trapped-ion processor.} All experiments are carried out on a trapped-ion quantum processor, as described in~\cite{PogorelovEtAl2021}. Sixteen $^{40}\text{Ca}^+$ ions are trapped in a macroscopic linear Paul trap, where electronic states are controlled via laser pulses. Each ion encodes one physical qubit into the electronic Zeeman levels $\ket{0} = \ket{4^2\text{S}_{1/2}, m_J=-1/2}$ and $\ket{1} = \ket{3^2\text{D}_{5/2}, m_J=-1/2}$, coupled via an optical quadrupole transition at a wavelength of 729\,nm. Coulomb interaction between the ions gives rise to collective motional modes, which are used to mediate entangling operations between any desired pair of qubits. The capability to selectively address individual as well as pairs of ions enables arbitrary single- and two-qubit interactions. The native gate set consists of (i) arbitrary rotations around an axis specified by~$\phi$ by an angle~$\Theta$ of the form $\exp\bigl(-\texttt{i}\tfrac{\Theta}{2}[\cos(\phi)X\!+\!\sin(\phi)Y]\bigr)$, (ii) $Z$ rotations $\exp(-\texttt{i}\tfrac{\Theta}{2}Z)$ corresponding to virtual phase updates implemented by adapting subsequent gates~\cite{mckay_efficient_z_2017}, and (iii) the two-qubit M{\o}lmer{\textendash}S{\o}rensen gate $\exp(\texttt{i}\tfrac{\pi}{4} X\!\otimes\! X)$~\cite{SoerensenMoelmer1999}, equivalent to a $\mathrm{CNOT}$ gate up to local rotations~\cite{maslov_basic_circuit_2017}.\\[-2mm]

{\noindent}\textbf{Encoding circuits.} The encoding circuits for the logical initial states of both code blocks were synthesised using the MQT-QECC toolkit~\cite{PehamEtAl2025} which searches for short Clifford circuits satisfying the stabiliser constraints, see Appendix~\ref{appendix:encoding}. Example preparation circuits for the logical states $\ket{\widebar{+\!0}}\suptiny{0}{0}{\mathrm{(SC)}}_{AB}$ and $\ket{\widebar{+\!+\!+}}\suptiny{0}{0}{\mathrm{(CC)}}_{ABC}$ of the SC and CC, respectively, are shown in Fig.~\ref{fig:circuits_diagrams_alt}. Similar circuits have been used in previous implementations of the CC~\cite{WangEtAl2023}. Neither encoder is fault-tolerant on its own: A single physical fault during the ten colour-code $\mathrm{CNOT}$s can propagate to a weight-two error logically equivalent to a $\widebar{Z}$ operator on one of the encoded qubits.

{\noindent}\textbf{Flag-based fault-tolerance.} We restore fault tolerance of the $[\![8,3,2]\!]$ encoder through the addition of auxiliary flag qubits~\cite{ChaoReichardt2018, ChamberlandBeverland2018}, used to check whether errors have propagated during the encoding~\cite{Postler2022, RyanAndersonEtAl2022}. Due to the QEC-code structure, the total parity of a chosen subset of physical qubits is known{\textemdash}deviations thus indicate a problem. Flag qubits are coupled to such subsets via CNOT gates and finally measured. Shots with a non-trivial outcome (a raised flag) can be removed by post-selection. Tracing single-qubit faults through the ten-CNOT CC encoder  
in Fig.~\ref{fig:circuits_diagrams_alt} identifies six weight-two errors that propagate to logical errors. They can all be detected by applying a flag to qubits $\{4,6,9,11\}$. The SC is prepared without a flag, as the complexity of the additional circuit outweighs that of the encoding circuit, introducing more errors than it can prevent. See Appendix~\ref{appendix:flag_FT} for more details on flags.\\[-5.5mm] 

%%%%%%%%%%%%%%%%%%%%%%%%%%%%%%%%%%%%%%%%%%%%%%%%%%%%%%%%%%%%%%%%%%%%%%%%%%%%%%%%
%%%%%%%%%%%%%%%%%%%%%%%%%%%%%%%%%%%%%%%%%%%%%%%%%%%%%%%%%%%%%%%%%%%%%%%%%%%%%%%%

\section{Lattice-Surgery Toolbox}\label{sec:lattice surgery toolbox}
\vspace*{-1.5mm}

{\noindent}We now describe protocols using fault-tolerantly implemented logical gates and lattice surgery that we employ to prepare complex logical states, the tools we use to verify GME and non-stabiliserness, as well as the $\widebar{\mathrm{CCZ}}$-based resource gadget for implementing continuous rotations on single logical qubits.\\[-2mm]  

{\noindent}\textbf{GHZ state}.\ A paradigmatic example for a state with GME is the three-qubit GHZ state $\ket{\mathrm{GHZ}} = \bigl(\ket{000} + \ket{111}\bigr)/{\sqrt 2}$, which we prepare for three logical qubits by applying a transversal Hadamard gate on the SC, followed by a smooth merge with the CC, shown in Fig.~\ref{fig:ghz_merged}~(a), and two fault-tolerant $\widebar{\mathrm{CNOT}}$ gates, see Fig.~\ref{fig:ghz_merged}~(b). Since the fidelity of every biseparable state with a pure (GME) state $\ket{\psi}$ is bounded by the square of the largest Schmidt coefficient $\lambda_{\mathrm{max}}$ of $\ket{\psi}$ across all bipartitions, a value $\mathcal{F}(\rho,\ket{\psi})=\bra{\psi}\rho\ket{\psi}>\lambda_{\mathrm{max}}^{2}$ detects GME, see, e.g.,~\cite{TothGuhne2005, FriisVitaglianoMalikHuber2019}. For the GHZ state one has $\lambda_{\mathrm{max}}=1/\sqrt{2}$ and the fidelity of the logical GHZ state $\ket{\widebar{\mathrm{GHZ}}}$ can be estimated from the seven expectation values of logical Pauli operators contributing non-trivially to its generalised Bloch decomposition, 
\begin{align}
\mathcal{F}\subtiny{0}{-1.5}{\widebar{\mathrm{GHZ}}} \nl=\nl \tfrac{1}{8}\bigl(1 \!+\! \langle \widebar{X}^{\otimes 3} \rangle \!+\! \!\sum\limits_{i<j} \langle \Zbar{i}\Zbar{j} \rangle 
{-}\langle \widebar{Y}\widebar{Y}\widebar{X} \rangle {-} \langle \widebar{Y}\widebar{X}\widebar{Y} \rangle {-} \langle \widebar{X}\widebar{Y}\widebar{Y} \rangle \bigr).
\label{eq:ghz_witness}\\[-6mm] \nonumber
\end{align}
We experimentally verify the preparation of $\ket{\widebar{\mathrm{GHZ}}}$ on the MC 
by estimating $\mathcal{F}\subtiny{0}{-1.5}{\widebar{\mathrm{GHZ}}}$ from measurements of the seven logical Pauli operators in Eq.~(\ref{eq:ghz_witness}). The resulting fidelities with and without flags on the CC are shown in Fig.~\ref{fig:ghz_merged}~(c). Using flags in the initialisation, we obtain an experimental fidelity of $\mathcal{F}\subtiny{0}{-1.5}{\widebar{\mathrm{GHZ}}}\suptiny{1}{1}{\mathrm{exp.}}=81.0(1.7)\%$, in good agreement with the numerical result $\mathcal{F}\subtiny{0}{-1.5}{\widebar{\mathrm{GHZ}}}\suptiny{1}{1}{\mathrm{sim.}}=81.8(3)\%$. With flags, the experimental fidelity is $\mathcal{F}\subtiny{0}{-1.5}{\widebar{\mathrm{GHZ}}}\suptiny{1}{1}{\mathrm{exp.}}=83.0(1.1)\%$, also consistent with the numerical simulation yielding $\mathcal{F}\subtiny{0}{-1.5}{\widebar{\mathrm{GHZ}}}\suptiny{1}{1}{\mathrm{sim.}}=83.4(2)\%$. Uncertainties refer to the standard deviation obtained from averaging over three independent fidelity measurements, each comprising 2500 shots per operator. The fidelities exceed the threshold of $\lambda_{\max}^2 = 0.5$ by 18{\textendash}30 standard deviations confirming logical GME in both cases.\\[-2.5mm]

%%%%%%%%%%%%%%%%%%%%%%%%%%%%%%%%%%%%%%%%%%%%%%%%%%%%%%%%%%%%%%%%%%%%%%%%%%%%%%%%

{\noindent}\textbf{Non-stabiliser GME state}.\ The hypergraph state~\cite{RossiHuberBrussMacchiavello2013} $\ket{\mathrm{CCZ}}=\mathrm{CCZ}\ket{+\!+\!+}$ is a non-stabiliser and hence magic state: It cannot be reached from $\ket{000}$ by Clifford unitaries. In addition to featuring GME, it thus represents a non-Clifford resource. To showcase the lattice-surgery toolbox, we use a universal gate set while preparing $\ket{\widebar{\mathrm{CCZ}}}$. As shown in  Fig.~\ref{fig:ghz_merged}~(b), the SC and CC are initialised in the states $\ket{\widebar{+0}}\suptiny{0}{0}{\mathrm{(SC)}}_{AB}$ and $\ket{\widebar{+\!+\!+}}\suptiny{0}{0}{\mathrm{(CC)}}_{ABC}$ respectively, using the physical circuit shown in Fig.~\ref{fig:circuits_diagrams_alt}, before the codes are merged. Finally, the transversal non-Clifford gate $\widebar{\mathrm{CCZ}}$ is applied on the MC. 

The non-stabiliserness of the state leads to a larger number of non-zero terms in the Bloch decomposition compared to $\ket{\widebar{\mathrm{GHZ}}}$, such that estimating the fidelity $\mathcal{F}\subtiny{0}{-1.5}{\widebar{\mathrm{CCZ}}}$ with $\ket{\widebar{\mathrm{CCZ}}}$ requires twenty-nine expectation values $\langle \overline{P}_i\rangle$ of logical Pauli operators $\overline{P}_i$, listed in Appendix~\ref{appendix:CCZ_state}. The largest Schmidt coefficient of $\ket{\widebar{\mathrm{CCZ}}}$ is $\lambda_{\max}=\sqrt{3}/2$, such that $\mathcal{F}\subtiny{0}{-1.5}{\widebar{\mathrm{CCZ}}}>3/4$ certifies GME. While the sampling overhead for the many individual expectation values is large, it allows us to go beyond GME certification, calculating the stabiliser norm
\begin{align}
    \mathcal D  &=\,\tfrac{1}{8}\sum_{i} |\langle \overline{P}_i\rangle|\\[-8.5mm]
    \nonumber
\end{align}
as a witness of non-stabiliserness when $\mathcal D>1$~\cite{howard_application_2017}. Moreover, $\mathcal{D}$ bounds the \emph{log-free robustness of magic} $\operatorname{LR} \ge \ln(\mathcal{D})$, a non-stabiliserness quantifier allowing the comparison of our $\ket{\widebar{\mathrm{CCZ}}}$ state with other realisations~\cite{HaugTarabunga2026}.

As shown in Fig.~\ref{fig:ghz_merged}~(c), the flag-based fidelity is $\mathcal{F}\subtiny{0}{-1.5}{\widebar{\mathrm{CCZ}}}\suptiny{1}{1}{\mathrm{exp.}}=76.1(7)\%$ exceeding the GME-certification threshold by one standard deviation, while simulations predict $\mathcal{F}\subtiny{0}{-1.5}{\widebar{\mathrm{CCZ}}}\suptiny{1}{1}{\mathrm{sim.}}=82.7(4)\%$. 
Without the flag, we obtain $\mathcal{F}\subtiny{0}{-1.5}{\widebar{\mathrm{CCZ}}}\suptiny{1}{1}{\mathrm{exp.}}=67.5(6)\%$ and $\mathcal{F}\subtiny{0}{-1.5}{\widebar{\mathrm{CCZ}}}\suptiny{1}{1}{\mathrm{sim.}}=74.2(3)\%$. 
The discrepancy to the simulation is likely due to the simplified noise model further discussed in Appendices~\ref{appendix:noise_model} and~\ref{appendix:CCZ_colour_only}. Using the flag-based data yields a stabiliser norm of $\mathcal{D}=1.396(15)$, which exceeds the stabiliser bound of $1$ by 26 standard deviations resulting in $\mathrm{LR}\ge0.334(11)$, thus certifying the non-stabiliserness of the logical state.\\[-2mm] 

%%%%%%%%%%%%%%%%%%%%%%%%%%%%%%%%%%%%%%%%%%%%%%%%%%%%%%%%%%%%%%%%%%%%%%%%%%%%%%%%

\begin{figure}[ht]
    \centering
    \includegraphics[width=0.9\columnwidth]{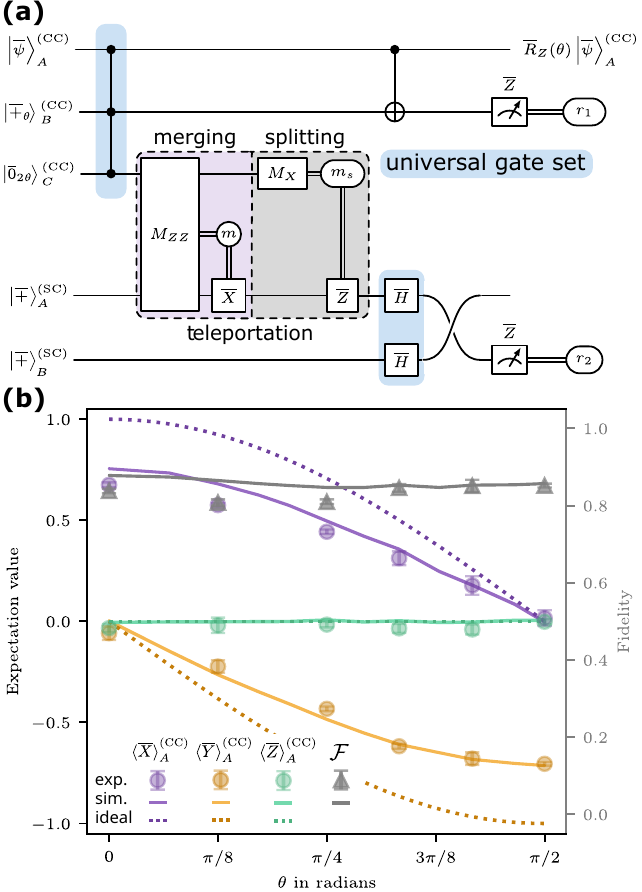}
    \vspace*{-1mm}
    \caption{\textbf{Arbitrary rotation $\widebar{R}_Z(\theta)$ via cross-code lattice surgery.}
    (a)~Teleportation-based rotation gadget: The initial state $\ket{\widebar{\psi}}\suptiny{0}{0}{\mathrm{(CC)}}_{A}$ is prepared together with two auxiliary states $\ket{\widebar{+}_\theta}\suptiny{0}{0}{\mathrm{(CC)}}_{B}$ and $\ket{\widebar{0}_{2\theta}}\suptiny{0}{0}{\mathrm{(CC)}}_{C}$. After applying 
    $\widebar{\mathrm{CCZ}}$ on the CC, the state of the $C$ qubit is teleported onto the SC via $Z$ merging and $X$ splitting before the application of $H$ and a $Z$ measurement. Finally, a $\widebar{\mathrm{CNOT}}$ is applied between CC qubits $A$ and $B$, and the latter is measured. Depending on the outcomes the final state of $A$ is accepted as correctly rotated or rejected. The combination of the $\widebar{\mathrm{CCZ}}$ and Hadamard gates demonstrates the use of a full universal gate set.
    (b)~Measured expectation values $\expval{\hspace*{-1pt}\widebar{X}\hspace*{-1pt}}\suptiny{0}{0}{\mathrm{(CC)}}_{A}$ (purple), $\expval{\hspace*{-1pt}\widebar{Y}\hspace*{-1pt}}\suptiny{0}{0}{\mathrm{(CC)}}_{A}$ (orange), $\expval{\hspace*{-1pt}\widebar{Z}\hspace*{-1pt}}\suptiny{0}{0}{\mathrm{(CC)}}_{A}$ (green), and fidelity (grey) vs.\ rotation angle $\theta$: experimental data (markers), numerical simulation (solid lines), and ideal curves (dotted lines). The experimental uncertainty corresponds to the standard deviation of three distinct measurement sets with 2500 shots per basis each. The average fidelity over the measured rotation angles is $83.4(3)\%$ post-selecting on successful rotations (see Appendix~\ref{appendix:rotation_gadget}).
    }
    \label{fig:rotationLS}
    \vspace*{-4mm}
\end{figure}

%%%%%%%%%%%%%%%%%%%%%%%%%%%%%%%%%%%%%%%%%%%%%%%%%%%%%%%%%%%%%%%%%%%%%%%%%%%%%%%%

{\noindent}\textbf{Arbitrary single-qubit rotations $\widebar{R}_Z(\theta)$} are essential for algorithms such as quantum phase estimation~\cite{Kitaev1995} and variational quantum algorithms~\cite{cerezo_variational_2021}. 
The implementation of continuous $Z$ rotations is typically approximated using a combination of Clifford and $T$ gates~\cite{ross_optimal_2016}, resulting in an error that depends on circuit depth. 
Here, we use a $\widebar{\mathrm{CCZ}}$-based resource gadget of constant depth as an alternative (adapted from~\cite{JerbiEtAl2023}, see Appendix~\ref{appendix:rotation_output}). It produces the rotation without such an inherent error but instead only with a {$75\%$} probability. However, wrong rotations are heralded and thus correctable. The gadget requires two resource qubits in the logical states $\ket{\widebar{+}_\theta}_B=\widebar{R}_Z(\theta)\ket{\widebar{+}}_B$ and $\ket{\widebar{0}_{2\theta}}_C=\widebar{R}_X(2\theta)\ket{\widebar{0}}_C$, which we prepare at the physical level. 
Initialising a rotated resource is considerably less demanding than performing a rotation on an arbitrary state, as the operation is not required to work on all possible input states. The full gadget circuit is shown in Fig.~\ref{fig:rotationLS}~(a) and contains a $\widebar{\mathrm{CCZ}}$ and a $\widebar{\mathrm{CNOT}}$ gate, which are readily available on the CC. As a proof of principle of the full universal gate set~$\{\widebar{H},\widebar{\mathrm{CCZ}}\}$, a final Hadamard gate, effectively changing the measurement basis, is implemented on the SC. The Hadamard gate's role becomes more central in the context of more complex circuits. Furthermore, to switch between codes we apply a merging and splitting procedure, as it would be required in an arbitrary logical circuit showcasing flexible access to the full universal gate set.

The rotation is performed on the initial logical state $\ket{\widebar{+}}\suptiny{0}{0}{\mathrm{(CC)}}_{A}$ and its output is characterised via state tomography after a successful rotation is heralded. The expectation values $\expval{\hspace*{-1pt}\widebar{X}\hspace*{-1pt}}\suptiny{0}{0}{\mathrm{(CC)}}_{A}$, $\expval{\hspace*{-1pt}\widebar{Y}\hspace*{-1pt}}\suptiny{0}{0}{\mathrm{(CC)}}_{A}$, and $\expval{\hspace*{-1pt}\widebar{Z}\hspace*{-1pt}}\suptiny{0}{0}{\mathrm{(CC)}}_{A}$ are shown in Fig.~\ref{fig:rotationLS}~(b) for different rotation angles~$\theta$. Here, experimental data is accompanied by numerical simulations showing good agreement. Overall, the process yields an average state fidelity of $83.4(3)\%$ confirming the implementation of arbitrary logical rotations.\\[-3mm]

%%%%%%%%%%%%%%%%%%%%%%%%%%%%%%%%%%%%%%%%%%%%%%%%%%%%%%%%%%%%%%%%%%%%%%%%%%%%%%%%

\section{Discussion and Conclusion}
\label{sec:discussion}\vspace*{-1.5mm}

{\noindent}We have demonstrated cross-code lattice surgery between a $[\![4,2,2]\!]$ SC and a $[\![8,3,2]\!]$ 3D CC on a trapped-ion quantum processor{\textemdash}to our knowledge, the first experimental demonstration of lattice surgery between codes with complementary, jointly universal, transversal gates. Using this primitive, we prepared logical three-qubit states on the merged $[\![12,4,2]\!]$ code: a GHZ state and the non-stabiliser state $\ket{\widebar{\mathrm{CCZ}}}$, with fidelities exceeding the respective thresholds for verifying GME in both states, and non-stabiliserness of $\ket{\widebar{\mathrm{CCZ}}}$. Finally, we used a $\widebar{\mathrm{CCZ}}$-based gadget to implement an arbitrary logical single-qubit rotation $\widebar{R}_Z(\theta)$, in which the lattice surgery-based teleportation makes use of the full transversally implemented universal gate set $\{\widebar{\mathrm{CCZ}}, \widebar{H}\}$ across codes within a single computation. These results realise elementary building blocks of a lattice-surgery toolbox for universal fault-tolerant quantum computation in the spirit of heterogeneous-code architectures that combine codes with complementary transversal gate sets~\cite{NelsonLandahlBaczewski2025}.

A complementary approach is to stay within a single code family, obtaining the non-Clifford resource by small-angle injection, as in the space-time-efficient analog-rotation (STAR) architecture~\cite{AkahoshiEtAl2024} and its transversal, high-rate quantum low-density parity-check (qLDPC) extension~\cite{IsmailEtAl2026}. More broadly, large-scale architectures typically employ lattice surgery as the measurement layer of Pauli-based computation and supply the non-Clifford resource through distilled, injected magic states~\cite{Litinski2019, WebsterEtAl2026, YoderEtAl2025}. The merge demonstrated here plays a different role. Rather than serving as a measurement primitive that consumes externally supplied magic, it generates the non-Clifford resource in the form of a transversal gate in situ, removing the magic-state factory that dominates the overhead of Pauli-based-computation architectures~\cite{Litinski2019}, at the cost of the presently non-scalable $[\![8,3,2]\!]$ $\widebar{\mathrm{CCZ}}$.

Most operations used here can be scaled fault-tolerantly: lattice surgery for larger code distances requires extending the measured boundary, flag-based preparation can generally make small fault sets detectable, and transversal or permutation-based Clifford operations persist in suitable code families. Indeed, Appendix~\ref{appendix:412 results} shows how performance can even be improved when replacing one of the codes, e.g., the $[[4,2,2]]$ with a $[[4,1,2]]$ SC. The one non-scalable ingredient in the present implementation is the specific transversal $\widebar{\mathrm{CCZ}}$ gate. The gate is fault-tolerant for the $[[8,3,2]]$ code but is not an instance of a transversal gate on a scalable code family. Scaling it would require replacing this primitive by a distance-growing code with a fault-tolerant non-Clifford gate, e.g., stacked 3D surface~\cite{VasmerBrowne2019} or gauge colour codes~\cite{Bombin2015} (see Appendix~\ref{appendix:transversality_merged}), leaving the lattice-surgery framework essentially unchanged.

The cross-code interface used here is a direct joint-stabiliser measurement between code blocks. Alternatively, heterogeneous blocks can be connected via a shared ancilla bus~\cite{SteinEtAl2024}. Either way, upscaling codes predominantly changes boundary engineering rather than protocol structure. For example, increasing the distance~$d$ of the SC, enlarges the merge boundary from a weight-4 stabiliser to a string of $O(d)$ stabilisers~\cite{HorsmanEtAl2012, LandahlRyanAnderson2014, BoedekerEtAl2026}. Similar changes would apply to higher-distance 3D CCs. In such a combined architecture, one block is typically assigned to store logical information, e.g., in a high-rate qLDPC memory~\cite{XuEtAl2023}, while a processor block, e.g., a 3D CC, supplies transversal non-Clifford gates. However, replacing the $[\![4,2,2]\!]$ block with a qLDPC memory is harder than just increasing distance: qLDPC codes lack the local 2D boundary structure that makes SC merges low-weight, and finding low-weight joint stabilisers becomes code specific~\cite{ChakrabortyGottesman2026}. In this sense the merge primitive is code-agnostic, while the boundary supporting it is not. Even without changing codes, the same primitive supports immediate multi-block extensions on current hardware: merging two $[\![4,2,2]\!]$ SC memories with an $[\![8,3,2]\!]$ CC processor (18 qubits, $[\![16,3,2]\!]$ merged code) implements a $\widebar{\mathrm{CZ}}$ between two teleported logical qubits, and three SC memories plus the CC processor ($\sim 23$ qubits) implements a transversal $\widebar{\mathrm{CCZ}}$ where all three logical qubits are drawn from SC memory.

Three near-term experiments would extend the present results in directions the protocols we have presented here have not fully probed yet. First, terminating the GHZ protocol with an $\widebar{X}$ split would yield a GHZ state with one logical qubit on the SC and two on the CC. This is the resource any heterogeneous-code computation must carry across the boundary at some stage, and a high-fidelity demonstration is well within the reach of present hardware. 
Second, the rotation gadget's terminal Hadamard acts immediately before measurement and could in principle be replaced by an $\widebar{X}$ measurement on the CC. Composing two $\widebar{\mathrm{CCZ}}$ stages separated by a transversal $\widebar{H}$ \emph{in the middle} of a logical computation{\textemdash}where the Hadamard acts on a state subsequently transformed by further gates{\textemdash}would isolate cross-code lattice surgery as a strict necessity rather than a chosen demonstration. Third, any sustained computation beyond Clifford gates alone invokes the merge primitive repeatedly. Characterising how merge errors accumulate and whether the merge fidelity is reproducible across cycles would need to be tested in a multi-cycle experiment. While many challenges remain, our work provides key milestones showing that cross-code lattice surgery may provide a foundation for future quantum computing architectures.

%%%%%%%%%%%%%%%%%%%%%%%%%%%%%%%%%%%%%%%%%%%%%%%%%%%%%%%%%%%%%%%%

\bibliographystyle{apsrev4-1fixed_with_article_titles_full_names_new}
\bibliography{references}

\begin{thebibliography}{85}%
\makeatletter
\providecommand \@ifxundefined [1]{%
 \@ifx{#1\undefined}
}%
\providecommand \@ifnum [1]{%
 \ifnum #1\expandafter \@firstoftwo
 \else \expandafter \@secondoftwo
 \fi
}%
\providecommand \@ifx [1]{%
 \ifx #1\expandafter \@firstoftwo
 \else \expandafter \@secondoftwo
 \fi
}%
\providecommand \natexlab [1]{#1}%
\providecommand \enquote  [1]{#1}%
\providecommand \bibnamefont  [1]{#1}%
\providecommand \bibfnamefont [1]{#1}%
\providecommand \citenamefont [1]{#1}%
\providecommand \href@noop [0]{\@secondoftwo}%
\providecommand \href [0]{\begingroup \@sanitize@url \@href}%
\providecommand \@href[1]{\@@startlink{#1}\@@href}%
\providecommand \@@href[1]{\endgroup#1\@@endlink}%
\providecommand \@sanitize@url [0]{\catcode `\\12\catcode `\$12\catcode
  `\&12\catcode `\#12\catcode `\^12\catcode `\_12\catcode `\%12\relax}%
\providecommand \@@startlink[1]{}%
\providecommand \@@endlink[0]{}%
\providecommand \url  [0]{\begingroup\@sanitize@url \@url }%
\providecommand \@url [1]{\endgroup\@href {#1}{\urlprefix }}%
\providecommand \urlprefix  [0]{URL }%
\providecommand \Eprint [0]{\href }%
\providecommand \doibase [0]{https://doi.org/}%
\providecommand \selectlanguage [0]{\@gobble}%
\providecommand \bibinfo  [0]{\@secondoftwo}%
\providecommand \bibfield  [0]{\@secondoftwo}%
\providecommand \translation [1]{[#1]}%
\providecommand \BibitemOpen [0]{}%
\providecommand \bibitemStop [0]{}%
\providecommand \bibitemNoStop [0]{.\EOS\space}%
\providecommand \EOS [0]{\spacefactor3000\relax}%
\providecommand \BibitemShut  [1]{\csname bibitem#1\endcsname}%
\let\auto@bib@innerbib\@empty
%</preamble>
\bibitem [{\citenamefont {Jozsa}\ and\ \citenamefont
  {Linden}(2003)}]{JozsaLinden2003}%
  \BibitemOpen
  \bibfield  {author} {\bibinfo {author} {\bibfnamefont {Richard}\ \bibnamefont
  {Jozsa}}\ and\ \bibinfo {author} {\bibfnamefont {Noah}\ \bibnamefont
  {Linden}},\ }\emph {\enquote {\bibinfo {title} {{On the role of entanglement
  in quantum-computational speed-up}},}\ }\href
  {https://doi.org/10.1098/rspa.2002.1097} {\bibfield  {journal} {\bibinfo
  {journal} {Proc. R. Soc. A}\ }\textbf {\bibinfo {volume} {459}},\ \bibinfo
  {pages} {2011{\textendash}2032} (\bibinfo {year} {2003})},\ \Eprint
  {http://arxiv.org/abs/quant-ph/0201143} {arXiv:quant-ph/0201143}\BibitemShut
  {NoStop}%
\bibitem [{\citenamefont {Preskill}(1997)}]{Preskill1997}%
  \BibitemOpen
  \bibfield  {author} {\bibinfo {author} {\bibfnamefont {John}\ \bibnamefont
  {Preskill}},\ }\emph {\enquote {\bibinfo {title} {{Fault-tolerant quantum
  computation}},}\ }in\ \href {https://doi.org/10.1142/9789812385253_0008}
  {\emph {\bibinfo {booktitle} {Introduction to Quantum Computation}}},\
  \bibinfo {editor} {edited by\ \bibinfo {editor} {\bibfnamefont {H.-K.}\
  \bibnamefont {Lo}}, \bibinfo {editor} {\bibfnamefont {Sandu}\ \bibnamefont
  {Popescu}}, \ and\ \bibinfo {editor} {\bibfnamefont {T.~P.}\ \bibnamefont
  {Spiller}}}\ (\bibinfo  {publisher} {World-Scientific},\ \bibinfo {year}
  {1997})\ Chap.~\bibinfo {chapter} {8}, p.\ \bibinfo {pages}
  {213{\textendash}269},\ \Eprint {http://arxiv.org/abs/quant-ph/9712048}
  {arXiv:quant-ph/9712048}\BibitemShut {NoStop}%
\bibitem [{\citenamefont {Nielsen}\ and\ \citenamefont
  {Chuang}(2010)}]{NielsenChuang2010}%
  \BibitemOpen
  \bibfield  {author} {\bibinfo {author} {\bibfnamefont {Michael~A.}\
  \bibnamefont {Nielsen}}\ and\ \bibinfo {author} {\bibfnamefont {Isaac~L.}\
  \bibnamefont {Chuang}},\ }\href {https://doi.org/10.1017/CBO9780511976667}
  {\emph {\bibinfo {title} {{Quantum Computation and Quantum Information}}}},\
  \bibinfo {edition} {{10th Anniversary Edition}}\ ed.\ (\bibinfo  {publisher}
  {Cambridge University Press},\ \bibinfo {address} {Cambridge, U.K.},\
  \bibinfo {year} {2010})\BibitemShut {NoStop}%
\bibitem [{\citenamefont {Lidar}\ and\ \citenamefont
  {Brun}(2013)}]{LidarBrun2013}%
  \BibitemOpen
  \bibfield  {author} {\bibinfo {author} {\bibfnamefont {Daniel~A.}\
  \bibnamefont {Lidar}}\ and\ \bibinfo {author} {\bibfnamefont {Todd~A.}\
  \bibnamefont {Brun}},\ }\href {https://doi.org/10.1017/CBO9781139034807}
  {\emph {\bibinfo {title} {{Quantum Error Correction}}}}\ (\bibinfo
  {publisher} {Cambridge University Press},\ \bibinfo {address} {Cambridge},\
  \bibinfo {year} {2013})\BibitemShut {NoStop}%
\bibitem [{\citenamefont {Eastin}\ and\ \citenamefont
  {Knill}(2009)}]{EastinKnill2009}%
  \BibitemOpen
  \bibfield  {author} {\bibinfo {author} {\bibfnamefont {Bryan}\ \bibnamefont
  {Eastin}}\ and\ \bibinfo {author} {\bibfnamefont {Emanuel}\ \bibnamefont
  {Knill}},\ }\emph {\enquote {\bibinfo {title} {{Restrictions on Transversal
  Encoded Quantum Gate Sets}},}\ }\href
  {https://doi.org/10.1103/PhysRevLett.102.110502} {\bibfield  {journal}
  {\bibinfo  {journal} {Phys. Rev. Lett.}\ }\textbf {\bibinfo {volume} {102}},\
  \bibinfo {pages} {110502} (\bibinfo {year} {2009})},\ \Eprint
  {http://arxiv.org/abs/0811.4262} {arXiv:0811.4262}\BibitemShut {NoStop}%
\bibitem [{\citenamefont {Bravyi}\ and\ \citenamefont
  {Kitaev}(2005)}]{BravyiKitaev2005}%
  \BibitemOpen
  \bibfield  {author} {\bibinfo {author} {\bibfnamefont {Sergey}\ \bibnamefont
  {Bravyi}}\ and\ \bibinfo {author} {\bibfnamefont {Alexei}\ \bibnamefont
  {Kitaev}},\ }\emph {\enquote {\bibinfo {title} {{Universal Quantum
  Computation with Ideal Clifford Gates and Noisy Ancillas}},}\ }\href
  {https://doi.org/10.1103/PhysRevA.71.022316} {\bibfield  {journal} {\bibinfo
  {journal} {Phys. Rev. A}\ }\textbf {\bibinfo {volume} {71}},\ \bibinfo
  {pages} {022316} (\bibinfo {year} {2005})},\ \Eprint
  {http://arxiv.org/abs/quant-ph/0403025} {arXiv:quant-ph/0403025}\BibitemShut
  {NoStop}%
\bibitem [{\citenamefont {Beverland}\ \emph {et~al.}(2021)\citenamefont
  {Beverland}, \citenamefont {Kubica},\ and\ \citenamefont
  {Svore}}]{BeverlandEtAl2021}%
  \BibitemOpen
  \bibfield  {author} {\bibinfo {author} {\bibfnamefont {Michael~E.}\
  \bibnamefont {Beverland}}, \bibinfo {author} {\bibfnamefont {Aleksander}\
  \bibnamefont {Kubica}}, \ and\ \bibinfo {author} {\bibfnamefont {Krysta~M.}\
  \bibnamefont {Svore}},\ }\emph {\enquote {\bibinfo {title} {{Cost of
  Universality: A Comparative Study of the Overhead of State Distillation and
  Code Switching with Color Codes}},}\ }\href
  {https://doi.org/10.1103/PRXQuantum.2.020341} {\bibfield  {journal} {\bibinfo
   {journal} {PRX Quantum}\ }\textbf {\bibinfo {volume} {2}},\ \bibinfo {pages}
  {020341} (\bibinfo {year} {2021})},\ \Eprint
  {http://arxiv.org/abs/2101.02211} {arXiv:2101.02211}\BibitemShut {NoStop}%
\bibitem [{\citenamefont {Gidney}\ \emph {et~al.}(2024)\citenamefont {Gidney},
  \citenamefont {Shutty},\ and\ \citenamefont {Jones}}]{GidneyEtAl2024}%
  \BibitemOpen
  \bibfield  {author} {\bibinfo {author} {\bibfnamefont {Craig}\ \bibnamefont
  {Gidney}}, \bibinfo {author} {\bibfnamefont {Noah}\ \bibnamefont {Shutty}}, \
  and\ \bibinfo {author} {\bibfnamefont {Cody}\ \bibnamefont {Jones}},\
  }\href@noop {} {\emph {\enquote {\bibinfo {title} {{Magic State Cultivation:
  Growing T States as Cheap as CNOT Gates}},}\ }}\Eprint
  {http://arxiv.org/abs/2409.17595} {arXiv:2409.17595} [quant-ph] (\bibinfo
  {year} {2024})\BibitemShut {NoStop}%
\bibitem [{\citenamefont {Bomb\'{\i}n}(2016)}]{Bombin2016}%
  \BibitemOpen
  \bibfield  {author} {\bibinfo {author} {\bibfnamefont {H\'{e}ctor}\
  \bibnamefont {Bomb\'{\i}n}},\ }\emph {\enquote {\bibinfo {title}
  {{Dimensional Jump in Quantum Error Correction}},}\ }\href
  {https://doi.org/10.1088/1367-2630/18/4/043038} {\bibfield  {journal}
  {\bibinfo  {journal} {New J. Phys.}\ }\textbf {\bibinfo {volume} {18}},\
  \bibinfo {pages} {043038} (\bibinfo {year} {2016})},\ \Eprint
  {http://arxiv.org/abs/1412.5079} {arXiv:1412.5079}\BibitemShut {NoStop}%
\bibitem [{\citenamefont {Butt}\ \emph {et~al.}(2024)\citenamefont {Butt},
  \citenamefont {Heu{\ss}en}, \citenamefont {Rispler},\ and\ \citenamefont
  {M{\"u}ller}}]{ButtEtAl2024}%
  \BibitemOpen
  \bibfield  {author} {\bibinfo {author} {\bibfnamefont {Friederike}\
  \bibnamefont {Butt}}, \bibinfo {author} {\bibfnamefont {Sascha}\ \bibnamefont
  {Heu{\ss}en}}, \bibinfo {author} {\bibfnamefont {Manuel}\ \bibnamefont
  {Rispler}}, \ and\ \bibinfo {author} {\bibfnamefont {Markus}\ \bibnamefont
  {M{\"u}ller}},\ }\emph {\enquote {\bibinfo {title} {{Fault-Tolerant Code
  Switching Protocols for Near-Term Quantum Processors}},}\ }\href
  {https://doi.org/10.1103/PRXQuantum.5.020345} {\bibfield  {journal} {\bibinfo
   {journal} {PRX Quantum}\ }\textbf {\bibinfo {volume} {5}},\ \bibinfo {pages}
  {020345} (\bibinfo {year} {2024})},\ \Eprint
  {http://arxiv.org/abs/2306.17686} {arXiv:2306.17686}\BibitemShut {NoStop}%
\bibitem [{\citenamefont {Pogorelov}\ \emph {et~al.}(2025)\citenamefont
  {Pogorelov}, \citenamefont {Butt}, \citenamefont {Postler}, \citenamefont
  {Marciniak}, \citenamefont {Schindler}, \citenamefont {M{\"u}ller},\ and\
  \citenamefont {Monz}}]{PogorelovEtAl2025}%
  \BibitemOpen
  \bibfield  {author} {\bibinfo {author} {\bibfnamefont {Ivan}\ \bibnamefont
  {Pogorelov}}, \bibinfo {author} {\bibfnamefont {Friederike}\ \bibnamefont
  {Butt}}, \bibinfo {author} {\bibfnamefont {Lukas}\ \bibnamefont {Postler}},
  \bibinfo {author} {\bibfnamefont {Christian~D.}\ \bibnamefont {Marciniak}},
  \bibinfo {author} {\bibfnamefont {Philipp}\ \bibnamefont {Schindler}},
  \bibinfo {author} {\bibfnamefont {Markus}\ \bibnamefont {M{\"u}ller}}, \ and\
  \bibinfo {author} {\bibfnamefont {Thomas}\ \bibnamefont {Monz}},\ }\emph
  {\enquote {\bibinfo {title} {{Experimental Fault-Tolerant Code Switching}},}\
  }\href {https://doi.org/10.1038/s41567-024-02727-2} {\bibfield  {journal}
  {\bibinfo  {journal} {Nat. Phys.}\ }\textbf {\bibinfo {volume} {21}},\
  \bibinfo {pages} {298{\textendash}303} (\bibinfo {year} {2025})},\ \Eprint
  {http://arxiv.org/abs/2403.13732} {arXiv:2403.13732}\BibitemShut {NoStop}%
\bibitem [{\citenamefont {Wu}\ \emph {et~al.}(2026)\citenamefont {Wu},
  \citenamefont {Zhong}, \citenamefont {Brun},\ and\ \citenamefont
  {Lidar}}]{WuZhongBrunLidar2026}%
  \BibitemOpen
  \bibfield  {author} {\bibinfo {author} {\bibfnamefont {Shixin}\ \bibnamefont
  {Wu}}, \bibinfo {author} {\bibfnamefont {Dawei}\ \bibnamefont {Zhong}},
  \bibinfo {author} {\bibfnamefont {Todd~A.}\ \bibnamefont {Brun}}, \ and\
  \bibinfo {author} {\bibfnamefont {Daniel~A.}\ \bibnamefont {Lidar}},\
  }\href@noop {} {\emph {\enquote {\bibinfo {title} {{Universal Weakly
  Fault-Tolerant Quantum Computation via Code Switching in the $[[8,3,2]]$
  Code}},}\ }}\Eprint {http://arxiv.org/abs/2603.15610} {arXiv:2603.15610}
  [quant-ph] (\bibinfo {year} {2026})\BibitemShut {NoStop}%
\bibitem [{\citenamefont {Horsman}\ \emph {et~al.}(2012)\citenamefont
  {Horsman}, \citenamefont {Fowler}, \citenamefont {Devitt},\ and\
  \citenamefont {Van~Meter}}]{HorsmanEtAl2012}%
  \BibitemOpen
  \bibfield  {author} {\bibinfo {author} {\bibfnamefont {Clare}\ \bibnamefont
  {Horsman}}, \bibinfo {author} {\bibfnamefont {Austin~G.}\ \bibnamefont
  {Fowler}}, \bibinfo {author} {\bibfnamefont {Simon}\ \bibnamefont {Devitt}},
  \ and\ \bibinfo {author} {\bibfnamefont {Rodney}\ \bibnamefont {Van~Meter}},\
  }\emph {\enquote {\bibinfo {title} {{Surface Code Quantum Computing by
  Lattice Surgery}},}\ }\href {https://doi.org/10.1088/1367-2630/14/12/123011}
  {\bibfield  {journal} {\bibinfo  {journal} {New J. Phys.}\ }\textbf {\bibinfo
  {volume} {14}},\ \bibinfo {pages} {123011} (\bibinfo {year} {2012})},\
  \Eprint {http://arxiv.org/abs/1111.4022} {arXiv:1111.4022}\BibitemShut
  {NoStop}%
\bibitem [{\citenamefont {Erhard}\ \emph {et~al.}(2021)\citenamefont {Erhard},
  \citenamefont {Poulsen~Nautrup}, \citenamefont {Meth}, \citenamefont
  {Postler}, \citenamefont {Stricker}, \citenamefont {Stadler}, \citenamefont
  {Negnevitsky}, \citenamefont {Ringbauer}, \citenamefont {Schindler},
  \citenamefont {Briegel}, \citenamefont {Blatt}, \citenamefont {Friis},\ and\
  \citenamefont {Monz}}]{ErhardEtAl2021}%
  \BibitemOpen
  \bibfield  {author} {\bibinfo {author} {\bibfnamefont {Alexander}\
  \bibnamefont {Erhard}}, \bibinfo {author} {\bibfnamefont {Hendrik}\
  \bibnamefont {Poulsen~Nautrup}}, \bibinfo {author} {\bibfnamefont {Michael}\
  \bibnamefont {Meth}}, \bibinfo {author} {\bibfnamefont {Lukas}\ \bibnamefont
  {Postler}}, \bibinfo {author} {\bibfnamefont {Roman}\ \bibnamefont
  {Stricker}}, \bibinfo {author} {\bibfnamefont {Martin}\ \bibnamefont
  {Stadler}}, \bibinfo {author} {\bibfnamefont {Vlad}\ \bibnamefont
  {Negnevitsky}}, \bibinfo {author} {\bibfnamefont {Martin}\ \bibnamefont
  {Ringbauer}}, \bibinfo {author} {\bibfnamefont {Philipp}\ \bibnamefont
  {Schindler}}, \bibinfo {author} {\bibfnamefont {Hans~J.}\ \bibnamefont
  {Briegel}}, \bibinfo {author} {\bibfnamefont {Rainer}\ \bibnamefont {Blatt}},
  \bibinfo {author} {\bibfnamefont {Nicolai}\ \bibnamefont {Friis}}, \ and\
  \bibinfo {author} {\bibfnamefont {Thomas}\ \bibnamefont {Monz}},\ }\emph
  {\enquote {\bibinfo {title} {{Entangling logical qubits with lattice
  surgery}},}\ }\href {https://doi.org/10.1038/s41586-020-03079-6} {\bibfield
  {journal} {\bibinfo  {journal} {Nature}\ }\textbf {\bibinfo {volume} {589}},\
  \bibinfo {pages} {220{\textendash}224} (\bibinfo {year} {2021})},\ \Eprint
  {http://arxiv.org/abs/2006.03071} {arXiv:2006.03071}\BibitemShut {NoStop}%
\bibitem [{\citenamefont {Poulsen~Nautrup}\ \emph {et~al.}(2017)\citenamefont
  {Poulsen~Nautrup}, \citenamefont {Friis},\ and\ \citenamefont
  {Briegel}}]{NautrupFrisBriegel2017}%
  \BibitemOpen
  \bibfield  {author} {\bibinfo {author} {\bibfnamefont {Hendrik}\ \bibnamefont
  {Poulsen~Nautrup}}, \bibinfo {author} {\bibfnamefont {Nicolai}\ \bibnamefont
  {Friis}}, \ and\ \bibinfo {author} {\bibfnamefont {Hans~J.}\ \bibnamefont
  {Briegel}},\ }\emph {\enquote {\bibinfo {title} {{Fault-Tolerant Interface
  Between Quantum Memories and Quantum Processors}},}\ }\href
  {https://doi.org/10.1038/s41467-017-01418-2} {\bibfield  {journal} {\bibinfo
  {journal} {Nat. Commun.}\ }\textbf {\bibinfo {volume} {8}},\ \bibinfo {pages}
  {1321} (\bibinfo {year} {2017})},\ \Eprint {http://arxiv.org/abs/1609.08062}
  {arXiv:1609.08062}\BibitemShut {NoStop}%
\bibitem [{\citenamefont {Litinski}(2019)}]{Litinski2019}%
  \BibitemOpen
  \bibfield  {author} {\bibinfo {author} {\bibfnamefont {Daniel}\ \bibnamefont
  {Litinski}},\ }\emph {\enquote {\bibinfo {title} {{A Game of Surface Codes:
  Large-Scale Quantum Computing with Lattice Surgery}},}\ }\href
  {https://doi.org/10.22331/q-2019-03-05-128} {\bibfield  {journal} {\bibinfo
  {journal} {Quantum}\ }\textbf {\bibinfo {volume} {3}},\ \bibinfo {pages}
  {128} (\bibinfo {year} {2019})},\ \Eprint {http://arxiv.org/abs/1808.02892}
  {arXiv:1808.02892}\BibitemShut {NoStop}%
\bibitem [{\citenamefont {T\'{o}th}\ and\ \citenamefont
  {G\"{u}hne}(2005)}]{TothGuhne2005}%
  \BibitemOpen
  \bibfield  {author} {\bibinfo {author} {\bibfnamefont {G\'{e}za}\
  \bibnamefont {T\'{o}th}}\ and\ \bibinfo {author} {\bibfnamefont {Otfried}\
  \bibnamefont {G\"{u}hne}},\ }\emph {\enquote {\bibinfo {title} {{Detecting
  Genuine Multipartite Entanglement with Two Local Measurements}},}\ }\href
  {https://doi.org/10.1103/PhysRevLett.94.060501} {\bibfield  {journal}
  {\bibinfo  {journal} {Phys. Rev. Lett.}\ }\textbf {\bibinfo {volume} {94}},\
  \bibinfo {pages} {060501} (\bibinfo {year} {2005})},\ \Eprint
  {http://arxiv.org/abs/quant-ph/0405165} {arXiv:quant-ph/0405165}\BibitemShut
  {NoStop}%
\bibitem [{\citenamefont {Friis}\ \emph {et~al.}(2019)\citenamefont {Friis},
  \citenamefont {Vitagliano}, \citenamefont {Malik},\ and\ \citenamefont
  {Huber}}]{FriisVitaglianoMalikHuber2019}%
  \BibitemOpen
  \bibfield  {author} {\bibinfo {author} {\bibfnamefont {Nicolai}\ \bibnamefont
  {Friis}}, \bibinfo {author} {\bibfnamefont {Giuseppe}\ \bibnamefont
  {Vitagliano}}, \bibinfo {author} {\bibfnamefont {Mehul}\ \bibnamefont
  {Malik}}, \ and\ \bibinfo {author} {\bibfnamefont {Marcus}\ \bibnamefont
  {Huber}},\ }\emph {\enquote {\bibinfo {title} {{Entanglement Certification
  From Theory to Experiment}},}\ }\href
  {https://doi.org/10.1038/s42254-018-0003-5} {\bibfield  {journal} {\bibinfo
  {journal} {Nat. Rev. Phys.}\ }\textbf {\bibinfo {volume} {1}},\ \bibinfo
  {pages} {72{\textendash}87} (\bibinfo {year} {2019})},\ \Eprint
  {http://arxiv.org/abs/1906.10929} {arXiv:1906.10929}\BibitemShut {NoStop}%
\bibitem [{\citenamefont {Bertlmann}\ and\ \citenamefont
  {Friis}(2023)}]{BertlmannFriis2023}%
  \BibitemOpen
  \bibfield  {author} {\bibinfo {author} {\bibfnamefont {Reinhold~A.}\
  \bibnamefont {Bertlmann}}\ and\ \bibinfo {author} {\bibfnamefont {Nicolai}\
  \bibnamefont {Friis}},\ }\href
  {https://doi.org/10.1093/oso/9780199683338.001.0001} {\emph {\bibinfo {title}
  {Modern Quantum Theory {\textendash} From Quantum Mechanics to Entanglement
  and Quantum Information}}}\ (\bibinfo  {publisher} {Oxford University
  Press},\ \bibinfo {address} {Oxford, U.K.},\ \bibinfo {year} {2023})\
  \bibinfo {note} {{C}hapter~18}\BibitemShut {NoStop}%
\bibitem [{\citenamefont {Epping}\ \emph {et~al.}(2017)\citenamefont {Epping},
  \citenamefont {Kampermann}, \citenamefont {Macchiavello},\ and\ \citenamefont
  {Bru{\ss}}}]{EppingKampermannMacchiavelloBruss2017}%
  \BibitemOpen
  \bibfield  {author} {\bibinfo {author} {\bibfnamefont {Michael}\ \bibnamefont
  {Epping}}, \bibinfo {author} {\bibfnamefont {Hermann}\ \bibnamefont
  {Kampermann}}, \bibinfo {author} {\bibfnamefont {Chiara}\ \bibnamefont
  {Macchiavello}}, \ and\ \bibinfo {author} {\bibfnamefont {Dagmar}\
  \bibnamefont {Bru{\ss}}},\ }\emph {\enquote {\bibinfo {title} {Multi-partite
  entanglement can speed up quantum key distribution in networks},}\ }\href
  {https://doi.org/10.1088/1367-2630/aa8487} {\bibfield  {journal} {\bibinfo
  {journal} {New J. Phys.}\ }\textbf {\bibinfo {volume} {19}},\ \bibinfo
  {pages} {093012} (\bibinfo {year} {2017})},\ \Eprint
  {http://arxiv.org/abs/1612.05585} {arXiv:1612.05585}\BibitemShut {NoStop}%
\bibitem [{\citenamefont {B{\"a}uml}\ and\ \citenamefont
  {Azuma}(2017)}]{BaeumlAzuma2017}%
  \BibitemOpen
  \bibfield  {author} {\bibinfo {author} {\bibfnamefont {Stefan}\ \bibnamefont
  {B{\"a}uml}}\ and\ \bibinfo {author} {\bibfnamefont {Koji}\ \bibnamefont
  {Azuma}},\ }\emph {\enquote {\bibinfo {title} {Fundamental limitation on
  quantum broadcast networks},}\ }\href
  {https://doi.org/10.1088/2058-9565/aa6d3c} {\bibfield  {journal} {\bibinfo
  {journal} {Quantum Sci. Technol.}\ }\textbf {\bibinfo {volume} {2}},\
  \bibinfo {pages} {024004} (\bibinfo {year} {2017})},\ \Eprint
  {http://arxiv.org/abs/1609.03994} {arXiv:1609.03994}\BibitemShut {NoStop}%
\bibitem [{\citenamefont {Pivoluska}\ \emph {et~al.}(2018)\citenamefont
  {Pivoluska}, \citenamefont {Huber},\ and\ \citenamefont
  {Malik}}]{PivoluskaHuberMalik2018}%
  \BibitemOpen
  \bibfield  {author} {\bibinfo {author} {\bibfnamefont {Matej}\ \bibnamefont
  {Pivoluska}}, \bibinfo {author} {\bibfnamefont {Marcus}\ \bibnamefont
  {Huber}}, \ and\ \bibinfo {author} {\bibfnamefont {Mehul}\ \bibnamefont
  {Malik}},\ }\emph {\enquote {\bibinfo {title} {Layered quantum key
  distribution},}\ }\href {https://doi.org/10.1103/PhysRevA.97.032312}
  {\bibfield  {journal} {\bibinfo  {journal} {Phys. Rev. A}\ }\textbf {\bibinfo
  {volume} {97}},\ \bibinfo {pages} {032312} (\bibinfo {year} {2018})},\
  \Eprint {http://arxiv.org/abs/1709.00377} {arXiv:1709.00377}\BibitemShut
  {NoStop}%
\bibitem [{\citenamefont {Ribeiro}\ \emph {et~al.}(2018)\citenamefont
  {Ribeiro}, \citenamefont {Murta},\ and\ \citenamefont
  {Wehner}}]{RibeiroMurtaWehner2018}%
  \BibitemOpen
  \bibfield  {author} {\bibinfo {author} {\bibfnamefont {J{\'e}r{\'e}my}\
  \bibnamefont {Ribeiro}}, \bibinfo {author} {\bibfnamefont {Gl{\'a}ucia}\
  \bibnamefont {Murta}}, \ and\ \bibinfo {author} {\bibfnamefont {Stephanie}\
  \bibnamefont {Wehner}},\ }\emph {\enquote {\bibinfo {title} {Fully
  device-independent conference key agreement},}\ }\href
  {https://doi.org/10.1103/PhysRevA.97.022307} {\bibfield  {journal} {\bibinfo
  {journal} {Phys. Rev. A}\ }\textbf {\bibinfo {volume} {97}},\ \bibinfo
  {pages} {022307} (\bibinfo {year} {2018})},\ \Eprint
  {http://arxiv.org/abs/1708.00798} {arXiv:1708.00798}\BibitemShut {NoStop}%
\bibitem [{\citenamefont {Yamasaki}\ \emph {et~al.}(2018)\citenamefont
  {Yamasaki}, \citenamefont {Pirker}, \citenamefont {Murao}, \citenamefont
  {D{\"u}r},\ and\ \citenamefont {Kraus}}]{YamasakiPirkerMuraoDuerKraus2018}%
  \BibitemOpen
  \bibfield  {author} {\bibinfo {author} {\bibfnamefont {Hayata}\ \bibnamefont
  {Yamasaki}}, \bibinfo {author} {\bibfnamefont {Alexander}\ \bibnamefont
  {Pirker}}, \bibinfo {author} {\bibfnamefont {Mio}\ \bibnamefont {Murao}},
  \bibinfo {author} {\bibfnamefont {Wolfgang}\ \bibnamefont {D{\"u}r}}, \ and\
  \bibinfo {author} {\bibfnamefont {Barbara}\ \bibnamefont {Kraus}},\ }\emph
  {\enquote {\bibinfo {title} {Multipartite entanglement outperforming
  bipartite entanglement under limited quantum system sizes},}\ }\href
  {https://doi.org/10.1103/PhysRevA.98.052313} {\bibfield  {journal} {\bibinfo
  {journal} {Phys. Rev. A}\ }\textbf {\bibinfo {volume} {98}},\ \bibinfo
  {pages} {052313} (\bibinfo {year} {2018})},\ \Eprint
  {http://arxiv.org/abs/1808.00005} {arXiv:1808.00005}\BibitemShut {NoStop}%
\bibitem [{\citenamefont {Raussendorf}\ and\ \citenamefont
  {Briegel}(2001)}]{RaussendorfBriegel2001}%
  \BibitemOpen
  \bibfield  {author} {\bibinfo {author} {\bibfnamefont {Robert}\ \bibnamefont
  {Raussendorf}}\ and\ \bibinfo {author} {\bibfnamefont {Hans~J.}\ \bibnamefont
  {Briegel}},\ }\emph {\enquote {\bibinfo {title} {{A One-Way Quantum
  Computer}},}\ }\href {https://doi.org/10.1103/PhysRevLett.86.5188} {\bibfield
   {journal} {\bibinfo  {journal} {Phys. Rev. Lett.}\ }\textbf {\bibinfo
  {volume} {86}},\ \bibinfo {pages} {5188{\textendash}5191} (\bibinfo {year}
  {2001})},\ \Eprint {http://arxiv.org/abs/quant-ph/0010033}
  {arXiv:quant-ph/0010033}\BibitemShut {NoStop}%
\bibitem [{\citenamefont {Briegel}\ and\ \citenamefont
  {Raussendorf}(2001)}]{BriegelRaussendorf2001}%
  \BibitemOpen
  \bibfield  {author} {\bibinfo {author} {\bibfnamefont {Hans~J.}\ \bibnamefont
  {Briegel}}\ and\ \bibinfo {author} {\bibfnamefont {Robert}\ \bibnamefont
  {Raussendorf}},\ }\emph {\enquote {\bibinfo {title} {{Persistent Entanglement
  in Arrays of Interacting Particles}},}\ }\href
  {https://doi.org/10.1103/PhysRevLett.86.910} {\bibfield  {journal} {\bibinfo
  {journal} {Phys. Rev. Lett.}\ }\textbf {\bibinfo {volume} {86}},\ \bibinfo
  {pages} {910{\textendash}913} (\bibinfo {year} {2001})},\ \Eprint
  {http://arxiv.org/abs/quant-ph/0004051} {arXiv:quant-ph/0004051}\BibitemShut
  {NoStop}%
\bibitem [{\citenamefont {Scott}(2004)}]{Scott2004}%
  \BibitemOpen
  \bibfield  {author} {\bibinfo {author} {\bibfnamefont {Andrew~J.}\
  \bibnamefont {Scott}},\ }\emph {\enquote {\bibinfo {title} {Multipartite
  entanglement, quantum-error-correcting codes, and entangling power of quantum
  evolutions},}\ }\href {https://doi.org/10.1103/PhysRevA.69.052330} {\bibfield
   {journal} {\bibinfo  {journal} {Phys. Rev. A}\ }\textbf {\bibinfo {volume}
  {69}},\ \bibinfo {pages} {052330} (\bibinfo {year} {2004})},\ \Eprint
  {http://arxiv.org/abs/quant-ph/0310137} {arXiv:quant-ph/0310137}\BibitemShut
  {NoStop}%
\bibitem [{\citenamefont {Bru{\ss}}\ and\ \citenamefont
  {Macchiavello}(2011)}]{BrussMacchiavello2011}%
  \BibitemOpen
  \bibfield  {author} {\bibinfo {author} {\bibfnamefont {Dagmar}\ \bibnamefont
  {Bru{\ss}}}\ and\ \bibinfo {author} {\bibfnamefont {Chiara}\ \bibnamefont
  {Macchiavello}},\ }\emph {\enquote {\bibinfo {title} {Multipartite
  entanglement in quantum algorithms},}\ }\href
  {https://doi.org/10.1103/PhysRevA.83.052313} {\bibfield  {journal} {\bibinfo
  {journal} {Phys. Rev. A}\ }\textbf {\bibinfo {volume} {83}},\ \bibinfo
  {pages} {052313} (\bibinfo {year} {2011})},\ \Eprint
  {http://arxiv.org/abs/1007.4179} {arXiv:1007.4179}\BibitemShut {NoStop}%
\bibitem [{\citenamefont {Friis}\ \emph {et~al.}(2018)\citenamefont {Friis},
  \citenamefont {Marty}, \citenamefont {Maier}, \citenamefont {Hempel},
  \citenamefont {Holz{\"a}pfel}, \citenamefont {Jurcevic}, \citenamefont
  {Plenio}, \citenamefont {Huber}, \citenamefont {Roos}, \citenamefont
  {Blatt},\ and\ \citenamefont {Lanyon}}]{FriisEtAl2018}%
  \BibitemOpen
  \bibfield  {author} {\bibinfo {author} {\bibfnamefont {Nicolai}\ \bibnamefont
  {Friis}}, \bibinfo {author} {\bibfnamefont {Oliver}\ \bibnamefont {Marty}},
  \bibinfo {author} {\bibfnamefont {Christine}\ \bibnamefont {Maier}}, \bibinfo
  {author} {\bibfnamefont {Cornelius}\ \bibnamefont {Hempel}}, \bibinfo
  {author} {\bibfnamefont {Milan}\ \bibnamefont {Holz{\"a}pfel}}, \bibinfo
  {author} {\bibfnamefont {Petar}\ \bibnamefont {Jurcevic}}, \bibinfo {author}
  {\bibfnamefont {Martin~B.}\ \bibnamefont {Plenio}}, \bibinfo {author}
  {\bibfnamefont {Marcus}\ \bibnamefont {Huber}}, \bibinfo {author}
  {\bibfnamefont {Christian}\ \bibnamefont {Roos}}, \bibinfo {author}
  {\bibfnamefont {Rainer}\ \bibnamefont {Blatt}}, \ and\ \bibinfo {author}
  {\bibfnamefont {Ben~P.}\ \bibnamefont {Lanyon}},\ }\emph {\enquote {\bibinfo
  {title} {{Observation of Entangled States of a Fully Controlled 20-Qubit
  System}},}\ }\href {https://doi.org/10.1103/PhysRevX.8.021012} {\bibfield
  {journal} {\bibinfo  {journal} {Phys. Rev. X}\ }\textbf {\bibinfo {volume}
  {8}},\ \bibinfo {pages} {021012} (\bibinfo {year} {2018})},\ \Eprint
  {http://arxiv.org/abs/1711.11092} {arXiv:1711.11092}\BibitemShut {NoStop}%
\bibitem [{\citenamefont {Canteri}\ \emph {et~al.}(2025)\citenamefont
  {Canteri}, \citenamefont {Bate}, \citenamefont {Mishra}, \citenamefont
  {Friis}, \citenamefont {Krutyanskiy},\ and\ \citenamefont
  {Lanyon}}]{CanteriEtAl2025}%
  \BibitemOpen
  \bibfield  {author} {\bibinfo {author} {\bibfnamefont {Marco}\ \bibnamefont
  {Canteri}}, \bibinfo {author} {\bibfnamefont {James}\ \bibnamefont {Bate}},
  \bibinfo {author} {\bibfnamefont {Ida}\ \bibnamefont {Mishra}}, \bibinfo
  {author} {\bibfnamefont {Nicolai}\ \bibnamefont {Friis}}, \bibinfo {author}
  {\bibfnamefont {Victor}\ \bibnamefont {Krutyanskiy}}, \ and\ \bibinfo
  {author} {\bibfnamefont {Benjamin~P.}\ \bibnamefont {Lanyon}},\ }\href@noop
  {} {\emph {\enquote {\bibinfo {title} {{Generation of Multipartite Photonic
  Entanglement Using a Trapped-Ion Quantum Processing Node}},}\ }}\Eprint
  {http://arxiv.org/abs/2510.15693} {arXiv:2510.15693} [quant-ph] (\bibinfo
  {year} {2025})\BibitemShut {NoStop}%
\bibitem [{\citenamefont {Rodriguez-Blanco}\ \emph {et~al.}(2021)\citenamefont
  {Rodriguez-Blanco}, \citenamefont {Bermudez}, \citenamefont {M\"{u}ller},\
  and\ \citenamefont {Shahandeh}}]{RodriguezBlancoEtAl2021}%
  \BibitemOpen
  \bibfield  {author} {\bibinfo {author} {\bibfnamefont {Adrian}\ \bibnamefont
  {Rodriguez-Blanco}}, \bibinfo {author} {\bibfnamefont {Alejandro}\
  \bibnamefont {Bermudez}}, \bibinfo {author} {\bibfnamefont {Markus}\
  \bibnamefont {M\"{u}ller}}, \ and\ \bibinfo {author} {\bibfnamefont {Farid}\
  \bibnamefont {Shahandeh}},\ }\emph {\enquote {\bibinfo {title} {{Efficient
  and Robust Certification of Genuine Multipartite Entanglement in Noisy
  Quantum Error Correction Circuits}},}\ }\href
  {https://doi.org/10.1103/PRXQuantum.2.020304} {\bibfield  {journal} {\bibinfo
   {journal} {PRX Quantum}\ }\textbf {\bibinfo {volume} {2}},\ \bibinfo {pages}
  {020304} (\bibinfo {year} {2021})},\ \Eprint
  {http://arxiv.org/abs/2010.02941} {arXiv:2010.02941}\BibitemShut {NoStop}%
\bibitem [{\citenamefont {Veitch}\ \emph {et~al.}(2014)\citenamefont {Veitch},
  \citenamefont {Hamed~Mousavian}, \citenamefont {Gottesman},\ and\
  \citenamefont {Emerson}}]{veitch_resource_2014}%
  \BibitemOpen
  \bibfield  {author} {\bibinfo {author} {\bibfnamefont {Victor}\ \bibnamefont
  {Veitch}}, \bibinfo {author} {\bibfnamefont {S.~A.}\ \bibnamefont
  {Hamed~Mousavian}}, \bibinfo {author} {\bibfnamefont {Daniel}\ \bibnamefont
  {Gottesman}}, \ and\ \bibinfo {author} {\bibfnamefont {Joseph}\ \bibnamefont
  {Emerson}},\ }\emph {\enquote {\bibinfo {title} {The resource theory of
  stabilizer quantum computation},}\ }\href
  {https://doi.org/10.1088/1367-2630/16/1/013009} {\bibfield  {journal}
  {\bibinfo  {journal} {New J. Phys.}\ }\textbf {\bibinfo {volume} {16}},\
  \bibinfo {pages} {013009} (\bibinfo {year} {2014})},\ \Eprint
  {http://arxiv.org/abs/1307.7171} {arXiv:1307.7171}\BibitemShut {NoStop}%
\bibitem [{\citenamefont {Lee}\ \emph {et~al.}(2026)\citenamefont {Lee},
  \citenamefont {Yuan}, \citenamefont {Chen}, \citenamefont {Tsubouchi},\ and\
  \citenamefont {Jiang}}]{LeeEtAl2025}%
  \BibitemOpen
  \bibfield  {author} {\bibinfo {author} {\bibfnamefont {Su-un}\ \bibnamefont
  {Lee}}, \bibinfo {author} {\bibfnamefont {Ming}\ \bibnamefont {Yuan}},
  \bibinfo {author} {\bibfnamefont {Senrui}\ \bibnamefont {Chen}}, \bibinfo
  {author} {\bibfnamefont {Kento}\ \bibnamefont {Tsubouchi}}, \ and\ \bibinfo
  {author} {\bibfnamefont {Liang}\ \bibnamefont {Jiang}},\ }\emph {\enquote
  {\bibinfo {title} {{Efficient Benchmarking of Logical Magic State}},}\ }\href
  {https://doi.org/10.1103/fwjt-mw2c} {\bibfield  {journal} {\bibinfo
  {journal} {Phys. Rev. Lett.}\ }\textbf {\bibinfo {volume} {136}},\ \bibinfo
  {pages} {050602} (\bibinfo {year} {2026})},\ \Eprint
  {http://arxiv.org/abs/2505.09687} {arXiv:2505.09687}\BibitemShut {NoStop}%
\bibitem [{\citenamefont {Daguerre}\ \emph {et~al.}(2025)\citenamefont
  {Daguerre}, \citenamefont {Blume-Kohout}, \citenamefont {Brown},
  \citenamefont {Hayes},\ and\ \citenamefont
  {Kim}}]{daguerre_experimental_2025}%
  \BibitemOpen
  \bibfield  {author} {\bibinfo {author} {\bibfnamefont {Lucas}\ \bibnamefont
  {Daguerre}}, \bibinfo {author} {\bibfnamefont {Robin}\ \bibnamefont
  {Blume-Kohout}}, \bibinfo {author} {\bibfnamefont {Natalie~C.}\ \bibnamefont
  {Brown}}, \bibinfo {author} {\bibfnamefont {David}\ \bibnamefont {Hayes}}, \
  and\ \bibinfo {author} {\bibfnamefont {Isaac~H.}\ \bibnamefont {Kim}},\
  }\emph {\enquote {\bibinfo {title} {{Experimental Demonstration of
  High-Fidelity Logical Magic States from Code Switching}},}\ }\href
  {https://doi.org/10.1103/dck4-x9c2} {\bibfield  {journal} {\bibinfo
  {journal} {Phys. Rev. X}\ }\textbf {\bibinfo {volume} {15}},\ \bibinfo
  {pages} {041008} (\bibinfo {year} {2025})},\ \Eprint
  {http://arxiv.org/abs/2506.14169} {arXiv:2506.14169}\BibitemShut {NoStop}%
\bibitem [{\citenamefont {Aharonov}(2003)}]{Aharonov2003}%
  \BibitemOpen
  \bibfield  {author} {\bibinfo {author} {\bibfnamefont {Dorit}\ \bibnamefont
  {Aharonov}},\ }\href@noop {} {\emph {\enquote {\bibinfo {title} {{A Simple
  Proof that Toffoli and Hadamard are Quantum Universal}},}\ }}\Eprint
  {http://arxiv.org/abs/quant-ph/0301040} {arXiv:quant-ph/0301040} [quant-ph]
  (\bibinfo {year} {2003})\BibitemShut {NoStop}%
\bibitem [{\citenamefont {Jerbi}\ \emph {et~al.}(2023)\citenamefont {Jerbi},
  \citenamefont {Fiderer}, \citenamefont {Poulsen~Nautrup}, \citenamefont
  {K\"{u}bler}, \citenamefont {Briegel},\ and\ \citenamefont
  {Dunjko}}]{JerbiEtAl2023}%
  \BibitemOpen
  \bibfield  {author} {\bibinfo {author} {\bibfnamefont {Sofiene}\ \bibnamefont
  {Jerbi}}, \bibinfo {author} {\bibfnamefont {Lukas~J.}\ \bibnamefont
  {Fiderer}}, \bibinfo {author} {\bibfnamefont {Hendrik}\ \bibnamefont
  {Poulsen~Nautrup}}, \bibinfo {author} {\bibfnamefont {Jonas~M.}\ \bibnamefont
  {K\"{u}bler}}, \bibinfo {author} {\bibfnamefont {Hans~J.}\ \bibnamefont
  {Briegel}}, \ and\ \bibinfo {author} {\bibfnamefont {Vedran}\ \bibnamefont
  {Dunjko}},\ }\emph {\enquote {\bibinfo {title} {{Quantum Machine Learning
  Beyond Kernel Methods}},}\ }\href
  {https://doi.org/10.1038/s41467-023-36159-y} {\bibfield  {journal} {\bibinfo
  {journal} {Nat. Commun.}\ }\textbf {\bibinfo {volume} {14}},\ \bibinfo
  {pages} {517} (\bibinfo {year} {2023})},\ \Eprint
  {http://arxiv.org/abs/2110.13162} {arXiv:2110.13162}\BibitemShut {NoStop}%
\bibitem [{\citenamefont {Gottesman}\ and\ \citenamefont
  {Chuang}(1999)}]{GottesmanChuang1999}%
  \BibitemOpen
  \bibfield  {author} {\bibinfo {author} {\bibfnamefont {Daniel}\ \bibnamefont
  {Gottesman}}\ and\ \bibinfo {author} {\bibfnamefont {Isaac~L.}\ \bibnamefont
  {Chuang}},\ }\emph {\enquote {\bibinfo {title} {{Demonstrating the Viability
  of Universal Quantum Computation Using Teleportation and Single-Qubit
  Operations}},}\ }\href {\doibase 10.1038/46503} {\bibfield  {journal}
  {\bibinfo  {journal} {Nature}\ }\textbf {\bibinfo {volume} {402}},\ \bibinfo
  {pages} {390{\textendash}393} (\bibinfo {year} {1999})},\ \Eprint
  {http://arxiv.org/abs/quant-ph/9908010} {arXiv:quant-ph/9908010}\BibitemShut
  {NoStop}%
\bibitem [{\citenamefont {Zhou}\ \emph {et~al.}(2000)\citenamefont {Zhou},
  \citenamefont {Leung},\ and\ \citenamefont {Chuang}}]{ZhouLeungChuang2000}%
  \BibitemOpen
  \bibfield  {author} {\bibinfo {author} {\bibfnamefont {Xinlan}\ \bibnamefont
  {Zhou}}, \bibinfo {author} {\bibfnamefont {Debbie~W.}\ \bibnamefont {Leung}},
  \ and\ \bibinfo {author} {\bibfnamefont {Isaac~L.}\ \bibnamefont {Chuang}},\
  }\emph {\enquote {\bibinfo {title} {{Methodology for Quantum Logic Gate
  Construction}},}\ }\href {\doibase 10.1103/PhysRevA.62.052316} {\bibfield
  {journal} {\bibinfo  {journal} {Phys. Rev. A}\ }\textbf {\bibinfo {volume}
  {62}},\ \bibinfo {pages} {052316} (\bibinfo {year} {2000})},\ \Eprint
  {http://arxiv.org/abs/quant-ph/0002039} {arXiv:quant-ph/0002039}\BibitemShut
  {NoStop}%
\bibitem [{\citenamefont {Hong}\ \emph {et~al.}(2024)\citenamefont {Hong},
  \citenamefont {Durso-Sabina}, \citenamefont {Hayes},\ and\ \citenamefont
  {Lucas}}]{hong_entangling_2024}%
  \BibitemOpen
  \bibfield  {author} {\bibinfo {author} {\bibfnamefont {Yifan}\ \bibnamefont
  {Hong}}, \bibinfo {author} {\bibfnamefont {Elijah}\ \bibnamefont
  {Durso-Sabina}}, \bibinfo {author} {\bibfnamefont {David}\ \bibnamefont
  {Hayes}}, \ and\ \bibinfo {author} {\bibfnamefont {Andrew}\ \bibnamefont
  {Lucas}},\ }\emph {\enquote {\bibinfo {title} {{Entangling Four Logical
  Qubits beyond Break-Even in a Nonlocal Code}},}\ }\href
  {https://doi.org/10.1103/PhysRevLett.133.180601} {\bibfield  {journal}
  {\bibinfo  {journal} {Phys. Rev. Lett.}\ }\textbf {\bibinfo {volume} {133}},\
  \bibinfo {pages} {180601} (\bibinfo {year} {2024})},\ \Eprint
  {http://arxiv.org/abs/2406.02666} {arXiv:2406.02666}\BibitemShut {NoStop}%
\bibitem [{\citenamefont {Bluvstein}\ \emph {et~al.}(2024)\citenamefont
  {Bluvstein}, \citenamefont {Evered}, \citenamefont {Geim}, \citenamefont
  {Li}, \citenamefont {Zhou}, \citenamefont {Manovitz}, \citenamefont {Ebadi},
  \citenamefont {Cain}, \citenamefont {Kalinowski}, \citenamefont {Hangleiter},
  \citenamefont {Bonilla~Ataides}, \citenamefont {Maskara}, \citenamefont
  {Cong}, \citenamefont {Gao}, \citenamefont {Sales~Rodriguez}, \citenamefont
  {Karolyshyn}, \citenamefont {Semeghini}, \citenamefont {Gullans},
  \citenamefont {Greiner}, \citenamefont {Vuletic},\ and\ \citenamefont
  {Lukin}}]{BluvsteinEtAl2024}%
  \BibitemOpen
  \bibfield  {author} {\bibinfo {author} {\bibfnamefont {Dolev}\ \bibnamefont
  {Bluvstein}}, \bibinfo {author} {\bibfnamefont {Simon~J.}\ \bibnamefont
  {Evered}}, \bibinfo {author} {\bibfnamefont {Alexandra~A.}\ \bibnamefont
  {Geim}}, \bibinfo {author} {\bibfnamefont {Sophie~H.}\ \bibnamefont {Li}},
  \bibinfo {author} {\bibfnamefont {Hengyun}\ \bibnamefont {Zhou}}, \bibinfo
  {author} {\bibfnamefont {Tom}\ \bibnamefont {Manovitz}}, \bibinfo {author}
  {\bibfnamefont {Sepehr}\ \bibnamefont {Ebadi}}, \bibinfo {author}
  {\bibfnamefont {Madelyn}\ \bibnamefont {Cain}}, \bibinfo {author}
  {\bibfnamefont {Marcin}\ \bibnamefont {Kalinowski}}, \bibinfo {author}
  {\bibfnamefont {Dominik}\ \bibnamefont {Hangleiter}}, \bibinfo {author}
  {\bibfnamefont {J.~Pablo}\ \bibnamefont {Bonilla~Ataides}}, \bibinfo {author}
  {\bibfnamefont {Nishad}\ \bibnamefont {Maskara}}, \bibinfo {author}
  {\bibfnamefont {Iris}\ \bibnamefont {Cong}}, \bibinfo {author} {\bibfnamefont
  {Xun}\ \bibnamefont {Gao}}, \bibinfo {author} {\bibfnamefont {Pedro}\
  \bibnamefont {Sales~Rodriguez}}, \bibinfo {author} {\bibfnamefont {Thomas}\
  \bibnamefont {Karolyshyn}}, \bibinfo {author} {\bibfnamefont {Giulia}\
  \bibnamefont {Semeghini}}, \bibinfo {author} {\bibfnamefont {Michael~J.}\
  \bibnamefont {Gullans}}, \bibinfo {author} {\bibfnamefont {Markus}\
  \bibnamefont {Greiner}}, \bibinfo {author} {\bibfnamefont {Vladan}\
  \bibnamefont {Vuletic}}, \ and\ \bibinfo {author} {\bibfnamefont
  {Mikhail~D.}\ \bibnamefont {Lukin}},\ }\emph {\enquote {\bibinfo {title}
  {{Logical Quantum Processor Based on Reconfigurable Atom Arrays}},}\ }\href
  {https://doi.org/10.1038/s41586-023-06927-3} {\bibfield  {journal} {\bibinfo
  {journal} {Nature}\ }\textbf {\bibinfo {volume} {626}},\ \bibinfo {pages}
  {58{\textendash}65} (\bibinfo {year} {2024})},\ \Eprint
  {http://arxiv.org/abs/2312.03982} {arXiv:2312.03982}\BibitemShut {NoStop}%
\bibitem [{\citenamefont {Besedin}\ \emph {et~al.}(2026)\citenamefont
  {Besedin}, \citenamefont {Kerschbaum}, \citenamefont {Knoll}, \citenamefont
  {Hesner}, \citenamefont {B{\"o}deker}, \citenamefont {Colmenarez},
  \citenamefont {Hofele}, \citenamefont {Lacroix}, \citenamefont {Hellings},
  \citenamefont {Swiadek}, \citenamefont {Flasby}, \citenamefont
  {Bahrami~Panah}, \citenamefont {Colao~Zanuz}, \citenamefont {M{\"u}ller},\
  and\ \citenamefont {Wallraff}}]{BesedinEtAl2025}%
  \BibitemOpen
  \bibfield  {author} {\bibinfo {author} {\bibfnamefont {Ilya}\ \bibnamefont
  {Besedin}}, \bibinfo {author} {\bibfnamefont {Michael}\ \bibnamefont
  {Kerschbaum}}, \bibinfo {author} {\bibfnamefont {Jonathan}\ \bibnamefont
  {Knoll}}, \bibinfo {author} {\bibfnamefont {Ian}\ \bibnamefont {Hesner}},
  \bibinfo {author} {\bibfnamefont {Lukas}\ \bibnamefont {B{\"o}deker}},
  \bibinfo {author} {\bibfnamefont {Luis}\ \bibnamefont {Colmenarez}}, \bibinfo
  {author} {\bibfnamefont {Luca}\ \bibnamefont {Hofele}}, \bibinfo {author}
  {\bibfnamefont {Nathan}\ \bibnamefont {Lacroix}}, \bibinfo {author}
  {\bibfnamefont {Christoph}\ \bibnamefont {Hellings}}, \bibinfo {author}
  {\bibfnamefont {Fran\c{c}ois}\ \bibnamefont {Swiadek}}, \bibinfo {author}
  {\bibfnamefont {Alexander}\ \bibnamefont {Flasby}}, \bibinfo {author}
  {\bibfnamefont {Mohsen}\ \bibnamefont {Bahrami~Panah}}, \bibinfo {author}
  {\bibfnamefont {Dante}\ \bibnamefont {Colao~Zanuz}}, \bibinfo {author}
  {\bibfnamefont {Markus}\ \bibnamefont {M{\"u}ller}}, \ and\ \bibinfo {author}
  {\bibfnamefont {Andreas}\ \bibnamefont {Wallraff}},\ }\emph {\enquote
  {\bibinfo {title} {{Lattice surgery realized on two distance-three repetition
  codes with superconducting qubits}},}\ }\href
  {https://doi.org/10.1038/s41567-025-03090-6} {\bibfield  {journal} {\bibinfo
  {journal} {Nat. Phys.}\ }\textbf {\bibinfo {volume} {22}},\ \bibinfo {pages}
  {189{\textendash}194} (\bibinfo {year} {2026})},\ \Eprint
  {http://arxiv.org/abs/2501.04612} {arXiv:2501.04612}\BibitemShut {NoStop}%
\bibitem [{\citenamefont {Wang}\ \emph {et~al.}(2026)\citenamefont {Wang},
  \citenamefont {Shen}, \citenamefont {Xie}, \citenamefont {Zhang},
  \citenamefont {Gao}, \citenamefont {Zhang}, \citenamefont {Zhu},
  \citenamefont {Jin}, \citenamefont {Zou}, \citenamefont {Wang}, \citenamefont
  {Cui}, \citenamefont {Bao}, \citenamefont {Zhu}, \citenamefont {Zhong},
  \citenamefont {Liu}, \citenamefont {Yang}, \citenamefont {Han}, \citenamefont
  {He}, \citenamefont {Shen}, \citenamefont {Wang}, \citenamefont {Huang},
  \citenamefont {Zhang}, \citenamefont {Zhou}, \citenamefont {Dong},
  \citenamefont {Deng}, \citenamefont {Wu}, \citenamefont {Song}, \citenamefont
  {Li}, \citenamefont {Wang}, \citenamefont {Song}, \citenamefont {Guo},
  \citenamefont {Zhang}, \citenamefont {Wang},\ and\ \citenamefont
  {Li}}]{WangEtAl2026}%
  \BibitemOpen
  \bibfield  {author} {\bibinfo {author} {\bibfnamefont {Yanzhe}\ \bibnamefont
  {Wang}}, \bibinfo {author} {\bibfnamefont {Fanhao}\ \bibnamefont {Shen}},
  \bibinfo {author} {\bibfnamefont {Haipeng}\ \bibnamefont {Xie}}, \bibinfo
  {author} {\bibfnamefont {Aosai}\ \bibnamefont {Zhang}}, \bibinfo {author}
  {\bibfnamefont {Yu}~\bibnamefont {Gao}}, \bibinfo {author} {\bibfnamefont
  {Chuanyu}\ \bibnamefont {Zhang}}, \bibinfo {author} {\bibfnamefont {Xuhao}\
  \bibnamefont {Zhu}}, \bibinfo {author} {\bibfnamefont {Feitong}\ \bibnamefont
  {Jin}}, \bibinfo {author} {\bibfnamefont {Yiren}\ \bibnamefont {Zou}},
  \bibinfo {author} {\bibfnamefont {Ning}\ \bibnamefont {Wang}}, \bibinfo
  {author} {\bibfnamefont {Zhengyi}\ \bibnamefont {Cui}}, \bibinfo {author}
  {\bibfnamefont {Zehang}\ \bibnamefont {Bao}}, \bibinfo {author}
  {\bibfnamefont {Zitian}\ \bibnamefont {Zhu}}, \bibinfo {author}
  {\bibfnamefont {Jiarun}\ \bibnamefont {Zhong}}, \bibinfo {author}
  {\bibfnamefont {Gongyu}\ \bibnamefont {Liu}}, \bibinfo {author}
  {\bibfnamefont {Jia-Nan}\ \bibnamefont {Yang}}, \bibinfo {author}
  {\bibfnamefont {Yihang}\ \bibnamefont {Han}}, \bibinfo {author}
  {\bibfnamefont {Yiyang}\ \bibnamefont {He}}, \bibinfo {author} {\bibfnamefont
  {Jiayuan}\ \bibnamefont {Shen}}, \bibinfo {author} {\bibfnamefont {Han}\
  \bibnamefont {Wang}}, \bibinfo {author} {\bibfnamefont {Jiahua}\ \bibnamefont
  {Huang}}, \bibinfo {author} {\bibfnamefont {Xinrong}\ \bibnamefont {Zhang}},
  \bibinfo {author} {\bibfnamefont {Sailang}\ \bibnamefont {Zhou}}, \bibinfo
  {author} {\bibfnamefont {Hang}\ \bibnamefont {Dong}}, \bibinfo {author}
  {\bibfnamefont {Jinfeng}\ \bibnamefont {Deng}}, \bibinfo {author}
  {\bibfnamefont {Yaozu}\ \bibnamefont {Wu}}, \bibinfo {author} {\bibfnamefont
  {Zixuan}\ \bibnamefont {Song}}, \bibinfo {author} {\bibfnamefont {Hekang}\
  \bibnamefont {Li}}, \bibinfo {author} {\bibfnamefont {Zhen}\ \bibnamefont
  {Wang}}, \bibinfo {author} {\bibfnamefont {Chao}\ \bibnamefont {Song}},
  \bibinfo {author} {\bibfnamefont {Qiujiang}\ \bibnamefont {Guo}}, \bibinfo
  {author} {\bibfnamefont {Pengfei}\ \bibnamefont {Zhang}}, \bibinfo {author}
  {\bibfnamefont {H.}~\bibnamefont {Wang}}, \ and\ \bibinfo {author}
  {\bibfnamefont {Ying}\ \bibnamefont {Li}},\ }\href@noop {} {\emph {\enquote
  {\bibinfo {title} {{A superconducting surface-code processor with
  lattice-surgery logical operations}},}\ }}\Eprint
  {http://arxiv.org/abs/2606.06598} {arXiv:2606.06598} [quant-ph] (\bibinfo
  {year} {2026})\BibitemShut {NoStop}%
\bibitem [{\citenamefont {Het\'enyi}\ and\ \citenamefont
  {Wootton}(2024)}]{HetenyiWootton2024}%
  \BibitemOpen
  \bibfield  {author} {\bibinfo {author} {\bibfnamefont {Bence}\ \bibnamefont
  {Het\'enyi}}\ and\ \bibinfo {author} {\bibfnamefont {James~R.}\ \bibnamefont
  {Wootton}},\ }\emph {\enquote {\bibinfo {title} {{Creating Entangled Logical
  Qubits in the Heavy-Hex Lattice with Topological Codes}},}\ }\href
  {https://doi.org/10.1103/PRXQuantum.5.040334} {\bibfield  {journal} {\bibinfo
   {journal} {PRX Quantum}\ }\textbf {\bibinfo {volume} {5}},\ \bibinfo {pages}
  {040334} (\bibinfo {year} {2024})},\ \Eprint
  {http://arxiv.org/abs/2404.15989} {arXiv:2404.15989}\BibitemShut {NoStop}%
\bibitem [{\citenamefont {Ryan-Anderson}\ \emph {et~al.}(2022)\citenamefont
  {Ryan-Anderson}, \citenamefont {Brown}, \citenamefont {Allman}, \citenamefont
  {Arkin}, \citenamefont {Asa-Attuah}, \citenamefont {Baldwin}, \citenamefont
  {Berg}, \citenamefont {Bohnet}, \citenamefont {Braxton}, \citenamefont
  {Burdick}, \citenamefont {Campora}, \citenamefont {Chernoguzov},
  \citenamefont {Esposito}, \citenamefont {Evans}, \citenamefont {Francois},
  \citenamefont {Gaebler}, \citenamefont {Gatterman}, \citenamefont {Gerber},
  \citenamefont {Gilmore}, \citenamefont {Gresh}, \citenamefont {Hall},
  \citenamefont {Hankin}, \citenamefont {Hostetter}, \citenamefont {Lucchetti},
  \citenamefont {Mayer}, \citenamefont {Myers}, \citenamefont {Neyenhuis},
  \citenamefont {Santiago}, \citenamefont {Sedlacek}, \citenamefont {Skripka},
  \citenamefont {Slattery}, \citenamefont {Stutz}, \citenamefont {Tait},
  \citenamefont {Tobey}, \citenamefont {Vittorini}, \citenamefont {Walker},\
  and\ \citenamefont {Hayes}}]{RyanAndersonEtAl2022}%
  \BibitemOpen
  \bibfield  {author} {\bibinfo {author} {\bibfnamefont {C.}~\bibnamefont
  {Ryan-Anderson}}, \bibinfo {author} {\bibfnamefont {N.~C.}\ \bibnamefont
  {Brown}}, \bibinfo {author} {\bibfnamefont {M.~S.}\ \bibnamefont {Allman}},
  \bibinfo {author} {\bibfnamefont {B.}~\bibnamefont {Arkin}}, \bibinfo
  {author} {\bibfnamefont {G.}~\bibnamefont {Asa-Attuah}}, \bibinfo {author}
  {\bibfnamefont {C.}~\bibnamefont {Baldwin}}, \bibinfo {author} {\bibfnamefont
  {J.}~\bibnamefont {Berg}}, \bibinfo {author} {\bibfnamefont {J.~G.}\
  \bibnamefont {Bohnet}}, \bibinfo {author} {\bibfnamefont {S.}~\bibnamefont
  {Braxton}}, \bibinfo {author} {\bibfnamefont {N.}~\bibnamefont {Burdick}},
  \bibinfo {author} {\bibfnamefont {J.~P.}\ \bibnamefont {Campora}}, \bibinfo
  {author} {\bibfnamefont {A.}~\bibnamefont {Chernoguzov}}, \bibinfo {author}
  {\bibfnamefont {J.}~\bibnamefont {Esposito}}, \bibinfo {author}
  {\bibfnamefont {B.}~\bibnamefont {Evans}}, \bibinfo {author} {\bibfnamefont
  {D.}~\bibnamefont {Francois}}, \bibinfo {author} {\bibfnamefont {J.~P.}\
  \bibnamefont {Gaebler}}, \bibinfo {author} {\bibfnamefont {T.~M.}\
  \bibnamefont {Gatterman}}, \bibinfo {author} {\bibfnamefont {J.}~\bibnamefont
  {Gerber}}, \bibinfo {author} {\bibfnamefont {K.}~\bibnamefont {Gilmore}},
  \bibinfo {author} {\bibfnamefont {D.}~\bibnamefont {Gresh}}, \bibinfo
  {author} {\bibfnamefont {A.}~\bibnamefont {Hall}}, \bibinfo {author}
  {\bibfnamefont {A.}~\bibnamefont {Hankin}}, \bibinfo {author} {\bibfnamefont
  {J.}~\bibnamefont {Hostetter}}, \bibinfo {author} {\bibfnamefont
  {D.}~\bibnamefont {Lucchetti}}, \bibinfo {author} {\bibfnamefont
  {K.}~\bibnamefont {Mayer}}, \bibinfo {author} {\bibfnamefont
  {J.}~\bibnamefont {Myers}}, \bibinfo {author} {\bibfnamefont
  {B.}~\bibnamefont {Neyenhuis}}, \bibinfo {author} {\bibfnamefont
  {J.}~\bibnamefont {Santiago}}, \bibinfo {author} {\bibfnamefont
  {J.}~\bibnamefont {Sedlacek}}, \bibinfo {author} {\bibfnamefont
  {T.}~\bibnamefont {Skripka}}, \bibinfo {author} {\bibfnamefont
  {A.}~\bibnamefont {Slattery}}, \bibinfo {author} {\bibfnamefont {R.~P.}\
  \bibnamefont {Stutz}}, \bibinfo {author} {\bibfnamefont {J.}~\bibnamefont
  {Tait}}, \bibinfo {author} {\bibfnamefont {R.}~\bibnamefont {Tobey}},
  \bibinfo {author} {\bibfnamefont {G.}~\bibnamefont {Vittorini}}, \bibinfo
  {author} {\bibfnamefont {J.}~\bibnamefont {Walker}}, \ and\ \bibinfo {author}
  {\bibfnamefont {D.}~\bibnamefont {Hayes}},\ }\href@noop {} {\emph {\enquote
  {\bibinfo {title} {{Implementing Fault-Tolerant Entangling Gates on the
  Five-Qubit Code and the Color Code}},}\ }}\Eprint
  {http://arxiv.org/abs/2208.01863} {arXiv:2208.01863} [quant-ph] (\bibinfo
  {year} {2022})\BibitemShut {NoStop}%
\bibitem [{\citenamefont {Nelson}\ \emph {et~al.}(2025)\citenamefont {Nelson},
  \citenamefont {Landahl},\ and\ \citenamefont
  {Baczewski}}]{NelsonLandahlBaczewski2025}%
  \BibitemOpen
  \bibfield  {author} {\bibinfo {author} {\bibfnamefont {Jacob~S.}\
  \bibnamefont {Nelson}}, \bibinfo {author} {\bibfnamefont {Andrew~J.}\
  \bibnamefont {Landahl}}, \ and\ \bibinfo {author} {\bibfnamefont {Andrew~D.}\
  \bibnamefont {Baczewski}},\ }\href@noop {} {\emph {\enquote {\bibinfo {title}
  {{A Small and Interesting Architecture for Early Fault-Tolerant Quantum
  Computers}},}\ }}\Eprint {http://arxiv.org/abs/2507.20387} {arXiv:2507.20387}
  [quant-ph] (\bibinfo {year} {2025})\BibitemShut {NoStop}%
\bibitem [{\citenamefont {Bluvstein}\ \emph {et~al.}(2026)\citenamefont
  {Bluvstein}, \citenamefont {Geim}, \citenamefont {Li}, \citenamefont
  {Evered}, \citenamefont {Bonilla~Ataides}, \citenamefont {Baranes},
  \citenamefont {Gu}, \citenamefont {Manovitz}, \citenamefont {Xu},
  \citenamefont {Kalinowski}, \citenamefont {Majidy}, \citenamefont {Kokail},
  \citenamefont {Maskara}, \citenamefont {Trapp}, \citenamefont {Stewart},
  \citenamefont {Hollerith}, \citenamefont {Zhou}, \citenamefont {Gullans},
  \citenamefont {Yelin}, \citenamefont {Greiner}, \citenamefont {Vuletic},
  \citenamefont {Cain},\ and\ \citenamefont {Lukin}}]{BluvsteinEtAl2025}%
  \BibitemOpen
  \bibfield  {author} {\bibinfo {author} {\bibfnamefont {Dolev}\ \bibnamefont
  {Bluvstein}}, \bibinfo {author} {\bibfnamefont {Alexandra~A.}\ \bibnamefont
  {Geim}}, \bibinfo {author} {\bibfnamefont {Sophie~H.}\ \bibnamefont {Li}},
  \bibinfo {author} {\bibfnamefont {Simon~J.}\ \bibnamefont {Evered}}, \bibinfo
  {author} {\bibfnamefont {J.~Pablo}\ \bibnamefont {Bonilla~Ataides}}, \bibinfo
  {author} {\bibfnamefont {Gefen}\ \bibnamefont {Baranes}}, \bibinfo {author}
  {\bibfnamefont {Andi}\ \bibnamefont {Gu}}, \bibinfo {author} {\bibfnamefont
  {Tom}\ \bibnamefont {Manovitz}}, \bibinfo {author} {\bibfnamefont {Muqing}\
  \bibnamefont {Xu}}, \bibinfo {author} {\bibfnamefont {Marcin}\ \bibnamefont
  {Kalinowski}}, \bibinfo {author} {\bibfnamefont {Shayan}\ \bibnamefont
  {Majidy}}, \bibinfo {author} {\bibfnamefont {Christian}\ \bibnamefont
  {Kokail}}, \bibinfo {author} {\bibfnamefont {Nishad}\ \bibnamefont
  {Maskara}}, \bibinfo {author} {\bibfnamefont {Elias~C.}\ \bibnamefont
  {Trapp}}, \bibinfo {author} {\bibfnamefont {Luke~M.}\ \bibnamefont
  {Stewart}}, \bibinfo {author} {\bibfnamefont {Simon}\ \bibnamefont
  {Hollerith}}, \bibinfo {author} {\bibfnamefont {Hengyun}\ \bibnamefont
  {Zhou}}, \bibinfo {author} {\bibfnamefont {Michael~J.}\ \bibnamefont
  {Gullans}}, \bibinfo {author} {\bibfnamefont {Susanne~F.}\ \bibnamefont
  {Yelin}}, \bibinfo {author} {\bibfnamefont {Markus}\ \bibnamefont {Greiner}},
  \bibinfo {author} {\bibfnamefont {Vladan}\ \bibnamefont {Vuletic}}, \bibinfo
  {author} {\bibfnamefont {Madelyn}\ \bibnamefont {Cain}}, \ and\ \bibinfo
  {author} {\bibfnamefont {Mikhail~D.}\ \bibnamefont {Lukin}},\ }\emph
  {\enquote {\bibinfo {title} {{A Fault-Tolerant Neutral-Atom Architecture for
  Universal Quantum Computation}},}\ }\href
  {https://doi.org/10.1038/s41586-025-09848-5} {\bibfield  {journal} {\bibinfo
  {journal} {Nature}\ }\textbf {\bibinfo {volume} {649}},\ \bibinfo {pages}
  {39{\textendash}46} (\bibinfo {year} {2026})},\ \Eprint
  {http://arxiv.org/abs/2506.20661} {arXiv:2506.20661}\BibitemShut {NoStop}%
\bibitem [{\citenamefont {Gottesman}(1997)}]{GottesmanPhD1997}%
  \BibitemOpen
  \bibfield  {author} {\bibinfo {author} {\bibfnamefont {Daniel}\ \bibnamefont
  {Gottesman}},\ }\emph {\bibinfo {title} {{Stabilizer Codes and Quantum Error
  Correction}}},\ \href@noop {} {Ph.D. thesis},\ \bibinfo  {school} {Caltech}
  (\bibinfo {year} {1997}),\ \Eprint {http://arxiv.org/abs/quant-ph/9705052}
  {arXiv:quant-ph/9705052}\BibitemShut {NoStop}%
\bibitem [{\citenamefont {Kitaev}(2003)}]{Kitaev2003}%
  \BibitemOpen
  \bibfield  {author} {\bibinfo {author} {\bibfnamefont {Alexei}\ \bibnamefont
  {Kitaev}},\ }\emph {\enquote {\bibinfo {title} {Fault-tolerant quantum
  computation by anyons},}\ }\href
  {https://doi.org/10.1016/S0003-4916(02)00018-0} {\bibfield  {journal}
  {\bibinfo  {journal} {Ann. Phys.}\ }\textbf {\bibinfo {volume} {303}},\
  \bibinfo {pages} {2{\textendash}30} (\bibinfo {year} {2003})},\ \Eprint
  {http://arxiv.org/abs/quant-ph/9707021} {arXiv:quant-ph/9707021}\BibitemShut
  {NoStop}%
\bibitem [{\citenamefont {Dennis}\ \emph {et~al.}(2002)\citenamefont {Dennis},
  \citenamefont {Kitaev}, \citenamefont {Landahl},\ and\ \citenamefont
  {Preskill}}]{DennisKitaevLandahlPreskill2002}%
  \BibitemOpen
  \bibfield  {author} {\bibinfo {author} {\bibfnamefont {Eric}\ \bibnamefont
  {Dennis}}, \bibinfo {author} {\bibfnamefont {Alexei}\ \bibnamefont {Kitaev}},
  \bibinfo {author} {\bibfnamefont {Andrew}\ \bibnamefont {Landahl}}, \ and\
  \bibinfo {author} {\bibfnamefont {John}\ \bibnamefont {Preskill}},\ }\emph
  {\enquote {\bibinfo {title} {Topological quantum memory},}\ }\href
  {https://doi.org/10.1063/1.1499754} {\bibfield  {journal} {\bibinfo
  {journal} {J. Math. Phys.}\ }\textbf {\bibinfo {volume} {43}},\ \bibinfo
  {pages} {4452{\textendash}4505} (\bibinfo {year} {2002})},\ \Eprint
  {http://arxiv.org/abs/quant-ph/0110143} {arXiv:quant-ph/0110143}\BibitemShut
  {NoStop}%
\bibitem [{\citenamefont {Fowler}\ \emph {et~al.}(2012)\citenamefont {Fowler},
  \citenamefont {Mariantoni}, \citenamefont {Martinis},\ and\ \citenamefont
  {Cleland}}]{FowlerMariantoniMartinisCleland2012}%
  \BibitemOpen
  \bibfield  {author} {\bibinfo {author} {\bibfnamefont {Austin~G.}\
  \bibnamefont {Fowler}}, \bibinfo {author} {\bibfnamefont {Matteo}\
  \bibnamefont {Mariantoni}}, \bibinfo {author} {\bibfnamefont {John~M.}\
  \bibnamefont {Martinis}}, \ and\ \bibinfo {author} {\bibfnamefont
  {Andrew~N.}\ \bibnamefont {Cleland}},\ }\emph {\enquote {\bibinfo {title}
  {{Surface codes: Towards practical large-scale quantum computation}},}\
  }\href {https://doi.org/10.1103/PhysRevA.86.032324} {\bibfield  {journal}
  {\bibinfo  {journal} {Phys. Rev. A}\ }\textbf {\bibinfo {volume} {86}},\
  \bibinfo {pages} {032324} (\bibinfo {year} {2012})},\ \Eprint
  {http://arxiv.org/abs/1208.0928} {arXiv:1208.0928}\BibitemShut {NoStop}%
\bibitem [{\citenamefont {Raussendorf}\ and\ \citenamefont
  {Harrington}(2007)}]{RaussendorfHarrington2007}%
  \BibitemOpen
  \bibfield  {author} {\bibinfo {author} {\bibfnamefont {Robert}\ \bibnamefont
  {Raussendorf}}\ and\ \bibinfo {author} {\bibfnamefont {Jim}\ \bibnamefont
  {Harrington}},\ }\emph {\enquote {\bibinfo {title} {{Fault-Tolerant Quantum
  Computation with High Threshold in Two Dimensions}},}\ }\href
  {https://doi.org/10.1103/PhysRevLett.98.190504} {\bibfield  {journal}
  {\bibinfo  {journal} {Phys. Rev. Lett.}\ }\textbf {\bibinfo {volume} {98}},\
  \bibinfo {pages} {190504} (\bibinfo {year} {2007})},\ \Eprint
  {http://arxiv.org/abs/quant-ph/0610082} {arXiv:quant-ph/0610082}\BibitemShut
  {NoStop}%
\bibitem [{\citenamefont {Kubica}\ \emph {et~al.}(2015)\citenamefont {Kubica},
  \citenamefont {Yoshida},\ and\ \citenamefont
  {Pastawski}}]{KubicaYoshidaPastawski2015}%
  \BibitemOpen
  \bibfield  {author} {\bibinfo {author} {\bibfnamefont {Aleksander}\
  \bibnamefont {Kubica}}, \bibinfo {author} {\bibfnamefont {Beni}\ \bibnamefont
  {Yoshida}}, \ and\ \bibinfo {author} {\bibfnamefont {Fernando}\ \bibnamefont
  {Pastawski}},\ }\emph {\enquote {\bibinfo {title} {{Unfolding the color
  code}},}\ }\href {https://doi.org/10.1088/1367-2630/17/8/083026} {\bibfield
  {journal} {\bibinfo  {journal} {New J. Phys.}\ }\textbf {\bibinfo {volume}
  {17}},\ \bibinfo {pages} {083026} (\bibinfo {year} {2015})},\ \Eprint
  {http://arxiv.org/abs/1503.02065} {arXiv:1503.02065}\BibitemShut {NoStop}%
\bibitem [{\citenamefont {Campbell}\ and\ \citenamefont
  {Howard}(2017{\natexlab{a}})}]{CampbellHoward2017a}%
  \BibitemOpen
  \bibfield  {author} {\bibinfo {author} {\bibfnamefont {Earl~T.}\ \bibnamefont
  {Campbell}}\ and\ \bibinfo {author} {\bibfnamefont {Mark}\ \bibnamefont
  {Howard}},\ }\emph {\enquote {\bibinfo {title} {{Unifying Gate Synthesis and
  Magic State Distillation}},}\ }\href
  {https://doi.org/10.1103/PhysRevLett.118.060501} {\bibfield  {journal}
  {\bibinfo  {journal} {Phys. Rev. Lett.}\ }\textbf {\bibinfo {volume} {118}},\
  \bibinfo {pages} {060501} (\bibinfo {year} {2017}{\natexlab{a}})},\ \Eprint
  {http://arxiv.org/abs/1606.01906} {arXiv:1606.01906}\BibitemShut {NoStop}%
\bibitem [{\citenamefont {Campbell}\ and\ \citenamefont
  {Howard}(2017{\natexlab{b}})}]{CampbellHoward2017b}%
  \BibitemOpen
  \bibfield  {author} {\bibinfo {author} {\bibfnamefont {Earl~T.}\ \bibnamefont
  {Campbell}}\ and\ \bibinfo {author} {\bibfnamefont {Mark}\ \bibnamefont
  {Howard}},\ }\emph {\enquote {\bibinfo {title} {Unified framework for magic
  state distillation and multiqubit gate synthesis with reduced resource
  cost},}\ }\href {https://doi.org/10.1103/PhysRevA.95.022316} {\bibfield
  {journal} {\bibinfo  {journal} {Phys. Rev. A}\ }\textbf {\bibinfo {volume}
  {95}},\ \bibinfo {pages} {022316} (\bibinfo {year} {2017}{\natexlab{b}})},\
  \Eprint {http://arxiv.org/abs/1606.01904} {arXiv:1606.01904}\BibitemShut
  {NoStop}%
\bibitem [{\citenamefont {Honciuc~Menendez}\ \emph {et~al.}(2024)\citenamefont
  {Honciuc~Menendez}, \citenamefont {Ray},\ and\ \citenamefont
  {Vasmer}}]{MenendezEtAl2023}%
  \BibitemOpen
  \bibfield  {author} {\bibinfo {author} {\bibfnamefont {Daniel}\ \bibnamefont
  {Honciuc~Menendez}}, \bibinfo {author} {\bibfnamefont {Annie}\ \bibnamefont
  {Ray}}, \ and\ \bibinfo {author} {\bibfnamefont {Michael}\ \bibnamefont
  {Vasmer}},\ }\emph {\enquote {\bibinfo {title} {{Implementing fault-tolerant
  non-Clifford gates using the $[\![8,3,2]\!]$ color code}},}\ }\href
  {https://doi.org/10.1103/PhysRevA.109.062438} {\bibfield  {journal} {\bibinfo
   {journal} {Phys. Rev. A}\ }\textbf {\bibinfo {volume} {109}},\ \bibinfo
  {pages} {062438} (\bibinfo {year} {2024})},\ \Eprint
  {http://arxiv.org/abs/2309.08663} {arXiv:2309.08663} [quant-ph]\BibitemShut
  {NoStop}%
\bibitem [{\citenamefont {Pogorelov}\ \emph {et~al.}(2021)\citenamefont
  {Pogorelov}, \citenamefont {Feldker}, \citenamefont {Marciniak},
  \citenamefont {Postler}, \citenamefont {Jacob}, \citenamefont
  {Krieglsteiner}, \citenamefont {Podlesnic}, \citenamefont {Meth},
  \citenamefont {Negnevitsky}, \citenamefont {Stadler}, \citenamefont
  {H{\"o}fer}, \citenamefont {W{\"a}chter}, \citenamefont {Lakhmanskiy},
  \citenamefont {Blatt}, \citenamefont {Schindler},\ and\ \citenamefont
  {Monz}}]{PogorelovEtAl2021}%
  \BibitemOpen
  \bibfield  {author} {\bibinfo {author} {\bibfnamefont {Ivan}\ \bibnamefont
  {Pogorelov}}, \bibinfo {author} {\bibfnamefont {Thomas}\ \bibnamefont
  {Feldker}}, \bibinfo {author} {\bibfnamefont {Christian~D.}\ \bibnamefont
  {Marciniak}}, \bibinfo {author} {\bibfnamefont {Lukas}\ \bibnamefont
  {Postler}}, \bibinfo {author} {\bibfnamefont {Georg}\ \bibnamefont {Jacob}},
  \bibinfo {author} {\bibfnamefont {Oliver}\ \bibnamefont {Krieglsteiner}},
  \bibinfo {author} {\bibfnamefont {Verena}\ \bibnamefont {Podlesnic}},
  \bibinfo {author} {\bibfnamefont {Michael}\ \bibnamefont {Meth}}, \bibinfo
  {author} {\bibfnamefont {Vlad}\ \bibnamefont {Negnevitsky}}, \bibinfo
  {author} {\bibfnamefont {Martin}\ \bibnamefont {Stadler}}, \bibinfo {author}
  {\bibfnamefont {Bernd}\ \bibnamefont {H{\"o}fer}}, \bibinfo {author}
  {\bibfnamefont {Christoph}\ \bibnamefont {W{\"a}chter}}, \bibinfo {author}
  {\bibfnamefont {Kirill}\ \bibnamefont {Lakhmanskiy}}, \bibinfo {author}
  {\bibfnamefont {Rainer}\ \bibnamefont {Blatt}}, \bibinfo {author}
  {\bibfnamefont {Philipp}\ \bibnamefont {Schindler}}, \ and\ \bibinfo {author}
  {\bibfnamefont {Thomas}\ \bibnamefont {Monz}},\ }\emph {\enquote {\bibinfo
  {title} {{Compact Ion-Trap Quantum Computing Demonstrator}},}\ }\href
  {https://doi.org/10.1103/PRXQuantum.2.020343} {\bibfield  {journal} {\bibinfo
   {journal} {PRX Quantum}\ }\textbf {\bibinfo {volume} {2}},\ \bibinfo {pages}
  {020343} (\bibinfo {year} {2021})},\ \Eprint
  {http://arxiv.org/abs/2101.11390} {arXiv:2101.11390}\BibitemShut {NoStop}%
\bibitem [{\citenamefont {McKay}\ \emph {et~al.}(2017)\citenamefont {McKay},
  \citenamefont {Wood}, \citenamefont {Sheldon}, \citenamefont {Chow},\ and\
  \citenamefont {Gambetta}}]{mckay_efficient_z_2017}%
  \BibitemOpen
  \bibfield  {author} {\bibinfo {author} {\bibfnamefont {David~C.}\
  \bibnamefont {McKay}}, \bibinfo {author} {\bibfnamefont {Christopher~J.}\
  \bibnamefont {Wood}}, \bibinfo {author} {\bibfnamefont {Sarah}\ \bibnamefont
  {Sheldon}}, \bibinfo {author} {\bibfnamefont {Jerry~M.}\ \bibnamefont
  {Chow}}, \ and\ \bibinfo {author} {\bibfnamefont {Jay~M.}\ \bibnamefont
  {Gambetta}},\ }\emph {\enquote {\bibinfo {title} {{Efficient $Z$ gates for
  quantum computing}},}\ }\href {https://doi.org/10.1103/PhysRevA.96.022330}
  {\bibfield  {journal} {\bibinfo  {journal} {Phys. Rev. A}\ }\textbf {\bibinfo
  {volume} {96}},\ \bibinfo {pages} {022330} (\bibinfo {year} {2017})},\
  \Eprint {http://arxiv.org/abs/1612.00858} {arXiv:1612.00858}\BibitemShut
  {NoStop}%
\bibitem [{\citenamefont {S{\o}rensen}\ and\ \citenamefont
  {M{\o}lmer}(1999)}]{SoerensenMoelmer1999}%
  \BibitemOpen
  \bibfield  {author} {\bibinfo {author} {\bibfnamefont {Anders}\ \bibnamefont
  {S{\o}rensen}}\ and\ \bibinfo {author} {\bibfnamefont {Klaus}\ \bibnamefont
  {M{\o}lmer}},\ }\emph {\enquote {\bibinfo {title} {{Quantum Computation with
  Ions in Thermal Motion}},}\ }\href
  {https://doi.org/10.1103/PhysRevLett.82.1971} {\bibfield  {journal} {\bibinfo
   {journal} {Phys. Rev. Lett.}\ }\textbf {\bibinfo {volume} {82}},\ \bibinfo
  {pages} {1971{\textendash}1974} (\bibinfo {year} {1999})},\ \Eprint
  {http://arxiv.org/abs/quant-ph/9810039} {arXiv:quant-ph/9810039}\BibitemShut
  {NoStop}%
\bibitem [{\citenamefont {Maslov}(2017)}]{maslov_basic_circuit_2017}%
  \BibitemOpen
  \bibfield  {author} {\bibinfo {author} {\bibfnamefont {Dmitri}\ \bibnamefont
  {Maslov}},\ }\emph {\enquote {\bibinfo {title} {{Basic circuit compilation
  techniques for an ion-trap quantum machine}},}\ }\href
  {https://doi.org/10.1088/1367-2630/aa5e47} {\bibfield  {journal} {\bibinfo
  {journal} {New J. Phys.}\ }\textbf {\bibinfo {volume} {19}},\ \bibinfo
  {pages} {023035} (\bibinfo {year} {2017})},\ \Eprint
  {http://arxiv.org/abs/1603.07678} {arXiv:1603.07678}\BibitemShut {NoStop}%
\bibitem [{\citenamefont {Peham}\ \emph {et~al.}(2025)\citenamefont {Peham},
  \citenamefont {Schmid}, \citenamefont {Berent}, \citenamefont {M\"{u}ller},\
  and\ \citenamefont {Wille}}]{PehamEtAl2025}%
  \BibitemOpen
  \bibfield  {author} {\bibinfo {author} {\bibfnamefont {Tom}\ \bibnamefont
  {Peham}}, \bibinfo {author} {\bibfnamefont {Ludwig}\ \bibnamefont {Schmid}},
  \bibinfo {author} {\bibfnamefont {Lucas}\ \bibnamefont {Berent}}, \bibinfo
  {author} {\bibfnamefont {Markus}\ \bibnamefont {M\"{u}ller}}, \ and\ \bibinfo
  {author} {\bibfnamefont {Robert}\ \bibnamefont {Wille}},\ }\emph {\enquote
  {\bibinfo {title} {{Automated Synthesis of Fault-Tolerant State Preparation
  Circuits for Quantum Error Correction Codes}},}\ }\href
  {https://doi.org/10.1103/PRXQuantum.6.020330} {\bibfield  {journal} {\bibinfo
   {journal} {PRX Quantum}\ }\textbf {\bibinfo {volume} {6}},\ \bibinfo {pages}
  {020330} (\bibinfo {year} {2025})},\ \Eprint
  {http://arxiv.org/abs/2408.11894} {arXiv:2408.11894}\BibitemShut {NoStop}%
\bibitem [{\citenamefont {Wang}\ \emph {et~al.}(2024)\citenamefont {Wang},
  \citenamefont {Simsek}, \citenamefont {Gatterman}, \citenamefont {Gerber},
  \citenamefont {Gilmore}, \citenamefont {Gresh}, \citenamefont {Hewitt},
  \citenamefont {Horst}, \citenamefont {Matheny}, \citenamefont {Mengle},
  \citenamefont {Neyenhuis},\ and\ \citenamefont {Criger}}]{WangEtAl2023}%
  \BibitemOpen
  \bibfield  {author} {\bibinfo {author} {\bibfnamefont {Yang}\ \bibnamefont
  {Wang}}, \bibinfo {author} {\bibfnamefont {Selwyn}\ \bibnamefont {Simsek}},
  \bibinfo {author} {\bibfnamefont {Thomas~M.}\ \bibnamefont {Gatterman}},
  \bibinfo {author} {\bibfnamefont {Justin~A.}\ \bibnamefont {Gerber}},
  \bibinfo {author} {\bibfnamefont {Kevin}\ \bibnamefont {Gilmore}}, \bibinfo
  {author} {\bibfnamefont {Dan}\ \bibnamefont {Gresh}}, \bibinfo {author}
  {\bibfnamefont {Nathan}\ \bibnamefont {Hewitt}}, \bibinfo {author}
  {\bibfnamefont {Chandler~V.}\ \bibnamefont {Horst}}, \bibinfo {author}
  {\bibfnamefont {Mitchell}\ \bibnamefont {Matheny}}, \bibinfo {author}
  {\bibfnamefont {Tanner}\ \bibnamefont {Mengle}}, \bibinfo {author}
  {\bibfnamefont {Brian}\ \bibnamefont {Neyenhuis}}, \ and\ \bibinfo {author}
  {\bibfnamefont {Ben}\ \bibnamefont {Criger}},\ }\emph {\enquote {\bibinfo
  {title} {{Fault-tolerant one-bit addition with the smallest interesting color
  code}},}\ }\href {https://doi.org/10.1126/sciadv.ado9024} {\bibfield
  {journal} {\bibinfo  {journal} {Sci. Adv.}\ }\textbf {\bibinfo {volume}
  {10}},\ \bibinfo {pages} {29} (\bibinfo {year} {2024})},\ \Eprint
  {http://arxiv.org/abs/2309.09893} {arXiv:2309.09893}\BibitemShut {NoStop}%
\bibitem [{\citenamefont {Chao}\ and\ \citenamefont
  {Reichardt}(2018)}]{ChaoReichardt2018}%
  \BibitemOpen
  \bibfield  {author} {\bibinfo {author} {\bibfnamefont {Rui}\ \bibnamefont
  {Chao}}\ and\ \bibinfo {author} {\bibfnamefont {Ben~W.}\ \bibnamefont
  {Reichardt}},\ }\emph {\enquote {\bibinfo {title} {{Quantum Error Correction
  with Only Two Extra Qubits}},}\ }\href
  {https://doi.org/10.1103/PhysRevLett.121.050502} {\bibfield  {journal}
  {\bibinfo  {journal} {Phys. Rev. Lett.}\ }\textbf {\bibinfo {volume} {121}},\
  \bibinfo {pages} {050502} (\bibinfo {year} {2018})},\ \Eprint
  {http://arxiv.org/abs/1705.02329} {arXiv:1705.02329}\BibitemShut {NoStop}%
\bibitem [{\citenamefont {Chamberland}\ and\ \citenamefont
  {Beverland}(2018)}]{ChamberlandBeverland2018}%
  \BibitemOpen
  \bibfield  {author} {\bibinfo {author} {\bibfnamefont {Christopher}\
  \bibnamefont {Chamberland}}\ and\ \bibinfo {author} {\bibfnamefont
  {Michael~E.}\ \bibnamefont {Beverland}},\ }\emph {\enquote {\bibinfo {title}
  {{Flag fault-tolerant error correction with arbitrary distance codes}},}\
  }\href {https://doi.org/10.22331/q-2018-02-08-53} {\bibfield  {journal}
  {\bibinfo  {journal} {Quantum}\ }\textbf {\bibinfo {volume} {2}},\ \bibinfo
  {pages} {53} (\bibinfo {year} {2018})},\ \Eprint
  {http://arxiv.org/abs/1708.02246} {arXiv:1708.02246}\BibitemShut {NoStop}%
\bibitem [{\citenamefont {Postler}\ \emph {et~al.}(2022)\citenamefont
  {Postler}, \citenamefont {Heu{\ss}en}, \citenamefont {Pogorelov},
  \citenamefont {Rispler}, \citenamefont {Feldker}, \citenamefont {Meth},
  \citenamefont {Marciniak}, \citenamefont {Stricker}, \citenamefont
  {Ringbauer}, \citenamefont {Blatt}, \citenamefont {Schindler}, \citenamefont
  {M{\"u}ller},\ and\ \citenamefont {Monz}}]{Postler2022}%
  \BibitemOpen
  \bibfield  {author} {\bibinfo {author} {\bibfnamefont {Lukas}\ \bibnamefont
  {Postler}}, \bibinfo {author} {\bibfnamefont {Sascha}\ \bibnamefont
  {Heu{\ss}en}}, \bibinfo {author} {\bibfnamefont {Ivan}\ \bibnamefont
  {Pogorelov}}, \bibinfo {author} {\bibfnamefont {Manuel}\ \bibnamefont
  {Rispler}}, \bibinfo {author} {\bibfnamefont {Thomas}\ \bibnamefont
  {Feldker}}, \bibinfo {author} {\bibfnamefont {Michael}\ \bibnamefont {Meth}},
  \bibinfo {author} {\bibfnamefont {Christian~D.}\ \bibnamefont {Marciniak}},
  \bibinfo {author} {\bibfnamefont {Roman}\ \bibnamefont {Stricker}}, \bibinfo
  {author} {\bibfnamefont {Martin}\ \bibnamefont {Ringbauer}}, \bibinfo
  {author} {\bibfnamefont {Rainer}\ \bibnamefont {Blatt}}, \bibinfo {author}
  {\bibfnamefont {Philipp}\ \bibnamefont {Schindler}}, \bibinfo {author}
  {\bibfnamefont {Markus}\ \bibnamefont {M{\"u}ller}}, \ and\ \bibinfo {author}
  {\bibfnamefont {Thomas}\ \bibnamefont {Monz}},\ }\emph {\enquote {\bibinfo
  {title} {{Demonstration of fault-tolerant universal quantum gate
  operations}},}\ }\href {https://doi.org/10.1038/s41586-022-04721-1}
  {\bibfield  {journal} {\bibinfo  {journal} {Nature}\ }\textbf {\bibinfo
  {volume} {605}},\ \bibinfo {pages} {675{\textendash}680} (\bibinfo {year}
  {2022})},\ \Eprint {http://arxiv.org/abs/2111.12654}
  {arXiv:2111.12654}\BibitemShut {NoStop}%
\bibitem [{\citenamefont {Rossi}\ \emph {et~al.}(2013)\citenamefont {Rossi},
  \citenamefont {Huber}, \citenamefont {Bru{\ss}},\ and\ \citenamefont
  {Macchiavello}}]{RossiHuberBrussMacchiavello2013}%
  \BibitemOpen
  \bibfield  {author} {\bibinfo {author} {\bibfnamefont {Matteo}\ \bibnamefont
  {Rossi}}, \bibinfo {author} {\bibfnamefont {Marcus}\ \bibnamefont {Huber}},
  \bibinfo {author} {\bibfnamefont {Dagmar}\ \bibnamefont {Bru{\ss}}}, \ and\
  \bibinfo {author} {\bibfnamefont {Chiara}\ \bibnamefont {Macchiavello}},\
  }\emph {\enquote {\bibinfo {title} {Quantum hypergraph states},}\ }\href
  {https://doi.org/10.1088/1367-2630/15/11/113022} {\bibfield  {journal}
  {\bibinfo  {journal} {New J. Phys.}\ }\textbf {\bibinfo {volume} {15}},\
  \bibinfo {pages} {113022} (\bibinfo {year} {2013})},\ \Eprint
  {http://arxiv.org/abs/1211.5554} {arXiv:1211.5554}\BibitemShut {NoStop}%
\bibitem [{\citenamefont {Howard}\ and\ \citenamefont
  {Campbell}(2017)}]{howard_application_2017}%
  \BibitemOpen
  \bibfield  {author} {\bibinfo {author} {\bibfnamefont {Mark}\ \bibnamefont
  {Howard}}\ and\ \bibinfo {author} {\bibfnamefont {Earl}\ \bibnamefont
  {Campbell}},\ }\emph {\enquote {\bibinfo {title} {{Application of a Resource
  Theory for Magic States to Fault-Tolerant Quantum Computing}},}\ }\href
  {https://doi.org/10.1103/PhysRevLett.118.090501} {\bibfield  {journal}
  {\bibinfo  {journal} {Phys. Rev. Lett.}\ }\textbf {\bibinfo {volume} {118}},\
  \bibinfo {pages} {090501} (\bibinfo {year} {2017})},\ \Eprint
  {http://arxiv.org/abs/1609.07488} {arXiv:1609.07488}\BibitemShut {NoStop}%
\bibitem [{\citenamefont {Haug}\ and\ \citenamefont
  {Tarabunga}(2026)}]{HaugTarabunga2026}%
  \BibitemOpen
  \bibfield  {author} {\bibinfo {author} {\bibfnamefont {Tobias}\ \bibnamefont
  {Haug}}\ and\ \bibinfo {author} {\bibfnamefont {Poetri~Sonya}\ \bibnamefont
  {Tarabunga}},\ }\emph {\enquote {\bibinfo {title} {Efficient witnessing and
  testing of magic in mixed quantum states},}\ }\href
  {https://doi.org/10.1038/s41534-026-01189-z} {\bibfield  {journal} {\bibinfo
  {journal} {npj Quantum Inf.}\ }\textbf {\bibinfo {volume} {12}},\ \bibinfo
  {pages} {40} (\bibinfo {year} {2026})},\ \Eprint
  {http://arxiv.org/abs/2504.18098} {arXiv:2504.18098}\BibitemShut {NoStop}%
\bibitem [{\citenamefont {Kitaev}(1995)}]{Kitaev1995}%
  \BibitemOpen
  \bibfield  {author} {\bibinfo {author} {\bibfnamefont {Alexei}\ \bibnamefont
  {Kitaev}},\ }\href@noop {} {\emph {\enquote {\bibinfo {title} {{Quantum
  measurements and the Abelian Stabilizer Problem}},}\ }}\Eprint
  {http://arxiv.org/abs/quant-ph/9511026} {arXiv:quant-ph/9511026} (\bibinfo
  {year} {1995})\BibitemShut {NoStop}%
\bibitem [{\citenamefont {Cerezo}\ \emph {et~al.}(2021)\citenamefont {Cerezo},
  \citenamefont {Arrasmith}, \citenamefont {Babbush}, \citenamefont {Benjamin},
  \citenamefont {Endo}, \citenamefont {Fujii}, \citenamefont {McClean},
  \citenamefont {Mitarai}, \citenamefont {Yuan}, \citenamefont {Cincio},\ and\
  \citenamefont {Coles}}]{cerezo_variational_2021}%
  \BibitemOpen
  \bibfield  {author} {\bibinfo {author} {\bibfnamefont {Marco}\ \bibnamefont
  {Cerezo}}, \bibinfo {author} {\bibfnamefont {Andrew}\ \bibnamefont
  {Arrasmith}}, \bibinfo {author} {\bibfnamefont {Ryan}\ \bibnamefont
  {Babbush}}, \bibinfo {author} {\bibfnamefont {Simon~C.}\ \bibnamefont
  {Benjamin}}, \bibinfo {author} {\bibfnamefont {Suguru}\ \bibnamefont {Endo}},
  \bibinfo {author} {\bibfnamefont {Keisuke}\ \bibnamefont {Fujii}}, \bibinfo
  {author} {\bibfnamefont {Jarrod~R.}\ \bibnamefont {McClean}}, \bibinfo
  {author} {\bibfnamefont {Kosuke}\ \bibnamefont {Mitarai}}, \bibinfo {author}
  {\bibfnamefont {Xiao}\ \bibnamefont {Yuan}}, \bibinfo {author} {\bibfnamefont
  {Lukasz}\ \bibnamefont {Cincio}}, \ and\ \bibinfo {author} {\bibfnamefont
  {Patrick~J.}\ \bibnamefont {Coles}},\ }\emph {\enquote {\bibinfo {title}
  {{Variational quantum algorithms}},}\ }\href
  {https://doi.org/10.1038/s42254-021-00348-9} {\bibfield  {journal} {\bibinfo
  {journal} {Nat. Rev. Phys.}\ }\textbf {\bibinfo {volume} {3}},\ \bibinfo
  {pages} {625{\textendash}644} (\bibinfo {year} {2021})},\ \Eprint
  {http://arxiv.org/abs/2012.09265} {arXiv:2012.09265}\BibitemShut {NoStop}%
\bibitem [{\citenamefont {Ross}\ and\ \citenamefont
  {Selinger}(2016)}]{ross_optimal_2016}%
  \BibitemOpen
  \bibfield  {author} {\bibinfo {author} {\bibfnamefont {Neil~J.}\ \bibnamefont
  {Ross}}\ and\ \bibinfo {author} {\bibfnamefont {Peter}\ \bibnamefont
  {Selinger}},\ }\emph {\enquote {\bibinfo {title} {{Optimal ancilla-free
  Clifford+T approximation of z-rotations}},}\ }\href
  {https://doi.org/10.26421/QIC16.11-12-1} {\bibfield  {journal} {\bibinfo
  {journal} {Quantum Inf. Comput.}\ }\textbf {\bibinfo {volume} {16}},\
  \bibinfo {pages} {901{\textendash}953} (\bibinfo {year} {2016})},\ \Eprint
  {http://arxiv.org/abs/1403.2975} {arXiv:1403.2975}\BibitemShut {NoStop}%
\bibitem [{\citenamefont {Akahoshi}\ \emph {et~al.}(2024)\citenamefont
  {Akahoshi}, \citenamefont {Maruyama}, \citenamefont {Oshima}, \citenamefont
  {Sato},\ and\ \citenamefont {Fujii}}]{AkahoshiEtAl2024}%
  \BibitemOpen
  \bibfield  {author} {\bibinfo {author} {\bibfnamefont {Yutaro}\ \bibnamefont
  {Akahoshi}}, \bibinfo {author} {\bibfnamefont {Kazunori}\ \bibnamefont
  {Maruyama}}, \bibinfo {author} {\bibfnamefont {Hirotaka}\ \bibnamefont
  {Oshima}}, \bibinfo {author} {\bibfnamefont {Shintaro}\ \bibnamefont {Sato}},
  \ and\ \bibinfo {author} {\bibfnamefont {Keisuke}\ \bibnamefont {Fujii}},\
  }\emph {\enquote {\bibinfo {title} {{Partially Fault-Tolerant Quantum
  Computing Architecture with Error-Corrected Clifford Gates and Space-Time
  Efficient Analog Rotations}},}\ }\href
  {https://doi.org/10.1103/PRXQuantum.5.010337} {\bibfield  {journal} {\bibinfo
   {journal} {PRX Quantum}\ }\textbf {\bibinfo {volume} {5}},\ \bibinfo {pages}
  {010337} (\bibinfo {year} {2024})},\ \Eprint
  {http://arxiv.org/abs/2303.13181} {arXiv:2303.13181}\BibitemShut {NoStop}%
\bibitem [{\citenamefont {Ismail}\ \emph {et~al.}(2026)\citenamefont {Ismail},
  \citenamefont {Chen}, \citenamefont {Zhao}, \citenamefont {Weiss},
  \citenamefont {Liu}, \citenamefont {Zhou}, \citenamefont {Wang},
  \citenamefont {Sornborger},\ and\ \citenamefont {Kornja{\v
  c}a}}]{IsmailEtAl2026}%
  \BibitemOpen
  \bibfield  {author} {\bibinfo {author} {\bibfnamefont {Refaat}\ \bibnamefont
  {Ismail}}, \bibinfo {author} {\bibfnamefont {I-Chi}\ \bibnamefont {Chen}},
  \bibinfo {author} {\bibfnamefont {Chen}\ \bibnamefont {Zhao}}, \bibinfo
  {author} {\bibfnamefont {Ronen}\ \bibnamefont {Weiss}}, \bibinfo {author}
  {\bibfnamefont {Fangli}\ \bibnamefont {Liu}}, \bibinfo {author}
  {\bibfnamefont {Hengyun}\ \bibnamefont {Zhou}}, \bibinfo {author}
  {\bibfnamefont {Sheng-Tao}\ \bibnamefont {Wang}}, \bibinfo {author}
  {\bibfnamefont {Andrew}\ \bibnamefont {Sornborger}}, \ and\ \bibinfo {author}
  {\bibfnamefont {Milan}\ \bibnamefont {Kornja{\v c}a}},\ }\emph {\enquote
  {\bibinfo {title} {{Transversal Architecture for Megaquop-Scale Quantum
  Simulation with Neutral Atoms}},}\ }\href {https://doi.org/10.1103/j2fw-ccmy}
  {\bibfield  {journal} {\bibinfo  {journal} {PRX Quantum}\ }\textbf {\bibinfo
  {volume} {7}},\ \bibinfo {pages} {020343} (\bibinfo {year} {2026})},\ \Eprint
  {http://arxiv.org/abs/2509.18294} {arXiv:2509.18294}\BibitemShut {NoStop}%
\bibitem [{\citenamefont {Webster}\ \emph {et~al.}(2026)\citenamefont
  {Webster}, \citenamefont {Berent}, \citenamefont {Chandra}, \citenamefont
  {Hockings}, \citenamefont {Baspin}, \citenamefont {Thomsen}, \citenamefont
  {Smith},\ and\ \citenamefont {Cohen}}]{WebsterEtAl2026}%
  \BibitemOpen
  \bibfield  {author} {\bibinfo {author} {\bibfnamefont {Paul}\ \bibnamefont
  {Webster}}, \bibinfo {author} {\bibfnamefont {Lucas}\ \bibnamefont {Berent}},
  \bibinfo {author} {\bibfnamefont {Omprakash}\ \bibnamefont {Chandra}},
  \bibinfo {author} {\bibfnamefont {Evan~T.}\ \bibnamefont {Hockings}},
  \bibinfo {author} {\bibfnamefont {Nou{\'e}dyn}\ \bibnamefont {Baspin}},
  \bibinfo {author} {\bibfnamefont {Felix}\ \bibnamefont {Thomsen}}, \bibinfo
  {author} {\bibfnamefont {Samuel~C.}\ \bibnamefont {Smith}}, \ and\ \bibinfo
  {author} {\bibfnamefont {Lawrence~Z.}\ \bibnamefont {Cohen}},\ }\href@noop {}
  {\emph {\enquote {\bibinfo {title} {{The Pinnacle Architecture: Reducing the
  Cost of Breaking RSA-2048 to 100,000 Physical Qubits Using Quantum LDPC
  Codes}},}\ }}\Eprint {http://arxiv.org/abs/2602.11457} {arXiv:2602.11457}
  [quant-ph] (\bibinfo {year} {2026})\BibitemShut {NoStop}%
\bibitem [{\citenamefont {Yoder}\ \emph {et~al.}(2025)\citenamefont {Yoder},
  \citenamefont {Schoute}, \citenamefont {Rall}, \citenamefont {Pritchett},
  \citenamefont {Gambetta}, \citenamefont {Cross}, \citenamefont {Carroll},\
  and\ \citenamefont {Beverland}}]{YoderEtAl2025}%
  \BibitemOpen
  \bibfield  {author} {\bibinfo {author} {\bibfnamefont {Theodore~J.}\
  \bibnamefont {Yoder}}, \bibinfo {author} {\bibfnamefont {Eddie}\ \bibnamefont
  {Schoute}}, \bibinfo {author} {\bibfnamefont {Patrick}\ \bibnamefont {Rall}},
  \bibinfo {author} {\bibfnamefont {Emily}\ \bibnamefont {Pritchett}}, \bibinfo
  {author} {\bibfnamefont {Jay~M.}\ \bibnamefont {Gambetta}}, \bibinfo {author}
  {\bibfnamefont {Andrew~W.}\ \bibnamefont {Cross}}, \bibinfo {author}
  {\bibfnamefont {Malcolm}\ \bibnamefont {Carroll}}, \ and\ \bibinfo {author}
  {\bibfnamefont {Michael~E.}\ \bibnamefont {Beverland}},\ }\href@noop {}
  {\emph {\enquote {\bibinfo {title} {{Tour de gross: A modular quantum
  computer based on bivariate bicycle codes}},}\ }}\Eprint
  {http://arxiv.org/abs/2506.03094} {arXiv:2506.03094} [quant-ph] (\bibinfo
  {year} {2025})\BibitemShut {NoStop}%
\bibitem [{\citenamefont {Vasmer}\ and\ \citenamefont
  {Browne}(2019)}]{VasmerBrowne2019}%
  \BibitemOpen
  \bibfield  {author} {\bibinfo {author} {\bibfnamefont {Michael}\ \bibnamefont
  {Vasmer}}\ and\ \bibinfo {author} {\bibfnamefont {Dan~E.}\ \bibnamefont
  {Browne}},\ }\emph {\enquote {\bibinfo {title} {{Three-dimensional surface
  codes: Transversal gates and fault-tolerant architectures}},}\ }\href
  {https://doi.org/10.1103/PhysRevA.100.012312} {\bibfield  {journal} {\bibinfo
   {journal} {Phys. Rev. A}\ }\textbf {\bibinfo {volume} {100}},\ \bibinfo
  {pages} {012312} (\bibinfo {year} {2019})},\ \Eprint
  {http://arxiv.org/abs/1801.04255} {arXiv:1801.04255}\BibitemShut {NoStop}%
\bibitem [{\citenamefont {Bomb\'{\i}n}(2015)}]{Bombin2015}%
  \BibitemOpen
  \bibfield  {author} {\bibinfo {author} {\bibfnamefont {H\'{e}ctor}\
  \bibnamefont {Bomb\'{\i}n}},\ }\emph {\enquote {\bibinfo {title} {{Gauge
  color codes: optimal transversal gates and gauge fixing in topological
  stabilizer codes}},}\ }\href {https://doi.org/10.1088/1367-2630/17/8/083002}
  {\bibfield  {journal} {\bibinfo  {journal} {New J. Phys.}\ }\textbf {\bibinfo
  {volume} {17}},\ \bibinfo {pages} {083002} (\bibinfo {year} {2015})},\
  \Eprint {http://arxiv.org/abs/1311.0879} {arXiv:1311.0879}\BibitemShut
  {NoStop}%
\bibitem [{\citenamefont {Stein}\ \emph {et~al.}(2025)\citenamefont {Stein},
  \citenamefont {Xu}, \citenamefont {Cross}, \citenamefont {Yoder},
  \citenamefont {Javadi-Abhari}, \citenamefont {Liu}, \citenamefont {Liu},
  \citenamefont {Zhou}, \citenamefont {Guinn}, \citenamefont {Ding},
  \citenamefont {Ding},\ and\ \citenamefont {Li}}]{SteinEtAl2024}%
  \BibitemOpen
  \bibfield  {author} {\bibinfo {author} {\bibfnamefont {Samuel}\ \bibnamefont
  {Stein}}, \bibinfo {author} {\bibfnamefont {Shifan}\ \bibnamefont {Xu}},
  \bibinfo {author} {\bibfnamefont {Andrew~W.}\ \bibnamefont {Cross}}, \bibinfo
  {author} {\bibfnamefont {Theodore~J.}\ \bibnamefont {Yoder}}, \bibinfo
  {author} {\bibfnamefont {Ali}\ \bibnamefont {Javadi-Abhari}}, \bibinfo
  {author} {\bibfnamefont {Chenxu}\ \bibnamefont {Liu}}, \bibinfo {author}
  {\bibfnamefont {Kun}\ \bibnamefont {Liu}}, \bibinfo {author} {\bibfnamefont
  {Zeyuan}\ \bibnamefont {Zhou}}, \bibinfo {author} {\bibfnamefont {Charles}\
  \bibnamefont {Guinn}}, \bibinfo {author} {\bibfnamefont {Yufei}\ \bibnamefont
  {Ding}}, \bibinfo {author} {\bibfnamefont {Yongshan}\ \bibnamefont {Ding}}, \
  and\ \bibinfo {author} {\bibfnamefont {Ang}\ \bibnamefont {Li}},\ }\emph
  {\enquote {\bibinfo {title} {{Architectures for Heterogeneous Quantum Error
  Correction Codes}},}\ }in\ \href {https://doi.org/10.1145/3676641.3716001}
  {\emph {\bibinfo {booktitle} {Proc. 30th ACM Int. Conf. on Architectural
  Support for Programming Languages and Operating Systems (ASPLOS '25), Vol.
  2}}}\ (\bibinfo  {publisher} {ACM},\ \bibinfo {year} {2025})\ \Eprint
  {http://arxiv.org/abs/2411.03202} {arXiv:2411.03202}\BibitemShut {NoStop}%
\bibitem [{\citenamefont {Landahl}\ and\ \citenamefont
  {Ryan-Anderson}(2014)}]{LandahlRyanAnderson2014}%
  \BibitemOpen
  \bibfield  {author} {\bibinfo {author} {\bibfnamefont {Andrew~J.}\
  \bibnamefont {Landahl}}\ and\ \bibinfo {author} {\bibfnamefont {Ciaran}\
  \bibnamefont {Ryan-Anderson}},\ }\href@noop {} {\emph {\enquote {\bibinfo
  {title} {{Quantum Computing by Color-Code Lattice Surgery}},}\ }}\Eprint
  {http://arxiv.org/abs/1407.5103} {arXiv:1407.5103} [quant-ph] (\bibinfo
  {year} {2014})\BibitemShut {NoStop}%
\bibitem [{\citenamefont {B{\"o}deker}\ \emph {et~al.}(2026)\citenamefont
  {B{\"o}deker}, \citenamefont {M{\'a}rton}, \citenamefont {Colmenarez},
  \citenamefont {Besedin}, \citenamefont {Wallraff},\ and\ \citenamefont
  {M{\"u}ller}}]{BoedekerEtAl2026}%
  \BibitemOpen
  \bibfield  {author} {\bibinfo {author} {\bibfnamefont {Lukas}\ \bibnamefont
  {B{\"o}deker}}, \bibinfo {author} {\bibfnamefont {{\'A}ron}\ \bibnamefont
  {M{\'a}rton}}, \bibinfo {author} {\bibfnamefont {Luis}\ \bibnamefont
  {Colmenarez}}, \bibinfo {author} {\bibfnamefont {Ilya}\ \bibnamefont
  {Besedin}}, \bibinfo {author} {\bibfnamefont {Andreas}\ \bibnamefont
  {Wallraff}}, \ and\ \bibinfo {author} {\bibfnamefont {Markus}\ \bibnamefont
  {M{\"u}ller}},\ }\href@noop {} {\emph {\enquote {\bibinfo {title} {{Lattice
  surgery for near-term experimental logical qubit entanglement creation in
  planar architectures}},}\ }}\Eprint {http://arxiv.org/abs/2606.15190}
  {arXiv:2606.15190} [quant-ph] (\bibinfo {year} {2026})\BibitemShut {NoStop}%
\bibitem [{\citenamefont {Xu}\ \emph {et~al.}(2024)\citenamefont {Xu},
  \citenamefont {Bonilla~Ataides}, \citenamefont {Pattison}, \citenamefont
  {Raveendran}, \citenamefont {Bluvstein}, \citenamefont {Wurtz}, \citenamefont
  {Vasic}, \citenamefont {Lukin}, \citenamefont {Jiang},\ and\ \citenamefont
  {Zhou}}]{XuEtAl2023}%
  \BibitemOpen
  \bibfield  {author} {\bibinfo {author} {\bibfnamefont {Qian}\ \bibnamefont
  {Xu}}, \bibinfo {author} {\bibfnamefont {J.~Pablo}\ \bibnamefont
  {Bonilla~Ataides}}, \bibinfo {author} {\bibfnamefont {Christopher~A.}\
  \bibnamefont {Pattison}}, \bibinfo {author} {\bibfnamefont {Nithin}\
  \bibnamefont {Raveendran}}, \bibinfo {author} {\bibfnamefont {Dolev}\
  \bibnamefont {Bluvstein}}, \bibinfo {author} {\bibfnamefont {Jonathan}\
  \bibnamefont {Wurtz}}, \bibinfo {author} {\bibfnamefont {Bane}\ \bibnamefont
  {Vasic}}, \bibinfo {author} {\bibfnamefont {Mikhail~D.}\ \bibnamefont
  {Lukin}}, \bibinfo {author} {\bibfnamefont {Liang}\ \bibnamefont {Jiang}}, \
  and\ \bibinfo {author} {\bibfnamefont {Hengyun}\ \bibnamefont {Zhou}},\
  }\emph {\enquote {\bibinfo {title} {{Constant-overhead fault-tolerant quantum
  computation with reconfigurable atom arrays}},}\ }\href
  {https://doi.org/10.1038/s41567-024-02479-z} {\bibfield  {journal} {\bibinfo
  {journal} {Nat. Phys.}\ }\textbf {\bibinfo {volume} {20}},\ \bibinfo {pages}
  {1084{\textendash}1090} (\bibinfo {year} {2024})},\ \Eprint
  {http://arxiv.org/abs/2308.08648} {arXiv:2308.08648}\BibitemShut {NoStop}%
\bibitem [{\citenamefont {Chakraborty}\ and\ \citenamefont
  {Gottesman}(2026)}]{ChakrabortyGottesman2026}%
  \BibitemOpen
  \bibfield  {author} {\bibinfo {author} {\bibfnamefont {Aranya}\ \bibnamefont
  {Chakraborty}}\ and\ \bibinfo {author} {\bibfnamefont {Daniel}\ \bibnamefont
  {Gottesman}},\ }\href@noop {} {\emph {\enquote {\bibinfo {title} {{No-Go
  Theorem on Fault Tolerant Gadgets for Multiple Logical Qubits}},}\ }}\Eprint
  {http://arxiv.org/abs/2602.13395} {arXiv:2602.13395} [quant-ph] (\bibinfo
  {year} {2026})\BibitemShut {NoStop}%
\bibitem [{\citenamefont {Campbell}(2016)}]{Campbell2016}%
  \BibitemOpen
  \bibfield  {author} {\bibinfo {author} {\bibfnamefont {Earl~T.}\ \bibnamefont
  {Campbell}},\ }\href@noop {} {\emph {\enquote {\bibinfo {title} {The smallest
  interesting colour code},}\ }}\bibinfo {note} {Accessed: 2026-06-12},\
  \bibinfo {howpublished} {Blog post,
  \url{https://earltcampbell.com/2016/09/26/the-smallest-interesting-colour-code/}}
  (\bibinfo {year} {2016})\BibitemShut {NoStop}%
\bibitem [{\citenamefont {Bravyi}\ and\ \citenamefont
  {K\"{o}nig}(2013)}]{BravyiKonig2013}%
  \BibitemOpen
  \bibfield  {author} {\bibinfo {author} {\bibfnamefont {Sergey}\ \bibnamefont
  {Bravyi}}\ and\ \bibinfo {author} {\bibfnamefont {Robert}\ \bibnamefont
  {K\"{o}nig}},\ }\emph {\enquote {\bibinfo {title} {{Classification of
  Topologically Protected Gates for Local Stabilizer Codes}},}\ }\href
  {https://doi.org/10.1103/PhysRevLett.110.170503} {\bibfield  {journal}
  {\bibinfo  {journal} {Phys. Rev. Lett.}\ }\textbf {\bibinfo {volume} {110}},\
  \bibinfo {pages} {170503} (\bibinfo {year} {2013})},\ \Eprint
  {http://arxiv.org/abs/1206.1609} {arXiv:1206.1609}\BibitemShut {NoStop}%
\bibitem [{\citenamefont {D{\"u}r}\ \emph {et~al.}(2000)\citenamefont
  {D{\"u}r}, \citenamefont {Vidal},\ and\ \citenamefont
  {Cirac}}]{DurVidalCirac2000}%
  \BibitemOpen
  \bibfield  {author} {\bibinfo {author} {\bibfnamefont {Wolfgang}\
  \bibnamefont {D{\"u}r}}, \bibinfo {author} {\bibfnamefont {Guifr{\'e}}\
  \bibnamefont {Vidal}}, \ and\ \bibinfo {author} {\bibfnamefont {J.~Ignacio}\
  \bibnamefont {Cirac}},\ }\emph {\enquote {\bibinfo {title} {{Three qubits can
  be entangled in two inequivalent ways}},}\ }\href
  {https://doi.org/10.1103/PhysRevA.62.062314} {\bibfield  {journal} {\bibinfo
  {journal} {Phys. Rev. A}\ }\textbf {\bibinfo {volume} {62}},\ \bibinfo
  {pages} {062314} (\bibinfo {year} {2000})},\ \Eprint
  {http://arxiv.org/abs/quant-ph/0005115} {arXiv:quant-ph/0005115}\BibitemShut
  {NoStop}%
\bibitem [{\citenamefont {Toshio}\ \emph {et~al.}(2025)\citenamefont {Toshio},
  \citenamefont {Akahoshi}, \citenamefont {Fujisaki}, \citenamefont {Oshima},
  \citenamefont {Sato},\ and\ \citenamefont {Fujii}}]{ToshioEtAl2025}%
  \BibitemOpen
  \bibfield  {author} {\bibinfo {author} {\bibfnamefont {Riki}\ \bibnamefont
  {Toshio}}, \bibinfo {author} {\bibfnamefont {Yutaro}\ \bibnamefont
  {Akahoshi}}, \bibinfo {author} {\bibfnamefont {Jun}\ \bibnamefont
  {Fujisaki}}, \bibinfo {author} {\bibfnamefont {Hirotaka}\ \bibnamefont
  {Oshima}}, \bibinfo {author} {\bibfnamefont {Shintaro}\ \bibnamefont {Sato}},
  \ and\ \bibinfo {author} {\bibfnamefont {Keisuke}\ \bibnamefont {Fujii}},\
  }\emph {\enquote {\bibinfo {title} {{Practical Quantum Advantage on Partially
  Fault-Tolerant Quantum Computer}},}\ }\href
  {https://doi.org/10.1103/PhysRevX.15.021057} {\bibfield  {journal} {\bibinfo
  {journal} {Phys. Rev. X}\ }\textbf {\bibinfo {volume} {15}},\ \bibinfo
  {pages} {021057} (\bibinfo {year} {2025})},\ \Eprint
  {http://arxiv.org/abs/2408.14848} {arXiv:2408.14848}\BibitemShut {NoStop}%
\end{thebibliography}%

%%%%%%%%%%%%%%%%%%%%%%%%%%%%%%%%%%%%%%%%%%%%%%%%%%%%%%%%%%%%%%%%%%%%%%%%%%%%%%%%%%%%%%%%%%%%%%%%%%%%

\vspace*{2mm}
{\noindent}\textbf{Data availability}. 
The datasets underlying this manuscript are available at \url{https://doi.org/10.5281/zenodo.21104533}.

{\noindent}\textbf{Acknowledgments}. 
T.A.\ thanks Tom Peham for valuable discussions on the use of MQT-QECC at QEC 2025, and Lukas Burgholzer for helpful correspondence regarding the toolkit. This project was funded in whole or in part by the Austrian Science Fund (FWF) [P 36478-N, DOI: \href{https://doi.org/10.55776/P36478}{10.55776/P36478}; 
ESP4563925, DOI: \href{https://doi.org/10.55776/ESP4563925}{10.55776/ESP4563925};
SFB BeyondC F7102 and F7109, DOI: \href{https://doi.org/10.55776/F71}{10.55776/F71}; 
WIT9503323, DOI: \href{https://doi.org/10.55776/WIT9503323}{10.55776/WIT9503323}]. 
For open access purposes, the authors have applied a CC BY public copyright license to any author-accepted manuscript version arising from this submission. This work was funded by the European Union (ERC, QuantAI, Project No. 101055129) and the European Union's Horizon Europe research and innovation program under Grant Agreement Number 101114305 ('MILLENION-SGA1'). 
This work was funded by the Intelligence Advanced Research Projects Activity (IARPA), under the Entangled Logical Qubits program through Cooperative Agreement Number W911NF-23-2-0216. 
Views and opinions expressed are however those of the author(s) only and do not necessarily reflect those of the European Union, the European Research Council or any other funding agency. 
Neither the European Union nor the granting authority can be held responsible for them. 
The work was funded by the Austrian Federal Ministry of Education, Science and Research via the Austrian Research Promotion Agency (FFG) through the flagship project FO999897481 (HPQC) and the project FO999914030 (MUSIQ) funded by the European Union{\textemdash}NextGenerationEU.\\

%%%%%%%%%%%%%%%%%%%%%%%%%%%%%%%%%%%%%%%%%%%%%%%%%%%%%%%%%%%%%%%%%%%%%%%%%%%%%%%%%%%%%%%%%%%%%%%%%%%%

{\noindent}\textbf{Author contributions}. 
A.S. performed the experiment and data analysis. 
T.A.\ contributed to the cross-code lattice-surgery theory, designed the flag-qubit fault-tolerance scheme, and performed the numerical simulations. 
P.R. contributed to the conceptualisation and coordination of the research, analysed, interpreted, and visualised results. 
H.P.N. contributed to the conceptualisation of the research. 
H.P.N., N.F. contributed to the cross-code lattice-surgery and fault-tolerance theory. 
M.M., Ch.D.M., R.F., I.P. contributed to the experimental apparatus. 
T.M., P.S., N.F. led the conceptualisation of the research and coordinated research activities across the institutions. 
M.M., A.S., P.S., T.A., P.R., N.F., H.P.N. wrote the initial manuscript. All authors contributed to revisions of the manuscript.

%%%%%%%%%%%%%%%%%%%%%%%%%%%%%%%%%%%%%%%%%%%%%%%%%%%%%%%%%%%%%%%%

%\clearpage
%\newpage
\hypertarget{sec:appendix}
\appendix

\section*{Appendix: Supplemental Information}

\renewcommand{\thesubsubsection}{A.\Roman{subsection}.\arabic{subsubsection}}
\renewcommand{\thesubsection}{A.\Roman{subsection}}
\renewcommand{\thesection}{}
\setcounter{equation}{0}
\numberwithin{equation}{section}
\setcounter{figure}{0}
\renewcommand{\theequation}{A.\arabic{equation}}
\renewcommand{\thefigure}{A.\arabic{figure}}

%%%%%%%%%%%%%%%%%%%%%%%%%%%%%%%%%%%%%%%%%%%%%%%%%%%%%%%%%%%%%%%%

\vspace*{-1mm}
{\noindent}In the appendix we provide additional details supporting the results of the main text. Section~\ref{appendix:lattice surgery and codes} specifies the QEC codes we consider, their logical operators and states, and the smooth-merge lattice-surgery procedure that joins them into the $[\![12,4,2]\!]$ merged code, including the Pauli-frame update across the merge (Sec.~\ref{appendix:pauli_frame}), and we show explicitly that the transversal character of the $\widebar{\mathrm{CZ}}$ and $\widebar{\mathrm{CCZ}}$ gates is preserved by the merge (Sec.~\ref{appendix:transversality_merged}). Section~\ref{appendix:CCZ_state} provides more details on the non-stabiliser $\ket{\widebar{\mathrm{CCZ}}}$ state{\textemdash}its form and GME threshold, its preparation on the merged code, and the 29-term Pauli decomposition used to estimate the fidelity $\mathcal{F}\subtiny{0}{-1.5}{\widebar{\mathrm{CCZ}}}$ and the stabiliser norm~$\mathcal{D}$. Section~\ref{appendix:rotation_gadget} expands on the rotation gadget: its measurement branches and success condition (Sec.~\ref{appendix:rotation_output}), the success-probability scaling and its relation to magic-state injection (Sec.~\ref{appendix:rotation_nesting}), and a bound on how coherent miscalibration of the two analog resource pulses propagates to the logical rotation angle (Sec.~\ref{appendix:rotation_coherent}). Section~\ref{appendix:flags} describes the MQT-QECC-derived encoding circuits (Sec.~\ref{appendix:encoding}) and the flag-qubit subroutine that restores fault tolerance of the colour-code preparation (Sec.~\ref{appendix:flag_FT}). Section~\ref{appendix:supplementary_results} reports complementary measurements: the $\ket{\widebar{\mathrm{CCZ}}}$ state prepared on the bare $[\![8,3,2]\!]$ colour code \emph{without} lattice surgery (Sec.~\ref{appendix:CCZ_colour_only}), and the full protocol set with the $[\![4,1,2]\!]$ sub-code in place of the $[\![4,2,2]\!]$ code (Sec.~\ref{appendix:412 results}). Finally, Sec.~\ref{appendix:physical} gives the physical-qubit circuit diagrams for all three protocols (Secs.~\ref{appendix:circuit_ghz}{\textendash}\ref{appendix:circuit_412_diff}), their decomposition into the trapped-ion native gate set (Sec.~\ref{appendix:transpilation}), and the depolarising-noise model used in all simulations (Sec.~\ref{appendix:noise_model}).

%%%%%%%%%%%%%%%%%%%%%%%%%%%%%%%%%%%%%%%%%%%%%%%%%%%%%%%%%%%%%%%%%%%%%%%%%%%%%%%%
%%%%%%%%%%%%%%%%%%%%%%%%%%%%%%%%%%%%%%%%%%%%%%%%%%%%%%%%%%%%%%%%%%%%%%%%%%%%%%%%

\vspace*{-2mm}

\subsection{Lattice Surgery between Surface Code and 3D Colour Code}\label{appendix:lattice surgery and codes}
\vspace*{-1mm}

{\noindent}In this section, we provide a more detailed description of the individual QEC codes we consider and the lattice-surgery protocol to merge them. We use subscripts to label distinct physical (Arabic numerals, $0,1,\ldots$) and logical qubits (upper-case letters $A,B,\ldots$), and superscripts in parentheses to indicate the particular QEC codes to which the logical operators pertain. In addition, we use horizontal bars above symbols for logical operators and states of logical qubits to better distinguish them from operators on and states of physical qubits, respectively. 
\vspace*{-1mm}

%%%%%%%%%%%%%%%%%%%%%%%%%%%%%%%%%%%%%%%%%%%%%%%%%%%%%%%%%%%%%%%%%%%%%%%%%%%%%%%%

\subsubsection{The $[\![4,2,2]\!]$ Surface Code}
\label{appendix:422}
\vspace*{-1mm}

{\noindent}The $[\![4,2,2]\!]$ code~\eczoo{stab_4_2_2} is a QEC code of four physical qubits, which we label by subscripts $0$, $1$, $2$, and $3$ on the corresponding physical-qubit operators. It encodes two logical qubits that we label~$A$ and $B$, respectively, and represents a version of the smallest patch of a rotated surface code (SC). 
With a distance of $d=2$ it allows for the detection but not the correction of any physical single-qubit error. 
The code space of the $[\![4,2,2]\!]$ code is stabilised by the set $\mathcal{S}\suptiny{1}{0}{\mathrm{(SC)}}=\left\langle\right.\! S\suptiny{1}{0}{\mathrm{(SC)}}_{0}, S\suptiny{1}{0}{\mathrm{(SC)}}_{1}\!\left.\right\rangle$ of stabiliser operators, generated by the two weight-four operators
\begin{subequations}
\begin{align}
    S\suptiny{1}{0}{\mathrm{(SC)}}_{0} &=\,X_0 X_1 X_2 X_3, 
    \\
    S\suptiny{1}{0}{\mathrm{(SC)}}_{1} &=\,Z_0 Z_1 Z_2 Z_3. 
    \label{eq:422 ZZZZ stab}
\end{align}
\end{subequations}
%%%%%%%%%%%%%%%%%%%%%%%%%%%%%%%%%%%%%%%%%%%%%%%%%%%%%%%%%%%%%%%%%%%%%%%%%%%%%%%%
\begin{figure}[!htbp]
    \centering
    %%%trim={<left> <lower> <right> <upper>}
    \includegraphics[width=\linewidth, trim={0 5cm 0 0},clip]{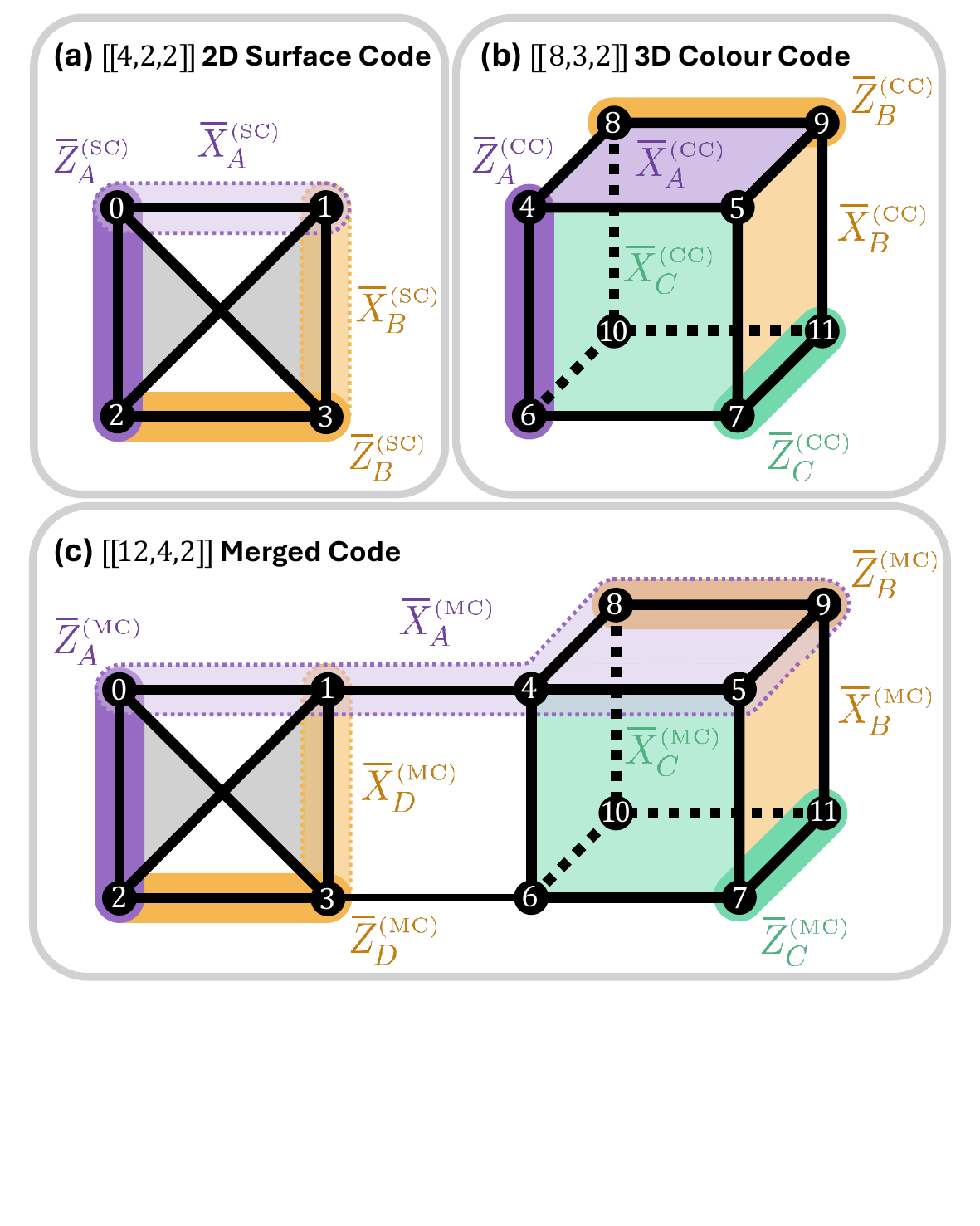}
    \caption{
    \textbf{Codes before and after lattice surgery.}
    \textbf{(a)} The $[\![4,2,2]\!]$ surface code~(SC), defined on four physical qubits (labelled $0$, $1$, $2$, and $3$) encodes two logical qubits with logical operators 
    $\Xbar{\!A}\suptiny{0}{0}{\mathrm{(SC)}}=X_0 X_1$, 
    $\Zbar{\!A}\suptiny{0}{0}{\mathrm{(SC)}}=Z_0 Z_2$, 
    and 
    $\Xbar{\!B}\suptiny{0}{0}{\mathrm{(SC)}}=X_1 X_3$, 
    $\Zbar{\!B}\suptiny{0}{0}{\mathrm{(SC)}}=Z_2 Z_3$, respectively. 
    \textbf{(b)} The $[\![8,3,2]\!]$ 3D colour code~(CC), defined on eight physical qubits (labelled $0,1,\ldots,7$) encodes three logical qubits with logical operators 
    $\Xbar{\!A}\suptiny{0}{0}{\mathrm{(CC)}}=X_4 X_5 X_8 X_9$, 
    $\Zbar{\!A}\suptiny{0}{0}{\mathrm{(CC)}}=Z_4 Z_6$, 
    $\Xbar{\!B}\suptiny{0}{0}{\mathrm{(CC)}}=X_5 X_7 X_9 X_{11}$, 
    $\Zbar{\!B}\suptiny{0}{0}{\mathrm{(CC)}}=Z_8 Z_9$, 
    and 
    $\Xbar{\!C}\suptiny{0}{0}{\mathrm{(CC)}}=X_4 X_5 X_6 X_7$, 
    $\Zbar{\!C}\suptiny{0}{0}{\mathrm{(CC)}}=Z_7 Z_{11}$, respectively.
    \textbf{(c)} The merged $[\![12,4,2]\!]$ code is defined on all twelve physical qubits and encodes four logical qubits. The four logical operators for the two previous~$A$ qubits are replaced by the pair $\Xbar{\!A}\suptiny{0}{0}{\mathrm{(MC)}}=X_0 X_1 X_4 X_5 X_8 X_9$ and $\Zbar{\!A}\suptiny{0}{0}{\mathrm{(MC)}}=\Zbar{\!A}\suptiny{0}{0}{\mathrm{(SC)}}=Z_0 Z_2$. The remaining logical operators are unchanged but relabelled to 
    $\Xbar{\!B}\suptiny{0}{0}{\mathrm{(MC)}}=\Xbar{\!B}\suptiny{0}{0}{\mathrm{(CC)}}$, 
    $\Zbar{\!B}\suptiny{0}{0}{\mathrm{(MC)}}=\Zbar{\!B}\suptiny{0}{0}{\mathrm{(CC)}}$, 
    $\Xbar{\!C}\suptiny{0}{0}{\mathrm{(MC)}}=\Xbar{\!C}\suptiny{0}{0}{\mathrm{(CC)}}$, 
    $\Zbar{\!C}\suptiny{0}{0}{\mathrm{(MC)}}=\Zbar{\!C}\suptiny{0}{0}{\mathrm{(CC)}}$, 
    and 
    $\Xbar{\!D}\suptiny{0}{0}{\mathrm{(MC)}}=\Xbar{\!B}\suptiny{0}{0}{\mathrm{(SC)}}$, 
    $\Zbar{\!D}\suptiny{0}{0}{\mathrm{(MC)}}=\Zbar{\!B}\suptiny{0}{0}{\mathrm{(SC)}}$. 
    }
    \label{fig:codes_overview}
\end{figure}
%%%%%%%%%%%%%%%%%%%%%%%%%%%%%%%%%%%%%%%%%%%%%%%%%%%%%%%%%%%%%%%%%%%%%%%%%%%%%%%%
As illustrated in Fig.~\ref{fig:codes_overview}~(a), the generators of the logical Pauli group (or just ``logical operators" for brevity from now on) for the two encoded logical qubits are
\begin{subequations}
\begin{align}
    \Xbar{\!A}\suptiny{0}{0}{\mathrm{(SC)}} &=\,X_0 X_1, \\
    \Zbar{\!A}\suptiny{0}{0}{\mathrm{(SC)}} &=\,Z_0 Z_2,\\
    \Xbar{\!B}\suptiny{0}{0}{\mathrm{(SC)}} &=\,X_1 X_3, \\
    \Zbar{\!B}\suptiny{0}{0}{\mathrm{(SC)}} &=\,Z_2 Z_3,
\end{align}
\end{subequations}
and the logical computational-basis states, i.e., the simultaneous eigenstates of $S\suptiny{1}{0}{\mathrm{(SC)}}_{0}$ and $S\suptiny{1}{0}{\mathrm{(SC)}}_{1}$ (with eigenvalues $+1$) as well as of $\Zbar{\!A}\suptiny{0}{0}{\mathrm{(SC)}}$ and $\Zbar{\!B}\suptiny{0}{0}{\mathrm{(SC)}}$ (with eigenvalues $\pm1$) are given by
\begin{subequations}
\begin{align}
\ket{\widebar{0}\widebar{0}}_{AB} &=\,
\tfrac{1}{\sqrt{2}}\bigl(\ket{0000}+\ket{1111}\bigr),\\
\ket{\widebar{0}\widebar{1}}_{AB} &=\,
\tfrac{1}{\sqrt{2}}\bigl(\ket{0101}+\ket{1010}\bigr),\\
%%
%\end{align}
%\end{subequations}\addtocounter{equation}{-1}
%\begin{subequations}\addtocounter{equation}{2}
%\begin{align}
\ket{\widebar{1}\widebar{0}}_{AB} &=\,
\tfrac{1}{\sqrt{2}}\bigl(\ket{0011}+\ket{1100}\bigr),\\
\ket{\widebar{1}\widebar{1}}_{AB} &=\,
\tfrac{1}{\sqrt{2}}\bigl(\ket{0110}+\ket{1001}\bigr).
\end{align}
\end{subequations}
The code is self-dual under $H^{\otimes 4}$, which acts transversally as the logical $\widebar{H}\otimes \widebar{H}$ up to a $\mathrm{SWAP}$ of the two logical qubits~$A$ and $B$. 
In the protocols using the $[\![4,2,2]\!]$ code that are presented in the main text we initialise logical qubit~$B$ in $\ket{\widebar{0}}$ before applying $\widebar{H}\otimes \widebar{H}$ transversally to prepare the logical state $\ket{\widebar{+}}$ for logical qubit~$A$. 

\vspace*{2mm}
{\noindent}\textit{The $[\![4,1,2]\!]$ sub-code.} The related $[\![4,1,2]\!]$ code, used in the supplementary measurements of Sec.~\ref{appendix:supplementary_results}, is obtained from $[\![4,2,2]\!]$ by gauge-fixing: by fixing the signs of both $Z_0 Z_1$ and $Z_2 Z_3$, the stabiliser $S_1\suptiny{0}{0}{(\mathrm{SC})}$ in Eq.~\eqref{eq:422 ZZZZ stab} is replaced by the two new stabilisers $Z_0 Z_1$ and $Z_2 Z_3$, which eliminates $\Xbar{B}\suptiny{0}{0}{(\mathrm{SC})}$ and $\Zbar{B}\suptiny{0}{0}{(\mathrm{SC})}$ as logical operators, leaving only $\Xbar{A}\suptiny{0}{0}{(\mathrm{SC})}$ and $\Zbar{A}\suptiny{0}{0}{(\mathrm{SC})}$. The corresponding logical states are
\begin{subequations}\label{eq:412_states}
\begin{align}
\ket{\widebar{0}}\suptiny{0}{0}{\mathrm{(SC)}} &= \tfrac{1}{\sqrt{2}}\bigl(\ket{0101} + \ket{1010}\bigr)\,, \\
\ket{\widebar{1}}\suptiny{0}{0}{\mathrm{(SC)}} &= \tfrac{1}{\sqrt{2}}\bigl(\ket{1001} + \ket{0110}\bigr)\,, \\
\ket{\widebar{+}}\suptiny{0}{0}{\mathrm{(SC)}} &= \tfrac{1}{2}\bigl(\ket{0101} + \ket{1010} + \ket{1001} + \ket{0110}\bigr)\,.
\end{align}
\end{subequations}
The $[\![4,1,2]\!]$ sub-code does not admit a code-space-preserving transversal Hadamard, but $H^{\otimes 4}$ remains a valid basis rotation immediately before measurement and is used as such in the rotation protocol of Sec.~\ref{appendix:412 results}.

%%%%%%%%%%%%%%%%%%%%%%%%%%%%%%%%%%%%%%%%%%%%%%%%%%%%%%%%%%%%%%%%%%%%%%%%%%%%%%%%

\subsubsection{The $[\![8,3,2]\!]$ 3D Colour Code}
\label{appendix:832}

{\noindent}The $[\![8,3,2]\!]$ colour code (CC)~\eczoo{stab_8_3_2}, sometimes referred to as the \textit{``smallest interesting colour code}"~\cite{Campbell2016}, encodes three logical qubits, $A$, $B$, and $C$, using eight physical qubits that we label $4,5,\ldots,11$, see Fig.~\ref{fig:codes_overview}~(b). Much like the previously discussed surface codes, it has a distance of $d=2$, thus allowing for the detection (but not correction) of a single error on any physical qubit. The stabiliser group 
\begin{align}
\mathcal{S}\suptiny{1}{0}{\mathrm{(CC)}}=\left\langle\right.\! 
S\suptiny{1}{0}{\mathrm{(CC)}}_{0}, 
S\suptiny{1}{0}{\mathrm{(CC)}}_{1}, 
S\suptiny{1}{0}{\mathrm{(CC)}}_{2}, 
S\suptiny{1}{0}{\mathrm{(CC)}}_{3}, 
S\suptiny{1}{0}{\mathrm{(CC)}}_{4}
\!\left.\right\rangle
\end{align}
of the $[\![8,3,2]\!]$ code is generated by the stabilisers
\begin{subequations}\label{eq:cc_stabilisers}
\begin{align}
    S\suptiny{1}{0}{\mathrm{(CC)}}_{0} &=\,X_4 X_5 X_6 X_7 X_8 X_9 X_{10} X_{11}, 
    \\
    S\suptiny{1}{0}{\mathrm{(CC)}}_{1} &=\,Z_4 Z_5 Z_6 Z_7,\\
    S\suptiny{1}{0}{\mathrm{(CC)}}_{2} &=\,Z_4 Z_5 Z_8 Z_9,\\
    S\suptiny{1}{0}{\mathrm{(CC)}}_{3} &=\,Z_8 Z_9 Z_{10} Z_{11},\\
    S\suptiny{1}{0}{\mathrm{(CC)}}_{4} &=\,Z_5 Z_7 Z_9 Z_{11}.
\end{align}
\end{subequations}
The three pairs of logical operators for the $[\![8,3,2]\!]$ code are 
\begin{subequations}
\begin{align}
    \Xbar{\!A}\suptiny{0}{0}{\mathrm{(CC)}} &=\,X_4 X_5 X_8 X_9, \\
    \Zbar{\!A}\suptiny{0}{0}{\mathrm{(CC)}} &=\,Z_4 Z_6,\\
    \Xbar{\!B}\suptiny{0}{0}{\mathrm{(CC)}} &=\,X_5 X_7 X_9 X_{11}, \\
    \Zbar{\!B}\suptiny{0}{0}{\mathrm{(CC)}} &=\,Z_8 Z_9,\\
    \Xbar{\!C}\suptiny{0}{0}{\mathrm{(CC)}} &=\,X_4 X_5 X_6 X_7, \\
    \Zbar{\!C}\suptiny{0}{0}{\mathrm{(CC)}} &=\,Z_7 Z_{11}.
\end{align}
\end{subequations}
The code admits transversal $\widebar{\mathrm{CZ}}$ gates on all pairs of logical qubits, and a transversal implementation of the non-Clifford gate $\widebar{\mathrm{CCZ}}$, while $\widebar{\mathrm{CNOT}}$ gates on all logical-qubit pairs can be implemented fault-tolerantly by permutations of physical qubits, a free relabelling on a trapped-ion processor that cannot spread errors, see Table~\ref{tab:transversal appendix}.

\begin{table}[ht!]
\centering
\setlength{\tabcolsep}{6pt}
\renewcommand{\arraystretch}{1.1}
\begin{tabular}{@{}llll@{}}
\toprule
\multicolumn{2}{c}{Transversal} & \multicolumn{2}{c}{Permutation} \\
\cmidrule(lr){1-2}\cmidrule(l){3-4}
Logical & Physical & Logical & Physical \\
\midrule
$\widebar{\mathrm{CZ}}_{AB}$ & $S_4^{\phantom{\dagger}} S_6^{\dagger} S_8^{\dagger} S_{10}^{\phantom{\dagger}}$ & $\widebar{\mathrm{CNOT}}_{AB}$ & $\mathrm{SWAP}_{6,10}\,\mathrm{SWAP}_{7,11}$ \\
$\widebar{\mathrm{CZ}}_{AC}$ & $S_4^{\phantom{\dagger}} S_5^{\dagger} S_8^{\dagger} S_9^{\phantom{\dagger}}$ & $\widebar{\mathrm{CNOT}}_{AC}$ & $\mathrm{SWAP}_{5,9}\,\mathrm{SWAP}_{7,11}$ \\
$\widebar{\mathrm{CZ}}_{BC}$ & $S_4^{\phantom{\dagger}} S_5^{\dagger} S_6^{\dagger} S_7^{\phantom{\dagger}}$ & $\widebar{\mathrm{CNOT}}_{BC}$ & $\mathrm{SWAP}_{5,6}\,\mathrm{SWAP}_{9,10}$ \\
\midrule
\multicolumn{4}{c}{$\widebar{\mathrm{CCZ}} = T_4^{\phantom{\dagger}} T_5^{\dagger} T_6^{\dagger} T_7^{\phantom{\dagger}} T_8^{\dagger} T_9^{\phantom{\dagger}} T_{10}^{\phantom{\dagger}} T_{11}^{\dagger}$ (transversal)} \\
\bottomrule
\end{tabular}
\caption{\textbf{Fault-tolerant gates on the $[\![8,3,2]\!]$ code.} Transversal gates are realised by applying single-qubit $S=\sqrt{Z}$, $T=\sqrt{S}$, and $T^{\dagger}$ to physical qubits. Permutation gates are realised by relabelling physical qubits and are free on hardware with all-to-all connectivity. Controlled-phase gates $\widebar{\mathrm{CZ}}_{ij}$ act on the physical qubits of the cube face in Fig.~\ref{fig:codes_overview}~(b) opposite to the cube face corresponding to $\Xbar{k}\suptiny{0}{0}{\mathrm{(CC)}}$ for $\{i,j,k\}=\{A,B,C\}$. The $\widebar{\mathrm{CCZ}}$ gate is realised by applying $T$ gates to physical qubits corresponding to four non-adjacent corners of the cube, and $T^{\dagger}$ to the four remaining qubits.
\label{tab:transversal appendix}}
\end{table}

Two logical states we use throughout are
\begin{subequations}\label{eq:832_states}
\begin{align}
\ket{\widebar{000}} &= \tfrac{1}{\sqrt{2}}\bigl(\ket{0\ldots 0} + \ket{1\ldots 1}\bigr)\,, \\
\ket{\widebar{+\!+\!+}} &= \tfrac{1}{4}\sum_{w \in \mathcal{C}} \ket{w}\,,
\end{align}
\end{subequations}
where $\mathcal{C}$ is the set of the sixteen $[\![8,3,2]\!]$ codewords.

%%%%%%%%%%%%%%%%%%%%%%%%%%%%%%%%%%%%%%%%%%%%%%%%%%%%%%%%%%%%%%%%%%%%%%%%%%%%%%%%

\vspace*{-1mm}
\subsubsection{The $[\![12,4,2]\!]$ Merged Code}\label{appendix:Z merge 12 4 2}
\vspace*{-1mm}

{\noindent}The $[\![4,2,2]\!]$ surface code and the $[\![8,3,2]\!]$ colour code can be joined via lattice surgery to obtain a $[\![12,4,2]\!]$ merged code (MC). Here, we consider a so-called smooth merge, corresponding to a measurement of the operator
\begin{align}
M_{ZZ}&:=\,\Zbar{\!A}\suptiny{0}{0}{\mathrm{(SC)}}\otimes\Zbar{\!A}\suptiny{0}{0}{\mathrm{(CC)}} \,=\,Z_0 Z_2 Z_4 Z_6.
\end{align}
This measurement can be realised by mapping the result to an auxiliary qubit prepared in the state $\ket{0}$ via four $\mathrm{CNOT}$ gates with control qubits $0$, $2$, $4$, and $6$ and the auxiliary qubit as the target, followed by a measurement of the auxiliary qubit in the computational basis. 
This results in a merged code whose stabiliser group $\mathcal{S}\suptiny{1}{0}{\mathrm{(MC)}}$ is generated by the union of the stabiliser generators of the two initial codes and the new stabiliser generator $S\suptiny{1}{0}{\mathrm{(MC)}}_{7}$. 
That is, the stabilisers $S\suptiny{1}{0}{\mathrm{(MC)}}_{i}$ with $i=0,1,\ldots,6$ of $\mathcal{S}\suptiny{1}{0}{\mathrm{(MC)}}$ are given by
\begin{subequations}\label{eq:merged code stabilisers 0-6}
\begin{align}
    S\suptiny{1}{0}{\mathrm{(MC)}}_{0} &=\,
    S\suptiny{1}{0}{\mathrm{(SC)}}_{0}\,=\,X_0 X_1 X_2 X_3, 
    \\
    S\suptiny{1}{0}{\mathrm{(MC)}}_{1} &=\,
    S\suptiny{1}{0}{\mathrm{(SC)}}_{1}\,=\,Z_0 Z_1 Z_2 Z_3,\\
    S\suptiny{1}{0}{\mathrm{(MC)}}_{2} &=\,
    S\suptiny{1}{0}{\mathrm{(CC)}}_{0}\,=\,X_4 X_5 X_6 X_7 X_8 X_9 X_{10} X_{11}, 
    \\
    S\suptiny{1}{0}{\mathrm{(MC)}}_{3} &=\,
    S\suptiny{1}{0}{\mathrm{(CC)}}_{1} \,=\,Z_4 Z_5 Z_6 Z_7,\\
    S\suptiny{1}{0}{\mathrm{(MC)}}_{4} &=\,
    S\suptiny{1}{0}{\mathrm{(CC)}}_{2} \,=\,Z_4 Z_5 Z_8 Z_9,\\
    S\suptiny{1}{0}{\mathrm{(MC)}}_{5} &=\,
    S\suptiny{1}{0}{\mathrm{(CC)}}_{3} \,=\,Z_8 Z_9 Z_{10} Z_{11},\\
    S\suptiny{1}{0}{\mathrm{(MC)}}_{6} &=\,
    S\suptiny{1}{0}{\mathrm{(CC)}}_{4} \,=\,Z_5 Z_7 Z_9 Z_{11}. 
\end{align}
\end{subequations}
When the outcome $m=+1$ is obtained for the measurement of the auxiliary qubit, the final stabiliser of the merged code is
\begin{align}
    S\suptiny{1}{0}{\mathrm{(MC)}}_{7} &=\,M_{ZZ}
    \,=\,Z_0 Z_2 Z_4 Z_6, 
\end{align}
whereas if the outcome $m=-1$ is obtained, the new stabiliser generator is $S\suptiny{1}{0}{\mathrm{(MC)}}_{7} =-M_{ZZ}$, but by applying a correction $\Xbar{\!A}\suptiny{1}{0}{\mathrm{(CC)}}$ one recovers the stabiliser group generated by the $S\suptiny{1}{0}{\mathrm{(MC)}}_{i}$ in Eq.~(\ref{eq:merged code stabilisers 0-6}) along with $S\suptiny{1}{0}{\mathrm{(MC)}}_{7} = M_{ZZ}$.

The merge eliminates one of the five initial logical qubits by combining the two initial $A$ qubits of the SC and the CC to a new logical $A$ qubit, replacing the initial logical operators $\Xbar{\!A}\suptiny{0}{0}{\mathrm{(SC)}}$, $\Zbar{\!A}\suptiny{0}{0}{\mathrm{(SC)}}$, $\Xbar{\!A}\suptiny{0}{0}{\mathrm{(CC)}}$, and $\Zbar{\!A}\suptiny{0}{0}{\mathrm{(CC)}}$ with the new logical operators
\begin{subequations}
\begin{align}
\Xbar{\!A}\suptiny{0}{0}{\mathrm{(MC)}} &=\,\Xbar{\!A}\suptiny{0}{0}{\mathrm{(SC)}}\otimes\Xbar{\!A}\suptiny{0}{0}{\mathrm{(CC)}}\,=\,
X_0 X_1 X_4 X_5 X_8 X_9,\\
\Zbar{\!A}\suptiny{0}{0}{\mathrm{(MC)}} &=\,\Zbar{\!A}\suptiny{0}{0}{\mathrm{(SC)}}\,=\,Z_0 Z_2,
\end{align}
\end{subequations}
while the remaining logical operators remain unchanged but are relabelled to 
$\Xbar{\!B}\suptiny{0}{0}{\mathrm{(MC)}}=\Xbar{\!B}\suptiny{0}{0}{\mathrm{(CC)}}$, 
$\Zbar{\!B}\suptiny{0}{0}{\mathrm{(MC)}}=\Zbar{\!B}\suptiny{0}{0}{\mathrm{(CC)}}$, 
$\Xbar{\!C}\suptiny{0}{0}{\mathrm{(MC)}}=\Xbar{\!C}\suptiny{0}{0}{\mathrm{(CC)}}$, 
$\Zbar{\!C}\suptiny{0}{0}{\mathrm{(MC)}}=\Zbar{\!C}\suptiny{0}{0}{\mathrm{(CC)}}$, and
$\Xbar{\!D}\suptiny{0}{0}{\mathrm{(MC)}}=\Xbar{\!B}\suptiny{0}{0}{\mathrm{(SC)}}$, 
$\Zbar{\!D}\suptiny{0}{0}{\mathrm{(MC)}}=\Zbar{\!B}\suptiny{0}{0}{\mathrm{(SC)}}$, see Fig.~\ref{fig:codes_overview}~(c). 
\vspace*{1mm}

{\noindent}\textit{Conditional outcome of the merge.} For initial logical states $\ket{\widebar{\psi}}\suptiny{0}{0}{\mathrm{(SC)}}_A = \alpha \ket{\widebar{0}}\suptiny{0}{0}{\mathrm{(SC)}}_A + \beta \ket{\widebar{1}}\suptiny{0}{0}{\mathrm{(SC)}}_A$ on the surface block and $\ket{\widebar{+}}\suptiny{0}{0}{\mathrm{(CC)}}_{A}$ on the colour block, the post-merge state of the merged $A$ qubit is
\begin{equation}\label{eq:merge_outcome}
\begin{aligned}
m_1 = +1 :\ & \alpha \ket{\widebar{0}}_M + \beta \ket{\widebar{1}}_M\,, \\
m_1 = -1 :\ & \alpha \ket{\widebar{01}} + \beta \ket{\widebar{10}}
\xrightarrow{\,\Xbar{A}\suptiny{0}{0}{(\mathrm{CC})}\,} \alpha \ket{\widebar{0}}_M + \beta \ket{\widebar{1}}_M\,.
\end{aligned}
\end{equation}
The conditional $\Xbar{A}\suptiny{0}{0}{(\mathrm{CC})}$ correction is implemented in software via Pauli-frame update, described in Sec.~\ref{appendix:pauli_frame}. 

%%%%%%%%%%%%%%%%%%%%%%%%%%%%%%%%%%%%%%%%%%%%%%%%%%%%%%%%%%%%%%%%%%%%%%%%%%%%%%%%

\subsubsection{Pauli-Frame Update across the Merge}\label{appendix:pauli_frame}

{\noindent}All conditional gates are implemented in post-processing trough either Pauli-frame update or post-selection. Conditional logical operations, which consist solely of single-qubit $X$ or $Z$ gates, are propagated to the measurement stage and updated according to their commutation relations with following Pauli gates. The measured bit string is subsequently updated in post-processing for each gate acting non-trivially after the measurement. In particular, a tracked $X$ gate induces a bit flip on the corresponding qubit when measured in the $Z$ or $Y$ basis, while a tracked $Z$ gate induces a bit flip for measurements performed in the $X$ or $Y$ basis.\\[-2mm]

The Pauli-frame update is straightforward for the GHZ-state preparation and the arbitrary rotation, Fig.~\ref{fig:ghz_merged}~(b) in Sec.~\ref{sec:lattice surgery toolbox} of the main text. The GHZ state. The GHZ state is realised through relabelling of physical qubits, such that no commutation of gates is required; the bit-string is adjusted in post-processing only if the ancilla qubit used for merging is measured in the $m=1$ state. The arbitrary rotation is slightly more involved. Here, the conditional merge and split operations act on the surface code and are followed by a transversal Hadamard gate, which interchanges the $X$ and $Z$ operations under propagation. The surface code is used to determine the state of the correction $r_2$, which is measured in the computational basis. Therefore, the merge correction induces only a phase and can be neglected, while the split correction induces a bit flip.\\[-2mm]

The Pauli-frame update becomes more involved for the CCZ-state protocol{\textemdash}see Fig.~\ref{fig:ghz_merged}~(b) in Sec.~\ref{sec:lattice surgery toolbox} of the main text{\textemdash}since the logical CCZ gate is decomposed into eight physical $T$ and $T^\dagger$ gates acting on the physical qubits associated to the colour code. These gates do not preserve the Pauli group under conjugation and therefore prevent a simple propagation of Pauli corrections within the Pauli frame. Conditional operations are implemented through post-selection rather than Pauli-frame tracking to circumvent this limitation, requiring each measurement configuration to be implemented twice; once with and once without conditional gates. During post-selection, only outcomes consistent with the corresponding branch are retained, while all measurement shots for which the merge ancilla yields the incorrect outcome are discarded, reducing the acceptance rate by approximately $50\%$. The final expectation values are then obtained by combining the two branches through a weighted average over the accepted shots in each branch.

%%%%%%%%%%%%%%%%%%%%%%%%%%%%%%%%%%%%%%%%%%%%%%%%%%%%%%%%%%%%%%%%%%%%%%%%%%%%%%%%

\subsubsection{\texorpdfstring{Transversality of $\widebar{\mathrm{CZ}}$ and $\widebar{\mathrm{CCZ}}$ on the Merged Code}{Transversality of Logical CZ and CCZ Gates on the Merged Code}}\label{appendix:transversality_merged}

{\noindent}The transversal $\widebar{\mathrm{CZ}}$ and $\widebar{\mathrm{CCZ}}$ gates of Table~\ref{tab:transversal appendix}, defined on the bare $[\![8,3,2]\!]$ colour code, continue to act as the corresponding logical gates after the lattice-surgery merge. This is what allows the non-Clifford resource state $\ket{\widebar{\mathrm{CCZ}}}$ to be prepared directly on the $[\![12,4,2]\!]$ merged code rather than on the colour block alone.\\[-2mm]

The argument rests on two facts. First, both gates are realised by single-qubit phase rotations applied only to the eight colour-code physical qubits $4,5,\ldots,11$: the $\widebar{\mathrm{CCZ}}$ by the layer
\begin{equation}
U_{\widebar{\mathrm{CCZ}}} = T_4^{\phantom{\dagger}} T_5^{\dagger} T_6^{\dagger} T_7^{\phantom{\dagger}} T_8^{\dagger} T_9^{\phantom{\dagger}} T_{10}^{\phantom{\dagger}} T_{11}^{\dagger},
\label{eq:ccz_layer_merged}
\end{equation}
and each $\widebar{\mathrm{CZ}}_{ij}$ by a layer of $S$ and $S^{\dagger}$ gates on the corresponding cube face. Being diagonal in the computational basis, both operators are products of $Z$-type single-qubit rotations and therefore commute with every $Z$-type stabiliser and logical operator of the merged code, in particular with the merge stabiliser $M_{ZZ} = Z_0 Z_2 Z_4 Z_6$ and with the surface-block stabilisers $S\suptiny{1}{0}{\mathrm{(MC)}}_{0} = X_0X_1X_2X_3$ and $S\suptiny{1}{0}{\mathrm{(MC)}}_{1} = Z_0Z_1Z_2Z_3$.\\[-2mm]

Second, because $U_{\widebar{\mathrm{CCZ}}}$ acts trivially on the four surface-code qubits $0,1,2,3$, it commutes with the spectator logical operators $\Xbar{\!D}\suptiny{0}{0}{\mathrm{(MC)}} = \Xbar{\!B}\suptiny{0}{0}{\mathrm{(SC)}}$ and $\Zbar{\!D}\suptiny{0}{0}{\mathrm{(MC)}} = \Zbar{\!B}\suptiny{0}{0}{\mathrm{(SC)}}$, and its action on the merged logical operators reduces to its action on the colour block. The merge replaces the two $A$ qubits of the initial codes by the single logical qubit $A$ of the merged code, with $\Xbar{\!A}\suptiny{0}{0}{\mathrm{(MC)}} = \Xbar{\!A}\suptiny{0}{0}{\mathrm{(SC)}} \otimes \Xbar{\!A}\suptiny{0}{0}{\mathrm{(CC)}}$ and $\Zbar{\!A}\suptiny{0}{0}{\mathrm{(MC)}} = \Zbar{\!A}\suptiny{0}{0}{\mathrm{(SC)}}$, while the $B$ and $C$ logical operators are inherited unchanged from the colour block. Under conjugation by $U_{\widebar{\mathrm{CCZ}}}$ the surface factor $\Xbar{\!A}\suptiny{0}{0}{\mathrm{(SC)}}$ commutes through untouched, and the colour factor $\Xbar{\!A}\suptiny{0}{0}{\mathrm{(CC)}}$ transforms exactly as it does on the bare $[\![8,3,2]\!]$ code. The transformation of the merged logical operators is therefore identical to that on the unmerged colour code, so $U_{\widebar{\mathrm{CCZ}}}$ implements the logical $\widebar{\mathrm{CCZ}}$ on the encoded qubits $A$, $B$, and $C$ of the merged code. The same reasoning applies to each transversal $\widebar{\mathrm{CZ}}_{ij}$.\\[-1mm] 

{\noindent}\textit{Beyond the distance-two regime.} The transversal gates used here are features of the distance-two codes and do not all persist as the codes grow. The transversal $\widebar{\mathrm{CCZ}}$ on $[\![8,3,2]\!]$ and the transversal $\widebar{\mathrm{CZ}}$ on $[\![4,2,2]\!]$ both originate in the hypercube code family $[\![2^D, D, 2]\!]$~\cite{KubicaYoshidaPastawski2015, CampbellHoward2017a}~\eczoo{hypercube_quantum}, which has a transversal $C^{D-1}Z$ on its $D$ encoded qubits at fixed distance two: $\widebar{\mathrm{CZ}}$ at $D=2$ and $\widebar{\mathrm{CCZ}}$ at $D=3$. The transversal $\widebar{H} \otimes \widebar{H}$ on $[\![4,2,2]\!]$ is independent of this construction: it is the self-duality of the $[\![4,2,2]\!]$ stabiliser group under $H^{\otimes 4}$, which exchanges the $X$ and $Z$ stabilisers up to a swap of the two encoded qubits. Higher-distance 3D colour codes do not retain a transversal $\widebar{\mathrm{CCZ}}$ on three encoded qubits: the tetrahedral codes (e.g.\ $[\![15,1,3]\!]$~\cite{KubicaYoshidaPastawski2015}~\eczoo{stab_15_1_3}) encode a single logical qubit with a transversal $T$, and a transversal $\widebar{\mathrm{CCZ}}$ on $k=3$ logical qubits is recovered only in stacked three-dimensional codes{\textemdash}stacked surface codes~\cite{VasmerBrowne2019} or in gauge colour codes via gauge fixing~\cite{Bombin2015}{\textemdash}consistent with the Bravyi{\textendash}K\"onig result that 2D topological codes admit only transversal Clifford gates~\cite{BravyiKonig2013}. The $\widebar{\mathrm{CNOT}}$-by-permutation of Table~\ref{tab:transversal appendix} is likewise specific to $[\![8,3,2]\!]$. What carries over to higher distance is not these small-code transversal gates but the lattice-surgery merge between complementary codes, as discussed in Sec.~\ref{sec:discussion} of the main text.

%%%%%%%%%%%%%%%%%%%%%%%%%%%%%%%%%%%%%%%%%%%%%%%%%%%%%%%%%%%%%%%%%%%%%%%%%%%%%%%%

\subsection{\texorpdfstring{Characterization of the $\ket{\widebar{\mathrm{CCZ}}}$ State}{Characterization of the Logical CCZ State}}\label{appendix:CCZ_state}

{\noindent}The state $\ket{\widebar{\mathrm{CCZ}}}$ is the non-stabiliser GME state verified in the main text. Here we describe its form, its preparation on the merged code, and the Pauli decomposition used to estimate its fidelity and stabiliser norm.

{\noindent}\textit{State form and GME threshold.}
Applying the transversal $\widebar{\mathrm{CCZ}}$ to the logical $\ket{\widebar{+\!+\!+}}$ state of the $[\![8,3,2]\!]$ code yields
\begin{equation}
\ket{\widebar{\mathrm{CCZ}}} \;=\; \widebar{\mathrm{CCZ}}\ket{\widebar{+\!+\!+}} \;=\; \frac{1}{2\sqrt{2}}\sum_{a,b,c\in\{0,1\}}(-1)^{abc}\ket{\widebar{abc}}.
\label{eq:CCZ_Zbasis}
\end{equation}
Expanded in the logical $\ket{\widebar{\pm}}$ basis,
\begin{align}
\ket{\widebar{\mathrm{CCZ}}} \;=\;& \tfrac{3}{4}\ket{\widebar{+\!+\!+}}
+ \tfrac{1}{4}\bigl(\ket{\widebar{+\!+\!-}}+\ket{\widebar{+\!-\!+}}+\ket{\widebar{-\!+\!+}}\bigr) \nonumber\\
&- \tfrac{1}{4}\bigl(\ket{\widebar{+\!-\!-}}+\ket{\widebar{-\!+\!-}}+\ket{\widebar{-\!-\!+}}\bigr)
+ \tfrac{1}{4}\ket{\widebar{-\!-\!-}},
\label{eq:CCZ_PMbasis}
\end{align}
showing its dominant overlap with $\ket{\widebar{+\!+\!+}}$. Across every bipartition of the three logical qubits the reduced state has Schmidt coefficients $\{\sqrt{3}/2,\,1/2\}$, so the largest squared Schmidt coefficient is $\lambda_{\max}^2 = 3/4$. Since the maximal fidelity attainable by any biseparable state equals $\lambda_{\max}^2$, a measured fidelity $\mathcal{F} > 3/4$ certifies GME. The state belongs to the GHZ SLOCC entanglement class~\cite{DurVidalCirac2000}{\textemdash}its three-tangle is nonzero, $\tau = 1/4${\textemdash}but unlike $\ket{\mathrm{GHZ}}$ it is a non-stabiliser (magic) state, which is what the stabiliser norm $\mathcal{D}$ of the main text witnesses.

\vspace*{2mm}
{\noindent}\textit{Preparation on the merged code.}
After $Z$ merging the surface code prepared in $\ket{\widebar{+}}\suptiny{0}{0}{\mathrm{(SC)}}$ with the colour code prepared in $\ket{\widebar{+\!+\!+}}\suptiny{0}{0}{\mathrm{(CC)}}$, the transversal $\widebar{\mathrm{CCZ}}$ (the $T/T^{\dagger}$ layer of Table~\ref{tab:transversal appendix} on the eight colour-code qubits) is applied on the merged code; the conditional $\Xbar{\!A}\suptiny{0}{0}{\mathrm{(CC)}}$ correction is applied when the merge ancilla outcome is $m_1 = -1$, as described in Sec.~\ref{appendix:pauli_frame}. The circuit is shown in Fig.~\ref{fig:ghz_merged}~(b) of the main text, and the fidelity of the resulting state is estimated via the Pauli decomposition below.

\begin{figure}[t!]
    \centering
    \includegraphics[width=\columnwidth]{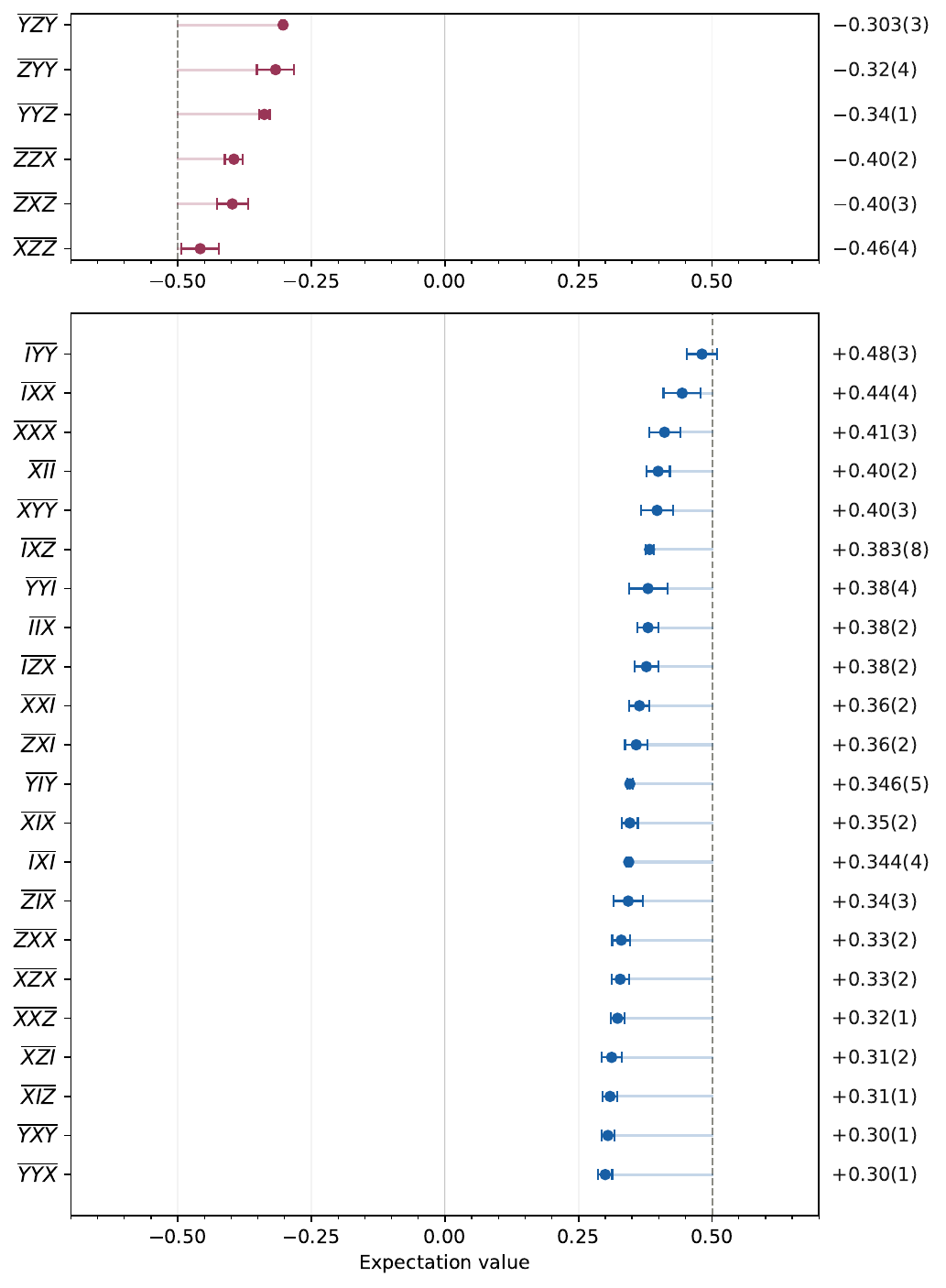}
    \vspace*{-6mm}
    \caption{\textbf{Pauli expectation values for $\ket{\widebar{\mathrm{CCZ}}}$ on the $[\![12,4,2]\!]$ merged code.} All 28 non-identity Pauli expectation values measured on the experimentally prepared $\ket{\widebar{\mathrm{CCZ}}}$ state as described in Sec.~\ref{sec:lattice surgery toolbox} and Fig.~\ref{fig:ghz_merged} of the main text. The 6 anti-correlated terms (top, ideal value $-1/2$) and 22 correlated terms (bottom, ideal value $+1/2$) are sorted within each group by their deviation from the ideal value. Error bars denote one standard deviation; the line connecting each point to the dashed ideal value indicates its deviation. The state fidelity is $F = 0.761(7)$ and the stabiliser witness is $\mathcal{D} = 1.396(15)$ in the half-FT configuration, exceeding the GME certification threshold $F > 3/4$. Approximately 2000{\textendash}2500 shots were taken per observable.}
    \label{fig:ccz_pauli_deviation}
\end{figure}

\vspace*{2mm}
{\noindent}\textit{Pauli decomposition.}
The fidelity of an experimental state $\rho$ with the ideal $\ket{\widebar{\mathrm{CCZ}}}$ state decomposes as~\cite{TothGuhne2005}
\begin{equation}
\mathcal{F}(\rho) \;=\; \frac{1}{8}\sum_{i,j,k=0}^{3} c_{ijk}\,\mathrm{Tr}\bigl(\rho\,\sigma_i\otimes\sigma_j\otimes\sigma_k\bigr),
\label{eq:fidelity_decomp}
\end{equation}
where $c_{ijk}$ are the Bloch coefficients of the target state. Of the $64$ Pauli terms, $29$ have nonzero $c_{ijk}$: the identity ($c_{000}=1$), six terms with ideal value $-1/2$, and $22$ with ideal value $+1/2$; the remaining $35$ vanish under the $S_3$ permutation symmetry of the state $\ket{\widebar{\mathrm{CCZ}}}$. The values are shown in Fig.~\ref{fig:ccz_pauli_deviation}.

\vspace*{2mm}
{\noindent}\textit{Clifford twirling as an alternative.}
An alternative to measuring all twenty-nine terms is Clifford twirling followed by a single-setting measurement of logical $\widebar{Z}\widebar{Z}\widebar{X}$, which concentrates the infidelity into a complementary bin and improves the sample complexity for fidelity estimation from $\mathcal{O}(1/\epsilon^2)$ to $\mathcal{O}(1/\epsilon)$~\cite{LeeEtAl2025}. This advantage is asymptotic: at the state fidelities and shot counts relevant here, our simulations show that the twirled estimate does not appreciably increase the margin of $\mathcal{F}$ above the GME threshold $3/4$ relative to the full $29$-term scan, while the full scan additionally yields the stabiliser norm $\mathcal{D}$ used to certify non-stabiliserness. We therefore measured the complete Pauli decomposition. The sample-complexity advantage of twirling will become significant at higher code distance, where it is the natural fidelity-estimation protocol.

%%%%%%%%%%%%%%%%%%%%%%%%%%%%%%%%%%%%%%%%%%%%%%%%%%%%%%%%%%%%%%%%%%%%%%%%%%%

\subsection{Rotation Gadget}\label{appendix:rotation_gadget}

{\noindent}This appendix details the rotation gadget used in the rotation protocol (Fig.~\ref{fig:rotationLS} in the main text): its four measurement branches and success condition (Sec.~\ref{appendix:rotation_output}), its success-probability scaling and relation to magic-state injection (Sec.~\ref{appendix:rotation_nesting}), and a bound on how coherent miscalibration of the two analog resource pulses propagates to the logical rotation angle (Sec.~\ref{appendix:rotation_coherent}). We label the three logical qubits by capital letters: $A$ is the data qubit, carrying the input $\ket{\widebar{\psi}}$ to be rotated, while $B$ and $C$ hold the rotated resource states $\ket{\widebar{+}_\theta}$ and $\ket{\widebar{0}_{2\theta}}$ that are consumed and measured. The gadget is adapted from the gate-teleportation construction of~\cite{JerbiEtAl2023}, where the rotation is teleported without encoding or lattice surgery; here it is realised at the logical level across two codes, drawing the non-Clifford $\widebar{\mathrm{CCZ}}$ from the $[\![8,3,2]\!]$ colour code and the terminal Hadamard from the $[\![4,2,2]\!]$ surface code through the cross-code merge.

%%%%%%%%%%%%%%%%%%%%%%%%%%%%%%%%%%%%%%%%%%%%%%%%%%%%%%%%%%%%%%%%%%%%%%%%%%%

\subsubsection{Gadget Output States}\label{appendix:rotation_output}

{\noindent}The goal of the gadget is to rotate the input state $\ket{\widebar{\psi}}$ by an angle $\theta$, returning $\widebar{R}_Z(\theta)\ket{\widebar{\psi}}=e^{-i\theta/2}\alpha\ket{\widebar{0}} + e^{i\theta/2}\beta\ket{\widebar{1}}$. The circuit of Fig.~\ref{fig:rotation_gadget} produces a superposition of four states,
\begin{equation}\label{eq:rotation_output}
\begin{aligned}
\ket{\widebar{\mathrm{gadget}}} \propto & \widebar{R}_Z(\theta)\ket{\widebar{\psi}}\bigl(\ket{\widebar{00}}\, e^{-i\theta}
+\ket{\widebar{01}}\, e^{i\theta}+\ket{\widebar{10}}\bigr) \\
&+ \widebar{R}_Z(-3\theta)\ket{\widebar{\psi}}\ket{\widebar{11}}\,,
\end{aligned}
\end{equation}
distinguished by measuring the resource qubits $B$ and $C$. For three of the four outcomes ($BC \in \{\ket{\widebar{00}}, \ket{\widebar{01}}, \ket{\widebar{10}}\}$) the data qubit $A$ is correctly rotated to $\widebar{R}_Z(\theta)\ket{\widebar{\psi}}$; the fourth outcome, $\ket{\widebar{11}}$, leaves $A$ with the wrong phase $-3\theta$. This wrong-phase outcome could in principle be corrected by re-applying the gadget with $\theta'=4\theta$; here we instead use it as a flag and discard the run.

%%%%%%%%%%%%%%%%%%%%%%%%%%%%%%%%%%%%%%%%%%%%%%%%%%%%%%%%%%%%%%%%%%%%%%%%%%%

\subsubsection{Success Probability and Analog-Rotation Injection}\label{appendix:rotation_nesting}

{\noindent}The $\theta'=4\theta$ correction is the first step of a general construction: iterating it across $N$ resource qubits raises the success probability to $1-2^{-N}$~\cite{JerbiEtAl2023}, the present two-qubit gadget being the $N=2$ instance ($1-2^{-2}=3/4$). Each added round requires one controlled-phase gate one level higher in the Clifford hierarchy, $\widebar{\mathrm{C}^{N}\mathrm{Z}}$; these are transversal on the hypercube code family $[\![2^{D},D,2]\!]$ (Sec.~\ref{appendix:transversality_merged}), with the $[\![8,3,2]\!]$ code ($D=3$) supplying the $\widebar{\mathrm{CCZ}}=\widebar{\mathrm{C}^{2}\mathrm{Z}}$ used here. This raises only the success probability, not the rotation fidelity conditional on success, which is set by physical noise on the unencoded resource states and Clifford gates. Since the hypercube codes are distance two (error-detecting only) and cost $2^{D}$ physical qubits, each added round halves the discard rate at an exponential qubit cost without improving the fidelity of the accepted output.

\begin{figure}[t!]
\centering
\begin{quantikz}[wire types={q,q,q}]
\lstick{\small$\ket{\widebar{\psi}}_A$}          & \ctrl{1} & \ctrl{1}  & \qw      & \qw       & \rstick{\small$\ket{\widebar{\psi}'}$}\qw \\
\lstick{\small$\ket{\widebar{+}_\theta}_B$}  & \targ{}  & \ctrl{1}  & \qw      & \meter{\widebar{Z}} \\
\lstick{\small$\ket{\widebar{0}_{2\theta}}_C$} & \qw      & \ctrl{-1} & \gate{\widebar{H}} & \meter{\widebar{Z}}
\end{quantikz}
\scriptsize{ \caption{\textbf{Rotation gadget.} The data qubit $A$ holds the input $\ket{\widebar{\psi}}$, and the
resource qubits $B$, $C$ are prepared in $\ket{\widebar{+}_\theta}=\widebar{R}_Z(\theta)\ket{\widebar{+}}$
and $\ket{\widebar{0}_{2\theta}}=\widebar{R}_X(2\theta)\ket{\widebar{0}}$. A logical
$\widebar{\mathrm{CNOT}}_{AB}$ is followed by a $\widebar{\mathrm{CCZ}}_{ABC}$ and a transversal
$\widebar{H}$ on $C$, after which $B$ and $C$ are measured in the logical $\widebar{Z}$ basis.
The three success outcomes ($BC\in\{\ket{\widebar{00}},\ket{\widebar{01}},\ket{\widebar{10}}\}$,
total probability $3/4$) leave $A$ in $\ket{\widebar{\psi}'}=\widebar{R}_Z(\theta)\ket{\widebar{\psi}}$;
the single failure outcome $\ket{\widebar{11}}$ ($1/4$) yields $\widebar{R}_Z(-3\theta)\ket{\widebar{\psi}}$,
which is heralded and discarded.
\label{fig:rotation_gadget}}}
\end{figure}

The gadget is complementary to the small-angle analog-rotation injection of the space-time-efficient analog-rotation (STAR) architecture~\cite{AkahoshiEtAl2024, ToshioEtAl2025}, whose magic state is prepared fault-tolerantly with a logical error that scales with the rotation angle. STAR is therefore advantageous only for small angles, whereas the gadget injects an exact angle; and STAR injects magic into a surface code, whereas the gadget consumes a $\widebar{\mathrm{CCZ}}$ already native to the colour code. This is not a fidelity advantage{\textemdash}at equal code distance STAR's fault-tolerant injection reaches a far lower error floor, and the gadget as demonstrated here is unencoded{\textemdash}but it makes the gadget a natural rotation primitive on a processor that already provides a transversal multi-controlled-$Z$.

%%%%%%%%%%%%%%%%%%%%%%%%%%%%%%%%%%%%%%%%%%%%%%%%%%%%%%%%%%%%%%%%%%%%%%%%%%%

\subsubsection{Bound on Coherent Miscalibration of Analog Pulses}\label{appendix:rotation_coherent}

{\noindent}The depolarising-noise simulations of Sec.~\ref{appendix:noise_model} ($p_1=0.005$, $p_2=0.015$) place a Pauli-stochastic error on every gate, including the analogue rotations $R_z(\theta)$ on qubit $B$ and $R_x(2\theta)$ on qubit $C$ that carry the rotation parameter through the gadget. This model is blind to \emph{coherent} over-rotation of those two pulses on two counts: a depolarising channel and a unitary over-rotation of the same average strength leave different signatures on the logical output, and $p_1$ is an average depolarising probability per Clifford gate, not the error of an arbitrary-angle single-qubit pulse. We close both gaps with the state-vector analysis below, which propagates a pulse miscalibration $\delta=(\delta_z,\delta_x)$ to the logical rotation angle of the gadget output.\\[2mm]

{\noindent}\textit{Propagation to the logical angle.} To leading order, $R_z(\theta)\!\to\!R_z(\theta+\delta_z)$ on qubit $B$ inserts $\exp(-\texttt{i}\,\delta_z \widebar{Z}_B/2)$ and $R_x(2\theta)\!\to\!R_x(2\theta+\delta_x)$ on qubit $C$ inserts $\exp(-\texttt{i}\,\delta_x \widebar{X}_C/2)$ before the gadget unitary $U \equiv \widebar{\mathrm{CNOT}}_{A\to B}\cdot\widebar{\mathrm{CCZ}}_{ABC}$; the first-order effect is fixed by how the generators $\widebar{Z}_B$, $\widebar{X}_C$ propagate onto qubit $A$. Since $\widebar{\mathrm{CCZ}}$ is diagonal, $\widebar{Z}_B$ commutes through it and the $\widebar{\mathrm{CNOT}}$ maps it to $\widebar{Z}_A\widebar{Z}_B$; projecting $\widebar{Z}_B=(-1)^{r_B}$ leaves a $\widebar{Z}_A$ byproduct of sign $(-1)^{r_B}$, independent of $r_C$. The generator $\widebar{X}_C$ instead picks up $\widebar{\mathrm{CCZ}}\,\widebar{X}_C\,\widebar{\mathrm{CCZ}} = \widebar{\mathrm{CZ}}_{AB}\,\widebar{X}_C$, which after the $\widebar{\mathrm{CNOT}}$ and the $\widebar{Z}_B=(-1)^{r_B}$ projection gives a $\widebar{Z}_A$ byproduct on the $r_B=0$ branches and none on the $r_B=1$ branches. The resulting per-branch first-order slopes $\kappa^{(z,x)}_{r_B r_C} = \partial\theta_{\mathrm{eff}}/\partial\delta_{z,x}$ are collected in Table~\ref{tab:kappa-pattern}. The protocol post-selects on $(r_B,r_C)\neq(0,0)$ without per-branch corrections, so the output is the equal-weight mixture of the three success branches; averaging the slopes over them gives
\begin{equation}
  \boxed{\;
    \begin{aligned}
      \delta\theta_{\mathrm{logical}}^{\mathrm{success}}
        &\;=\; \tfrac{1}{3}\bigl(\delta_z + \delta_x\bigr) + \mathcal{O}(\delta^2),\\[2pt]
      \langle \widebar{Z}_A \rangle^{\mathrm{success}} &\;=\; 0 \text{ identically.}
    \end{aligned}
  \;}
  \label{eq:rotation-sensitivity}
\end{equation}
The factor $1/3$ is the cost of averaging over the three success branches: recording $(r_B,r_C)$ and propagating the corresponding $\widebar{Z}_A$ into the downstream Pauli frame would make every branch contribute with $\kappa=1$, so the logical angle would track the physical miscalibration faithfully.

\begin{table}[h]
  \centering
  \begin{tabular}{l|cccc}
    & $(0,0)$ & $(0,1)$ & $(1,0)$ & $(1,1)$ \\
    & failure & \multicolumn{3}{c}{$\leftarrow$ success $\rightarrow$} \\
    \hline
    $\kappa_z = \partial\theta_{\mathrm{eff}}/\partial\delta_z$ & $-1$ & $-1$ & $+1$ & $+1$ \\
    $\kappa_x = \partial\theta_{\mathrm{eff}}/\partial\delta_x$ & $-1$ & $+1$ & $\phantom{+}0$ & $\phantom{+}0$ \\
  \end{tabular}
  \caption{Per-branch first-order sensitivity of the logical rotation angle to the two physical-pulse miscalibrations. The column label $(r_B,r_C)$ encodes the $\widebar{Z}_B$ measurement on qubit $B$ and the $\widebar{X}_C$ measurement on qubit $C$ ($0$ for the $+1$ eigenvalue); $(0,0)$ is the post-selected-out failure branch. Both axes sum to $+1$ over the three success branches, giving the common blind-post-selected average $\kappa_z=\kappa_x=+1/3$ through different per-branch patterns.}
  \label{tab:kappa-pattern}
\end{table}

\vspace*{2mm}
{\noindent}\textit{Vanishing of $\langle \widebar{Z}_A \rangle$.} The $\widebar{Z}_A$ tilt vanishes not only at first order but exactly, on every branch and at all orders in $\delta$. After projection, the qubit-$A$ amplitudes carry all $\delta$ dependence in their phases and in the relative sign of a single $\sin(\theta+\delta_x/2)$ term, never in their moduli, so $|c_0|^2 = |c_1|^2 = 1/2$ on every branch by the Pythagorean identity and $\langle \widebar{Z}_A \rangle = |c_0|^2 - |c_1|^2 = 0$. State-vector simulation across $\delta \in \{0, \pm0.1, \pm0.2, \pm0.4\}$~rad confirms $|\langle \widebar{Z}_A \rangle| < 10^{-16}$.

\vspace*{2mm}
{\noindent}\textit{Comparison with the data.} Equation~\eqref{eq:rotation-sensitivity} makes two falsifiable predictions for the six-angle scan of Fig.~\ref{fig:rotationLS} of the main text: an exactly zero $\langle \widebar{Z}_A \rangle$, and an analog-pulse contribution to the angle shift of at most $p_1/3 \approx 0.10^\circ$. The measured $\langle \widebar{Z}_A \rangle$ is consistent with zero throughout (largest deviation $-0.053 \pm 0.041$ at $\theta=5\pi/12$, $1.3\sigma$; mean $-0.021 \pm 0.017$). The measured angle shifts agree with the depolarising simulation within $1\sigma$ at every angle except $\theta=\pi/3$, where a $+8.4\pm2.9^\circ$ residual ($2.9\sigma$) remains; reproducing it from analog-pulse miscalibration alone would require $|\delta_z+\delta_x|\approx 0.42$~rad, $\sim\!80\times$ the RB-bounded value and inconsistent with independent gate calibrations. The residual is therefore either a $\sim\!3\sigma$ fluctuation or a sub-leading channel outside the scope of this bound (coherent error on the entangling gates inside the transversal $\widebar{\mathrm{CCZ}}$, governed by $p_2$, or correlated errors across the eight $T/T^\dagger$ pulses). In either case, coherent miscalibration of the two analog pulses is not the origin of the residual.

%%%%%%%%%%%%%%%%%%%%%%%%%%%%%%%%%%%%%%%%%%%%%%%%%%%%%%%%%%%%%%%%%%%%%%%%%%%

\subsection{Encoding Circuits and Flag-Qubit Fault Tolerance}\label{appendix:flags}

%%%%%%%%%%%%%%%%%%%%%%%%%%%%%%%%%%%%%%%%%%%%%%%%%%%%%%%%%%%%%%%%%%%%%%%%%%%

\subsubsection{Encoding Circuits via MQT-QECC}\label{appendix:encoding}

{\noindent}The non-fault-tolerant encoding circuits for both code blocks are obtained from the MQT-QECC toolkit~\cite{PehamEtAl2025}, which synthesises Clifford circuits preparing the target logical state from the all-zero physical state by encoding the stabiliser constraints as a Boolean satisfiability (SAT) instance and searching over sequences of CNOT and $H$ gates with a SAT solver. For the $[\![8,3,2]\!]$ CC (Sec.~\ref{appendix:832}) the toolkit returns a depth-four circuit of ten CNOTs and four Hadamards preparing $\ket{\widebar{+\!+\!+}}\suptiny{0}{0}{\mathrm{(CC)}}$; For the $[\![4,2,2]\!]$ SC (Sec.~\ref{appendix:422}) a two-CNOT, two-Hadamard circuit preparing $\ket{\widebar{+0}}\suptiny{0}{0}{\mathrm{(SC)}}$. Both encoders are shown explicitly in the circuits of Figs.~\ref{fig:circuit_ghz}{\textendash}\ref{fig:circuit_rotation}; The $[\![4,1,2]\!]$ sub-code of Sec.~\ref{appendix:412 results} uses the same surface-block encoder with the spectator logical qubit gauged out (Fig.~\ref{fig:circuit_412_encoder}).

The toolkit offers several synthesis strategies{\textemdash}gate-count-optimal and depth-optimal SAT-based searches, and a faster heuristic search{\textemdash}which for a given code can return inequivalent circuits with different gate counts, depths, and qubit assignments. For the $[\![8,3,2]\!]$ state preparation used here, the gate-count-optimal and heuristic searches both return ten-CNOT circuits, differing only in which qubits carry the Hadamards and in the ordering of gates. Its ten-CNOT structure coincides, up to relabelling, with the colour-code encoder of Wang \emph{et al.}~\cite{WangEtAl2023}, so the fault-propagation analysis and flag construction of Sec.~\ref{appendix:flag_FT} apply directly.

Both encoders are non-fault-tolerant by construction. A single physical fault during the ten colour-code CNOTs can propagate to a weight-two $Z$ error, which is dangerous because the code has distance $d=2$ and its logical $\widebar{Z}$ operators are weight-two (Sec.~\ref{appendix:832}). The surface block is prepared by one of two short encoders{\textemdash}a two-CNOT circuit for $\ket{\widebar{+0}}\suptiny{0}{0}{\mathrm{(SC)}}$ and a three-CNOT circuit for $\ket{\widebar{00}}\suptiny{0}{0}{\mathrm{(SC)}}${\textemdash}and we omit its flag in both cases. Flagging the surface block would require measuring a weight-four stabiliser of the $[\![4,2,2]\!]$ code, which costs four data-to-ancilla CNOTs, more than the encoder itself; at $d=2$ this overhead introduces more error than it removes, and both simulation and hardware confirm that it lowers the overall fidelity. We restore fault tolerance (in the sense defined in Sec.~\ref{sec:background} of the main text) only on the colour-code preparation, where it is most needed, by appending a flag-qubit measurement, as described next.

%%%%%%%%%%%%%%%%%%%%%%%%%%%%%%%%%%%%%%%%%%%%%%%%%%%%%%%%%%%%%%%%%%%%%%%%%%%

\subsubsection{Flag-Qubit Fault Tolerance}\label{appendix:flag_FT}

{\noindent}\textit{Dangerous errors.} An $X$ or $Y$ fault on the control qubit of a CNOT in the ten-CNOT encoder of Sec.~\ref{appendix:encoding} propagates as a $Z$ error through the subsequent CNOTs. Tracing all single-qubit faults through the circuit yields six weight-two $Z$ errors at the encoder output that are each stabiliser-equivalent to a logical $\widebar{Z}$ operator of the $[\![8,3,2]\!]$ code (Table~\ref{tab:flags}). These are the only dangerous faults: a weight-one $Z$ error anticommutes with at least one $Z$ stabiliser and is caught by stabiliser readout in any subsequent protocol, while the logical $\widebar{X}$ operators of the code are weight four, so no propagated weight-two $X$ error can be equivalent to a logical operator. The six errors fall into two logical classes, four equivalent to $\Zbar{\!C}\suptiny{0}{0}{\mathrm{(CC)}}$ and two to $\Zbar{\!B}\suptiny{0}{0}{\mathrm{(CC)}}$.

{\noindent}\textit{Admissible flags.} A flag $X_S$ measured on a single ancilla qubit anticommutes with $Z_i Z_j$ exactly when $|\{i,j\} \cap S|$ is odd, flipping the ancilla outcome so that the faulty shot can be post-selected out. Detecting all six errors requires odd intersection with each of the six pairs. The four pairs $\{4,8\}$, $\{5,9\}$, $\{6,10\}$, $\{7,11\}$ force $S$ to contain exactly one qubit from each, and the two further pairs $\{8,9\}$ and $\{10,11\}$ then single out four admissible flags: $\{4,6,9,11\}$, $\{4,7,9,10\}$, $\{5,6,8,11\}$, and $\{5,7,8,10\}$. The corresponding $Z$ operators are all elements of the code's $Z$-stabiliser group; for the flag we adopt, $Z_4 Z_6 Z_9 Z_{11} = (Z_4 Z_5 Z_6 Z_7)(Z_5 Z_7 Z_9 Z_{11})$ is the product of two of the four $Z$-stabiliser generators of Eq.~(\ref{eq:cc_stabilisers}). We use $S_\star = \{4,6,9,11\}$, which gives the highest preparation fidelity in numerical simulation; the other three flags detect the same six errors and perform within ${\sim}1\%$. The flag $S_{\mathrm{Wang}} = \{6,7,8,9\}$ used by~\cite{WangEtAl2023} with the same encoder contains both qubits of the pair $\{8,9\}$ and neither of $\{10,11\}$, so it misses both $\Zbar{\!B}\suptiny{0}{0}{\mathrm{(CC)}}$-class errors and detects only four of the six (Table~\ref{tab:flags}). The logical observable read out in the one-bit-addition algorithm of~\cite{WangEtAl2023} is sensitive only to the $\Zbar{\!C}\suptiny{0}{0}{\mathrm{(CC)}}$ class, so four-of-six detection suffices there; the fidelity witnesses measured here sample both classes and require full detection.

\begin{table}[ht!]
\centering\small
\setlength{\tabcolsep}{6pt}
\renewcommand{\arraystretch}{1.2}
\begin{tabular}{lccc}
\toprule
Error & Class & $S_{\mathrm{Wang}}$ & $S_\star$ \\
\midrule
$Z_4 Z_8$ & $\Zbar{\!C}\suptiny{0}{0}{\mathrm{(CC)}}$ & \textcolor{green!50!black}{$\checkmark$} & \textcolor{green!50!black}{$\checkmark$} \\
$Z_5 Z_9$ & $\Zbar{\!C}\suptiny{0}{0}{\mathrm{(CC)}}$ & \textcolor{green!50!black}{$\checkmark$} & \textcolor{green!50!black}{$\checkmark$} \\
$Z_6 Z_{10}$ & $\Zbar{\!C}\suptiny{0}{0}{\mathrm{(CC)}}$ & \textcolor{green!50!black}{$\checkmark$} & \textcolor{green!50!black}{$\checkmark$} \\
$Z_7 Z_{11}$ & $\Zbar{\!C}\suptiny{0}{0}{\mathrm{(CC)}}$ & \textcolor{green!50!black}{$\checkmark$} & \textcolor{green!50!black}{$\checkmark$} \\
$Z_8 Z_9$ & $\Zbar{\!B}\suptiny{0}{0}{\mathrm{(CC)}}$ & \textcolor{red}{$\times$} & \textcolor{green!50!black}{$\checkmark$} \\
$Z_{10} Z_{11}$ & $\Zbar{\!B}\suptiny{0}{0}{\mathrm{(CC)}}$ & \textcolor{red}{$\times$} & \textcolor{green!50!black}{$\checkmark$} \\
\midrule
Detected & & 4/6 & \textbf{6/6} \\
\bottomrule
\end{tabular}
\caption{\textbf{Flag detection of the six dangerous weight-two $Z$ errors.} A flag $X_S$ detects $Z_i Z_j$ when $|\{i,j\} \cap S|$ is odd (\textcolor{green!50!black}{$\checkmark$}) and misses it otherwise (\textcolor{red}{$\times$}). The optimal flag $S_\star = \{4,6,9,11\}$ detects all six; the three further admissible flags $\{4,7,9,10\}$, $\{5,6,8,11\}$, and $\{5,7,8,10\}$ detect the same six and are omitted for brevity. The flag $S_{\mathrm{Wang}} = \{6,7,8,9\}$ used by~\cite{WangEtAl2023} with the same encoder detects only the four $\Zbar{\!C}\suptiny{0}{0}{\mathrm{(CC)}}$-class errors.}
\label{tab:flags}
\end{table}

\vspace*{2mm}
{\noindent}\textit{Flag type.} The two error classes call for complementary flag types: an $X_S$ flag catches propagated $Z$ errors, while a $Z_S$ flag catches propagated $X$ errors. For this encoder only $Z$ errors propagate to logical errors, so the $X_{S_\star}$ flag is the one that restores fault tolerance; a simultaneous $Z_{S_\star}$ flag would only catch $X$ errors that are already detected by stabiliser readout, at the cost of additional post-selection. The choice between the two is further constrained by the prepared state, since $X_{S_\star}$ and $Z_{S_\star}$ differ in status: $Z_{S_\star} = Z_4 Z_6 Z_9 Z_{11}$ is a product of $Z$-stabiliser generators and hence a stabiliser, reading $+1$ deterministically on every code-space state, whereas $X_{S_\star}$ is a logical operator, deterministic only on states that are eigenstates of the corresponding logical $\widebar{X}$. For the CCZ and GHZ protocols the colour code is prepared in $\ket{\widebar{+\!+\!+}}_C$, an eigenstate of all logical $\widebar{X}_i$, so the $X_{S_\star}$ flag is deterministic and we use it. For the rotation protocol the third encoded qubit is prepared in $\ket{\widebar{0}}$ rather than $\ket{\widebar{+}}$, on which $X_{S_\star}$ measures $\pm 1$ at random (${\sim}50\%$ false rejection); we use the $Z_{S_\star}$ flag instead, which remains deterministic as a stabiliser and catches the dual $X$-error class.

%%%%%%%%%%%%%%%%%%%%%%%%%%%%%%%%%%%%%%%%%%%%%%%%%%%%%%%%%%%%%%%%%%%%%%%%%%%

\vspace*{-1mm}
\subsection{Supplementary Measurement Results}\label{appendix:supplementary_results}
\vspace*{-1mm}

{\noindent}In this section we discuss experiments going beyond the results demonstrated in the main text: We implement the logical $\widebar{\text{CNOT}}$ gates on the $[\![8,3,2]\!]$ and $[\![12,4,2]\!]$ code by decomposing the physical SWAP gates into three CNOT gates each. Furthermore, we discuss replacing the $[\![4,2,2]\!]$ with the $[\![4,1,2]\!]$ sub-code and characterise state preparation of the $[\![4,1,2]\!]$ code, the $[\![8,3,2]\!]$ code and the merged $[\![12,3,2]\!]$ code.

%%%%%%%%%%%%%%%%%%%%%%%%%%%%%%%%%%%%%%%%%%%%%%%%%%%%%%%%%%%%%%%%%%%%%%%%%%%

\vspace*{-1mm}
\subsubsection{SWAP Decomposition}
\vspace*{-1mm}

{\noindent}The all-to-all connectivity of the experiment allows the relabelling of qubits, a tool which can be used to implement the logical $\widebar{\text{CNOT}}$ gates without additional performance cost. Relabelling of qubits was used for the creation of the GHZ state and the arbitrary rotation in the main text. Here, we compare those results to decomposing physical SWAP gates into three physical CNOT gates.

In the main text we report a fidelity of $\mathcal{F}\subtiny{0}{-1.5}{\widebar{\mathrm{GHZ}}}\suptiny{1}{1}{\mathrm{exp.}}=81.0(1.7)\%$, in good agreement with the numerical result $\mathcal{F}\subtiny{0}{-1.5}{\widebar{\mathrm{GHZ}}}\suptiny{1}{1}{\mathrm{sim.}}=81.8(3)\%$, for the flagged approach. Decomposing the two logical $\widebar{\text{CNOT}}$ gates into six physical CNOT gates reduces the fidelity to $\mathcal{F}\subtiny{0}{-1.5}{\widebar{\mathrm{GHZ}}}\suptiny{1}{1}{\mathrm{exp.}}=70(5)\%$, in good agreement with the numerical result $\mathcal{F}\subtiny{0}{-1.5}{\widebar{\mathrm{GHZ}}}\suptiny{1}{1}{\mathrm{sim.}}=73.6(2)\%$. The reduction in fidelity corresponds to the infidelity gained due to the application of six additional CNOT gates. Additional infidelity can be acquired due to incoherence gained by additional idling of qubits which is not captured in the simplified numerical depolarising noise model. Similarly, we report an average state fidelity of $83.4(3)\%$ for the arbitrary rotation in the main text. Decomposing the logical $\widebar{\text{CNOT}}$ gate into six physical CNOT gates reduces the average fidelity to  $80.1(4) \%$, in good agreement with the simulated value of $80.3(2) \%$. In both cases, the flag corresponding to fault-tolerant initialisation is triggered approximately in $13\%$ of the shots and the rotation is implemented successfully with a probability of approximately $3/4$.

%%%%%%%%%%%%%%%%%%%%%%%%%%%%%%%%%%%%%%%%%%%%%%%%%%%%%%%%%%%%%%%%%%%%%%%%%%%

\vspace*{-1mm}
\subsubsection{\texorpdfstring{$\ket{\widebar{\mathrm{CCZ}}}$ on the $[\![8,3,2]\!]$ Code Alone}{Logical CCZ State on the $[\![8,3,2]\!]$ Code Alone}}\label{appendix:CCZ_colour_only}
\vspace*{-1mm}

\begin{figure}
    \centering
    \includegraphics[width=0.8\columnwidth]{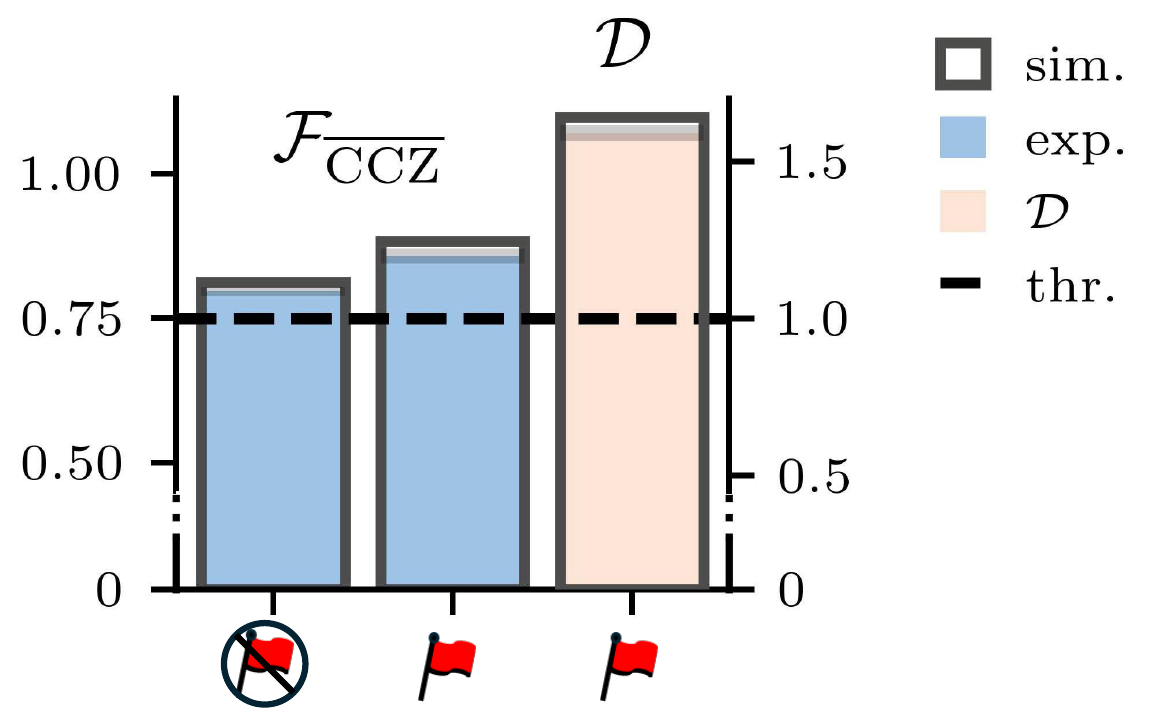}\vspace*{-4mm}
    \caption{\textbf{GME on} $[\![8,3,2]\!]$: Fidelity (blue) and stabiliser witness (purple) for the logical $\overline{\text{CCZ}}$ gate applied on the colour code only. Numerical simulation using a simple depolarising model is depicted in the grey outlines. GME and magic is demonstrated since both witnesses exceed their respective threshold of $3/4$ (fidelity) and 1 (stabiliser witness). The error bars are given as shaded region and depict the standard deviation of 3 independent measurement runs, whereas each stabiliser term consists of 7500 shots total.}
    \label{fig:CCZ_832}
\end{figure}

{\noindent}Here we implement the logical $\widebar{\mathrm{CCZ}}$ gate on the $[\![8,3,2]\!]$ colour code only: a simplification related to the experiment demonstrated in the main-text. The initialisation of the colour code is the same as for the experiment in the main text, where all three logical qubits $A$, $B$, and $C$ are initialised in $+$. The transversal $\overline{\text{CCZ}}$ gate is applied and all 29 stabilisers given in Eq.~\eqref{eq:fidelity_decomp} are measured. Both, the fidelity and the stabiliser witness are demonstrated in Fig.~\ref{fig:CCZ_832}. The fidelity is $79.6(8)\%$ without the flag (the non-fault-tolerant preparation) and $85.8(1.3)\%$ with the flag of Appendix~\ref{appendix:flag_FT} (the fault-tolerant preparation, in the sense of Sec.~\ref{sec:background} of the main text); both exceed the $3/4$ threshold given by the square of the largest Schmidt coefficient $\lambda_{\max}^2$. This improvement compared to the merged case in main text, directly follows from the reduced number of two-qubit gates, which are required to initialise the surface code and apply the merging. Using the same data yields a stabiliser norm of $\mathcal{D}=1.59(3)$, exceeding the threshold by 20 standard deviations.   

\vspace*{-2mm}
%%%%%%%%%%%%%%%%%%%%%%%%%%%%%%%%%%%%%%%%%%%%%%%%%%%%%%%%%%%%%%%%%%%%%%%%%%%
\subsubsection{$[\![12,3,2]\!]$ Results from Merging $[\![8,3,2]\!]$ with $[\![4,1,2]\!]$ Code}\label{appendix:412 results}
\vspace*{-1.5mm}

{\noindent}Replacing the $[\![4,2,2]\!]$ memory block of the main-text experiments with the $[\![4,1,2]\!]$ sub-code (Sec.~\ref{appendix:422}) reduces state-preparation and measurement errors at the cost of giving up the code-space-preserving transversal Hadamard, which becomes a pre-measurement basis change. The experimental results are summarised below.

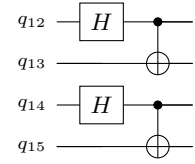
\begin{figure}[t!]
    \centering
    \begin{quantikz}[thin lines, row sep={0.55cm,between origins}, column sep=0.30cm]
    \lstick{\scriptsize$q_{12}$} & \gate{H} & \ctrl{1} & \qw \\
    \lstick{\scriptsize$q_{13}$} & \qw      & \targ{}  & \qw \\
    \lstick{\scriptsize$q_{14}$} & \gate{H} & \ctrl{1} & \qw \\
    \lstick{\scriptsize$q_{15}$} & \qw      & \targ{}  & \qw
    \end{quantikz}
    \caption{\textbf{$[\![4,1,2]\!]$ encoder used in the supplementary measurements of Sec.~\ref{appendix:412 results}.} This circuit replaces the $[\![4,2,2]\!]$ encoder in the surface-block region of Figs.~\ref{fig:circuit_ghz}{\textendash}\ref{fig:circuit_rotation}. Two CNOTs and two Hadamards; prepares $\ket{\widebar{+}}\suptiny{0}{0}{\mathrm{(SC)}}$ with the spectator logical qubit $B$ gauged out. The merge operator $M_{ZZ} = Z_0 Z_2 Z_4 Z_6$ and the rest of each protocol are unchanged.}
    \label{fig:circuit_412_encoder}
\end{figure}

\vspace*{2mm}
{\noindent}\textit{State-preparation diagnostics.} We demonstrate the state preparation of the $[\![4,1,2]\!]$ and the $[\![8,3,2]\!]$ codes in this section. In order to prepare the code a subset of qubits is selected from the 16 ion register. The $[\![8,3,2]\!]$ code is initialised on the eight physical qubit on one end of the chain (ions 0-8) while the $[\![4,1,2]\!]$ code is initialised on the other end (12{\textendash}15). Flag qubits are the direct neighbours of the data qubits, while qubit 10 is used as an ancilla if needed. Both codes are initialised such that all logical qubits are in $+$. The respective quantum circuits used for state preparation are those of Figs.~\ref{fig:circuit_ghz}{\textendash}\ref{fig:circuit_rotation}, with the $[\![4,1,2]\!]$ surface-block encoder given in Fig.~\ref{fig:circuit_412_encoder}.

%%%%%%%%%%%%%%%%%%%%%%%%%%%%%%%%%%%%%%%%%%%%%%%%%
%%%%%%%%%%%%%%%%%%%%%%%%%%%%%%%%%%%%%%%%%%%%%%%%%

\begin{figure*}[!t!]
    \centering
    \includegraphics{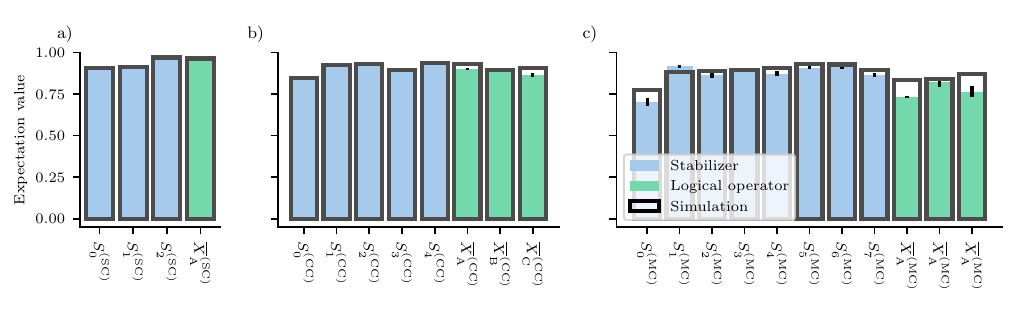}
    \vspace*{-8.0mm}
    \caption{\textbf{State preparation.} (a) Expectation value of the stabiliser generators and the logical $X$ operators for state preparation of the $[\![4,1,2]\!]$ code. The coloured bars represent measured data while the grey outline shows numerical simulations. The error bars are the standard deviation over four distinct measurement runs. (b) Expectation values of the stabiliser generators and logical $X$ operators for the $[\![8,3,2]\!]$ code. (c) Expectation value of the stabiliser generators and logical $X$ operators of the merged code.}
    \label{fig:appendix_state_prep}
\end{figure*}

The expectation values of the stabiliser generators and logical $X$ operators are used to determine if a state was initialised correctly. Those are demonstrated for the $[\![4,1,2]\!]$ code in Fig.~\ref{fig:appendix_state_prep}~(a). The measured stabiliser generators are shown as light blue bars, the logical $X$ operator is shown as blue bar while the numerical simulation is shown as thick grey outline. The error of the data is the standard deviation of four distinct measurement runs. Similar, Fig.~\ref{fig:appendix_state_prep}~(b) displays the stabiliser generators and logical $X$ operators for the $[\![8,3,2]\!]$ code.

Figure~\ref{fig:appendix_state_prep}~(c) depicts the stabiliser generators and logical $X$ operators when merging both codes. While the simulation and measured data are in good agreement when initialising both codes independently, the simulation is above the data for the merged code. The reason for this can be explained by the simplified depolarising-noise model and the variation of the two-qubit gate fidelity over distinct measurement runs. In addition, errors can spread during the merging of both codes, reducing the expectation value of the logical operators. 

\vspace*{2mm}
{\noindent}\textit{$\ket{\widebar{\mathrm{CCZ}}}$ on the merged  $[\![12, 3, 2]\!]$ code.} We demonstrate a simplification on the $\ket{\widebar{\mathrm{CCZ}}}$ on the merged code, consisting of the $[\![8,3,2]\!]$ colour code and the $[\![4,1,2]\!]$ surface code, by gauge fixing on of the two logical qubits of the $[\![4,2,2]\!]$ surface code. We initialise the surface code in $+$ and teleport the logical qubit onto the surface code via smooth merge. In contrast to the main text, where we evaluate both outcomes of the merge via post selection and the weighted average, we here evaluate only the zero branch of the merging. 

\begin{figure}[!h]
    \centering
    \includegraphics[width=0.8\linewidth]{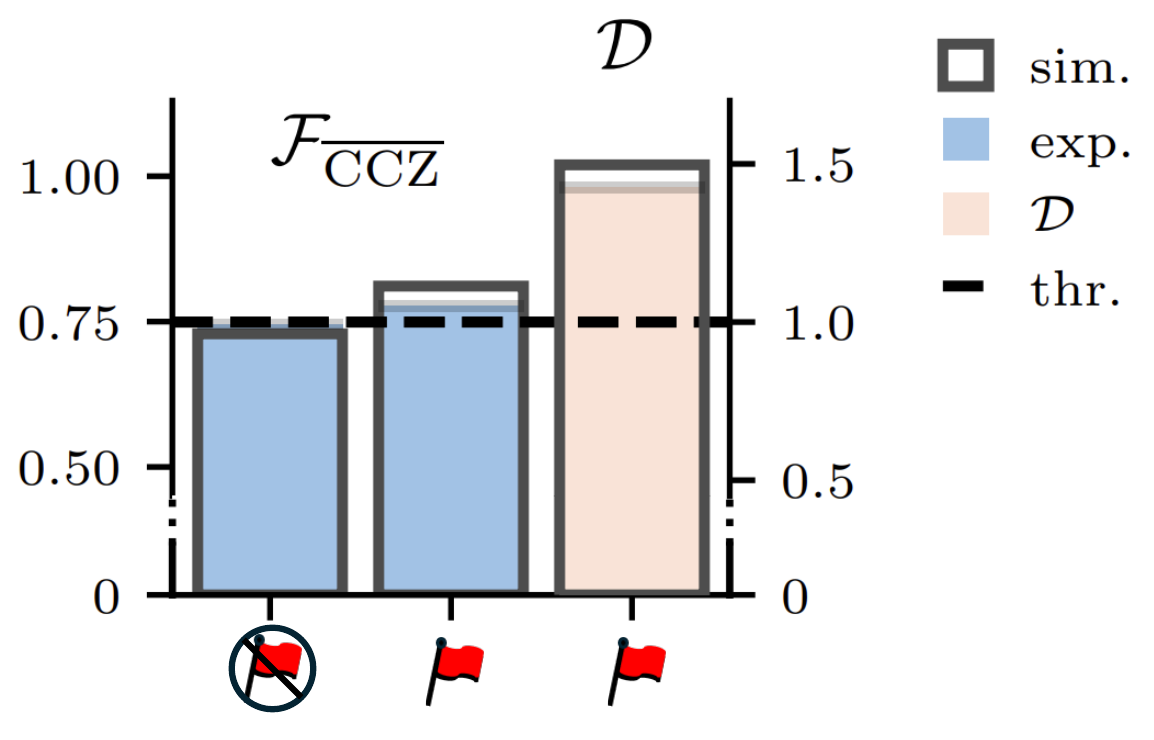}
    \vspace*{-5mm}
    \caption{\textbf{Fidelity and stabiliser witness for merged $[\![12,3,2]\!]$ code.} Measured (blue) and simulated (grey outlines, depolarising-noise model) fidelities $\mathcal{F}\subtiny{0}{-1.5}{\widebar{\mathrm{CCZ}}}\suptiny{1}{1}{\mathrm{exp.}}$ and stabiliser norm (peach) $\mathcal{D}$ for flagged and unflagged circuits, shown relative to their respective thresholds (dashed line) of 0.75 for GME and 1 for non-stabiliserness. The data represents only the zero branch (merge ancilla measured in zero) of the merging. The experimental uncertainty corresponds to 3 independent measurement sets with 2500 shot each.}
    \label{fig:CCZ_832_412}
\end{figure}

We measure the expectation value $\braket{\widebar{P}_i}$ of the logical twenty-nine non-zero Pauli operators $\widebar{P}_i$ listed in Sec.~\ref{appendix:CCZ_state} to calculate the fidelity $\mathcal{F}$ and stabiliser norm $\mathcal{D}$. As shown in Fig.~\ref{fig:CCZ_832_412} the measured fidelity using a flag is $\mathcal{F}\subtiny{0}{-1.5}{\widebar{\mathrm{CCZ}}}\suptiny{1}{1}{\mathrm{exp.}}=77.6(1.0)\%$ exceeding the GME-certification threshold of $3/4$ by two sigma, while the simulations predict $\mathcal{F}\subtiny{0}{-1.5}{\widebar{\mathrm{CCZ}}}\suptiny{1}{1}{\mathrm{sim.}}=81.1(5)\%$. Without the flag, we obtain $\mathcal{F}\subtiny{0}{-1.5}{\widebar{\mathrm{CCZ}}}\suptiny{1}{1}{\mathrm{exp.}}=74.5(9)\%$ and $\mathcal{F}\subtiny{0}{-1.5}{\widebar{\mathrm{CCZ}}}\suptiny{1}{1}{\mathrm{sim.}}=72.9(4)\%$. The discrepancy between simulation and experiment can be likely explained by the simple depolarising noise model which assumes depolarising noise as sole noise source. The same data using flags results in a stabiliser norm of $\mathcal{D}=1.427(19)$ and bounds the log-free robustness of magic $\text{LR}\ge\ln{\left(\mathcal{D}\right)}=0.355(13)$. Both, the fidelity and the stabiliser norm, agree with the results of the main text, where the logical $\widebar{\text{CCZ}}$ gate is applied on the merged $[\![8,3,2]\!]$ colour code and $[\![4,2,2]\!]$ surface code within their standard deviations. This is expected since the complexity of the physical circuit is equivalent for both experiments.\\

\vspace*{2mm}
{\noindent}\textit{Arbitrary rotation $R_Z(\theta)$ with $[\![4,1,2]\!]$ code.} We demonstrate the arbitrary rotation  using the $[\![4,1,2]\!]$ code instead of the $[\![4,2,2]\!]$ code, constituting to a simplification compared to the main text. The transversal Hadamard gate shown in the main text required for the implementation of the arbitrary rotation constitutes here to a basis change before measurement since the $[\![4,1,2]\!]$ code omits a transversal Hadamard gate, as explained in Sec.~\ref{appendix:422}.

\begin{figure}[!h]
    \centering
    \includegraphics[width=\linewidth]{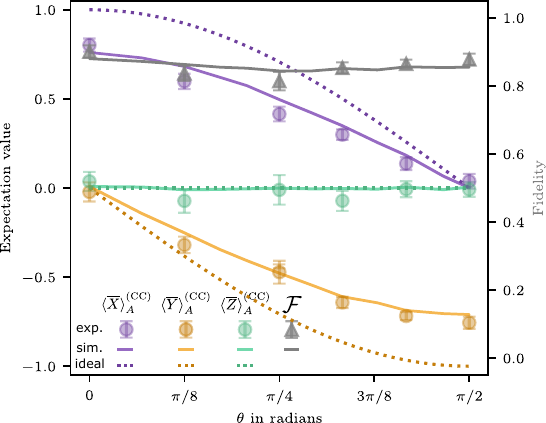}
    \caption{\textbf{Arbitrary logical rotation on the $[\![12,3,2]\!]$ code.} Measured expectation values $\expval{\hspace*{-1pt}\widebar{X}\hspace*{-1pt}}\suptiny{0}{0}{\mathrm{(CC)}}_{A}$ (purple), $\expval{\hspace*{-1pt}\widebar{Y}\hspace*{-1pt}}\suptiny{0}{0}{\mathrm{(CC)}}_{A}$ (orange), $\expval{\hspace*{-1pt}\widebar{Z}\hspace*{-1pt}}\suptiny{0}{0}{\mathrm{(CC)}}_{A}$ (green), and fidelity (grey) vs.\ rotation angle $\theta$: experimental data (markers), numerical simulation (solid lines), and ideal curves (dotted lines). The experimental uncertainty corresponds to the standard deviation of three independent measurement sets with 2500 shots each. The average fidelity over the measured rotation angles is $\mathcal{F}=86.4(7) \%$ post selecting on successful rotations.}
    \label{fig:arb_rot_412}
\end{figure}

The rotation is characterised by performing single-qubit logical state tomography on the rotated qubit. Figure~\ref{fig:arb_rot_412} shows the expectation values $\expval{\hspace*{-1pt}\widebar{X}\hspace*{-1pt}}\suptiny{0}{0}{\mathrm{(CC)}}_{A}$, $\expval{\hspace*{-1pt}\widebar{Y}\hspace*{-1pt}}\suptiny{0}{0}{\mathrm{(CC)}}_{A}$, and $\expval{\hspace*{-1pt}\widebar{Z}\hspace*{-1pt}}\suptiny{0}{0}{\mathrm{(CC)}}_{A}$, and the fidelity $\mathcal{F}$ for different rotation angles $\theta$. Here, experimental data is accompanied by numerical simulations showing good agreement. Overall, we show an average state fidelity of $\mathcal{F}\suptiny{1}{1}{\mathrm{exp.}}=86.4(7)\%$ and $\mathcal{F}\suptiny{1}{1}{\mathrm{sim.}}=85.6(1)\%$ confirming the implementation of arbitrary logical rotations. These results show an improvement of $3\%$ over the results presented in the main text. This improvement is likely due to the difference in circuit design, since the number of entangling gates is the same for both realizations.

%%%%%%%%%%%%%%%%%%%%%%%%%%%%%%%%%%%%%%%%%%%%%%%%%%%%%%%%%%%%%%%%%%%%%%%%%%%

\vspace*{-2.5mm}
\subsection{Physical Implementation: Circuits, Transpilation, and Noise Model}\label{appendix:physical}
\vspace*{-2mm}

{\noindent}This section gives the physical-qubit-level circuit diagrams for the three protocols whose logical-level descriptions appear in the main text. The diagrams in Figs.~\ref{fig:circuit_ghz}{\textendash}\ref{fig:circuit_rotation} are written in terms of single-qubit Clifford gates, $T$ and $T^{\dagger}$ gates, and CNOTs{\textemdash}one level above the trapped-ion native gate set. All three full circuits use the $[\![4,2,2]\!]$ surface code as the memory block; the $[\![4,1,2]\!]$ variants of Sec.~\ref{appendix:412 results} differ only in the surface-block encoder, given separately in Fig.~\ref{fig:circuit_412_encoder}. The $[\![8,3,2]\!]$ encoder is the ten-CNOT MQT-QECC output of Sec.~\ref{appendix:encoding}; the flag-qubit subroutine measures $X_{S_\star}$ and/or $Z_{S_\star}$ on $S_\star = \{4, 6, 9, 11\}$ (Sec.~\ref{appendix:flag_FT}); the merge ancilla measures $M_{ZZ} = Z_0 Z_2 Z_4 Z_6$ (Sec.~\ref{appendix:Z merge 12 4 2}). The decomposition of each CNOT and Hadamard into the trapped-ion native gate set $\{R_x, R_y, R_z, \mathrm{MS}\}$ is summarised in Sec.~\ref{appendix:transpilation}. Resource counts are quoted in each figure caption and in Table~\ref{tab:resources}.
Section~\ref{appendix:noise_model} closes this section with the depolarising-noise model applied at the gate level in all simulations.

%%%%%%%%%%%%%%%%%%%%%%%%%%%%%%%%%%%%%%%%%%%%%%%%%%%%%%%%%%%%%%%%%%%%%%%%%%%
\vspace*{-2mm}
\subsubsection{GHZ on the Merged Code}\label{appendix:circuit_ghz}
\vspace*{-2mm}

{\noindent}The compiled circuit for the logical GHZ protocol of Fig.~\ref{fig:ghz_merged}~(b) of the main text is shown in Fig.~\ref{fig:circuit_ghz}. The two logical $\widebar{\mathrm{CNOT}}$s acting on the merged code after lattice surgery are implemented by qubit relabelling, $\mathrm{SWAP}_{1,5}\,\mathrm{SWAP}_{2,6}$, at zero gate cost.

%%%%%%%%%%%%%%%%%%%%%%%%%%%%%%%%%%%%%%%%%%%%%%%%%%%%%%%%%%%%%%%%%%%%%%%%%%%
\vspace*{-2mm}
\subsubsection{\texorpdfstring{$\widebar{\mathrm{CCZ}}$ on the Merged Code}{Logical CCZ Gate on the Merged Code}}\label{appendix:circuit_ccz}
\vspace*{-2mm}

{\noindent}The compiled circuit for the $\ket{\widebar{\mathrm{CCZ}}}$ protocol of Fig.~\ref{fig:ghz_merged}~(b) of the main text is shown in Fig.~\ref{fig:circuit_ccz}. The transversal $\widebar{\mathrm{CCZ}}$ is realised by $T$ on $\{0, 3, 5, 6\}$ and $T^{\dagger}$ on $\{1, 2, 4, 7\}$.

%%%%%%%%%%%%%%%%%%%%%%%%%%%%%%%%%%%%%%%%%%%%%%%%%%%%%%%%%%%%%%%%%%%%%%%%%%%
\vspace*{-2mm}
\subsubsection{\texorpdfstring{$\widebar{R}_Z(\theta)$ via Cross-Code Lattice Surgery}{Logical Single-Qubit Rotation via Cross-Code Lattice Surgery}}\label{appendix:circuit_rotation}
\vspace*{-2mm}

{\noindent}The compiled circuit for the rotation gadget of Fig.~\ref{fig:rotationLS} of the main text is shown in Fig.~\ref{fig:circuit_rotation}. The $\widebar{\mathrm{CNOT}}_{12}$ in the gadget is realised by qubit relabelling; the merge and split sub-blocks are highlighted by coloured dashed boundaries.

%%%%%%%%%%%%%%%%%%%%%%%%%%%%%%%%%%%%%%%%%%%%%%%%%%%%%%%%%%%%%%%%%%%%%%%%%%%
\vspace*{-2mm}
\subsubsection{Modifications for the $[\![4,1,2]\!]$ Surface Block}\label{appendix:circuit_412_diff}
\vspace*{-2mm}

{\noindent}In the supplementary measurements of Sec.~\ref{appendix:412 results}, the $[\![4,2,2]\!]$ surface block of Figs.~\ref{fig:circuit_ghz}{\textendash}\ref{fig:circuit_rotation} is replaced by the $[\![4,1,2]\!]$ sub-code (Sec.~\ref{appendix:422}). The only change to the compiled circuits is the surface-block encoder, shown in Fig.~\ref{fig:circuit_412_encoder}: The spectator logical qubit $B$ is gauged out, so only $\ket{\widebar{+}}\suptiny{0}{0}{\mathrm{(SC)}}$ is prepared. The merge operator $M_{ZZ} = Z_0 Z_2 Z_4 Z_6$ is unchanged at the physical level, and the transversal $H^{\otimes 4}$ appearing in the rotation circuit is read as a pre-measurement basis change implementing the $\widebar{X}$-basis readout of the surface block rather than as a code-space-preserving logical Hadamard.

%%%%%%%%%%%%%%%%%%%%%%%%%%%%%%%%%%%%%%%%%%%%%%%%%%%%%%%%%%%%%%%%%%%%%%%%%%%

\vspace*{-2mm}
\subsubsection{Transpilation to Trapped-Ion Native Gates}\label{appendix:transpilation}
\vspace*{-2mm}

\begin{figure}[t!]
\centering
\textbf{(a)}\;\;
\begin{quantikz}[wire types={q,q}, row sep=0.35cm]
\lstick{$c$} & \ctrl{1} & \qw \\
\lstick{$t$} & \targ{}  & \qw
\end{quantikz}
\;\;$=$\;\;
\begin{quantikz}[wire types={q,q}, row sep=0.35cm, column sep=0.22cm]
\lstick{$c$} & \gate{R_y(\text{-}\tfrac{\pi}{2})} & \gate[2]{\mathrm{MS}(\tfrac{\pi}{2})} & \gate{R_x(\text{-}\tfrac{\pi}{2})} & \gate{R_y(\tfrac{\pi}{2})} & \qw \\
\lstick{$t$} & \qw                                &                                       & \gate{R_x(\text{-}\tfrac{\pi}{2})} & \qw                         & \qw
\end{quantikz}

\vspace{0.6em}

\textbf{(b)}\;\;
\begin{quantikz}[wire types={q}, row sep=0.35cm]
& \gate{H} & \qw
\end{quantikz}
\;\;$=$\;\;
$\left\{
\vcenter{\hbox{%
\begin{tabular}{@{}l@{}}
\begin{quantikz}[wire types={q}, row sep=0.35cm, column sep=0.22cm]
& \gate{R_z(\pi)} & \gate{R_y(\tfrac{\pi}{2})} & \qw
\end{quantikz}\\[4pt]
\begin{quantikz}[wire types={q}, row sep=0.35cm, column sep=0.22cm]
& \gate{R_y(\tfrac{\pi}{2})} & \gate{R_x(\pi)} & \qw
\end{quantikz}
\end{tabular}}}\right.$
\caption{\textbf{Decomposition of CNOT and Hadamard into the trapped-ion native gate set}~\cite{maslov_basic_circuit_2017}. \textbf{(a)}~CNOT is realised as a single M\o{}lmer{\textendash}S\o{}rensen gate $\mathrm{MS}(\pi/2) = \exp(\texttt{i}\pi/4\, X{\otimes}X)$ flanked by four single-qubit $\pi/2$ rotations on the control and target. \textbf{(b)}~Hadamard admits two equivalent native decompositions, $H = R_y(\tfrac{\pi}{2})\,R_z(\pi)$ (top) and $H = R_x(\pi)\,R_y(\tfrac{\pi}{2})$ (bottom); the transpiler selects whichever minimises the overall pulse count of the surrounding circuit, since the virtual $R_z(\pi)$ of the first is free whereas the explicit $R_x(\pi)$ of the second can merge with a neighbouring single-qubit rotation. All identities hold up to a global phase. The remaining single-qubit gates appearing in the protocol circuits{\textemdash}$T$, $T^{\dagger}$, and $S${\textemdash}are diagonal in the $Z$ basis and are implemented as virtual $Z$ rotations in software at zero gate cost~\cite{mckay_efficient_z_2017}.}
\label{fig:transpilation}
\end{figure}
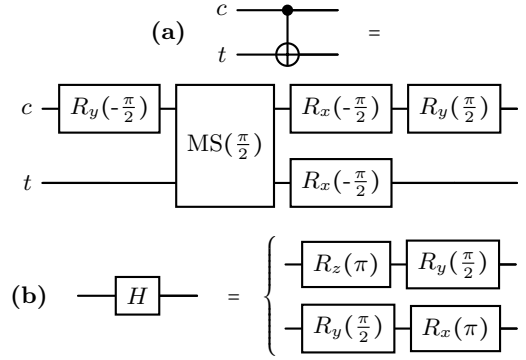

{\noindent}The trapped-ion processor of~\cite{PogorelovEtAl2021} natively implements arbitrary single-qubit rotations around the $X$ and $Y$ axes of the Bloch sphere, $R_x(\theta) = \exp(-\texttt{i}\theta X/2)$ and $R_y(\theta) = \exp(-\texttt{i}\theta Y/2)$, on individually addressed ions; virtual $Z$ rotations $R_z(\theta) = \exp(-\texttt{i}\theta Z/2)$ implemented in software at zero gate cost~\cite{mckay_efficient_z_2017}; and the M\o{}lmer{\textendash}S\o{}rensen two-qubit entangling gate $\mathrm{MS}(\theta) = \exp(\texttt{i}\theta\, X{\otimes}X / 2)$, with $\mathrm{MS}(\pi/2) = \exp(\texttt{i}\pi/4\, X{\otimes}X)$ the maximally entangling instance~\cite{SoerensenMoelmer1999}. The CNOT and Hadamard gates appearing in Figs.~\ref{fig:circuit_ghz}{\textendash}\ref{fig:circuit_rotation} are decomposed into this native set as shown in Fig.~\ref{fig:transpilation}, with one $\mathrm{MS}(\pi/2)$ pulse plus four single-qubit rotations per CNOT and two single-qubit rotations per Hadamard. The non-Clifford gates $T$, $T^{\dagger}$, and $S$ appearing in the transversal $\widebar{\mathrm{CCZ}}$ are diagonal in the computational basis and are realised as virtual $Z$ rotations at zero gate cost. Consequently, the entangling-gate counts quoted in Table~\ref{tab:resources} correspond directly to the number of $\mathrm{MS}(\pi/2)$ pulses executed on the trap.\\

\begin{table}[ht!]
\centering\small
\setlength{\tabcolsep}{12pt}\renewcommand{\arraystretch}{1.15}
\begin{tabular}{lcc}
\toprule
Protocol (merged code)            & Qubits     & CNOTs      \\
\midrule
$\ket{\widebar{\mathrm{GHZ}}}$    & $13\,(+1)$ & $17\,(+4)$ \\
$\ket{\widebar{\mathrm{CCZ}}}$    & $13\,(+2)$ & $16\,(+8)$ \\
$\widebar{R}_Z(\theta)$ rotation  & $14\,(+1)$ & $20\,(+4)$ \\
\bottomrule
\end{tabular}
\caption{\textbf{Resource counts per protocol.} Number of physical qubits and CNOT gates on the merged
$[\![12,4,2]\!]$ code. The CNOT count equals the number of $\mathrm{MS}(\pi/2)$ pulses executed on the trap,
since $T$, $T^{\dagger}$ and $S$ are realised as virtual $Z$ rotations at zero gate cost
(Sec.~\ref{appendix:transpilation}). Counts outside the parentheses exclude the flag checks; the
parenthesised value is the additional cost of flagging, each weight-four flag adding one ancilla qubit
and four CNOTs{\textemdash}one flag for the GHZ and rotation protocols, two for the CCZ protocol.}
\label{tab:resources}
\end{table}

%%%%%%%%%%%%%%%%%%%%%%%%%%%%%%%%%%%%%%%%%%%%%%%%%%%%%%%%%%%%%%%%%%%%%%%%%%%

\begin{figure*}[t!]
\centering
\begin{adjustbox}{max width=\textwidth,max totalheight=0.30\textheight}
\begin{quantikz}[row sep={0.55cm,between origins}, column sep=0.20cm]
\lstick{\scriptsize $0$} &   &   &   &   &   &   &   &   &   &   &   &   &   &   & \gate{H}\gategroup[4,steps=5,style={dashed,rounded corners=2pt,draw=blue!55!cyan,fill=blue!55!cyan,fill opacity=0.10,line width=0.7pt},background,label style={label position=above,anchor=south,yshift=0.10cm,text=blue!55!cyan}]{\scriptsize $\ket{\widebar{0}\widebar{0}}\!\to\!\ket{\widebar{+}\widebar{+}}$} & \ctrl{2} & \ctrl{1} & \ctrl{3} & \gate{H} &   & \gategroup[14,steps=7,style={dashed,rounded corners=2pt,draw=green!55!black,fill=green!55!black,fill opacity=0.10,line width=0.7pt},background,label style={label position=above,anchor=south,yshift=0.10cm,text=green!55!black}]{\scriptsize Merge $\widebar{Z}_{A}\suptiny{0}{0}{\mathrm{CC}}\!\otimes\!\widebar{Z}_{A}\suptiny{0}{0}{\mathrm{SC}}$} &   & \ctrl{4} &   &   &   &   &   &   &   &   & \meter{}\gategroup[14,steps=1,style={dashed,rounded corners=2pt,draw=gray!60,fill=gray!60,fill opacity=0.10,line width=0.7pt},background,label style={label position=above,anchor=south,yshift=0.10cm,text=gray!60}]{\scriptsize Readout} \\
\lstick{\scriptsize $1$} &   &   &   &   &   &   &   &   &   &   &   &   &   &   &   &   & \targ{} &   & \gate{H} &   &   &   &   &   &   &   &   &   &   &   &   & \meter{} \\
\lstick{\scriptsize $2$} &   &   &   &   &   &   &   &   &   &   &   &   &   &   &   & \targ{} &   &   & \gate{H} &   &   &   &   & \ctrl{2} &   &   &   &   &   &   &   & \meter{} \\
\lstick{\scriptsize $3$} &   &   &   &   &   &   &   &   &   &   &   &   &   &   &   &   &   & \targ{} & \gate{H} &   &   &   &   &   &   &   &   &   &   &   &   & \meter{} \\
\lstick{\scriptsize $a_0$} &   &   &   &   &   &   &   &   &   &   &   &   &   &   &   &   &   &   &   &   & \targ{} & \targ{} & \targ{} & \targ{} & \meter{} & \setwiretype{c} & \ctrl{9} &  \setwiretype{n} &   &   &   &   \\
\lstick{\scriptsize $a_1$} &   &   &   &   &   &   &   &   & \targ{}\gategroup[9,steps=5,style={dashed,rounded corners=2pt,draw=teal!70,fill=teal!70,fill opacity=0.10,line width=0.7pt},background,label style={label position=below,anchor=north,yshift=-0.30cm,text=teal!70}]{\scriptsize Colour $Z$ flag} & \targ{} & \targ{} & \targ{} & \meter{} &   &   &   &   &   &   &   &   &   &   &   &   &   &   &   &   &   &   &   \\
\lstick{\scriptsize $4$} & \gategroup[8,steps=7,style={dashed,rounded corners=2pt,draw=blue!70,fill=blue!70,fill opacity=0.10,line width=0.7pt},background,label style={label position=below,anchor=north,yshift=-0.30cm,text=blue!70}]{\scriptsize Colour $\ket{\widebar{+}\widebar{0}\widebar{0}}$} &   &   & \targ{} &   &   &   &   & \ctrl{-1} &   &   &   &   &   &   &   &   &   &   &   & \ctrl{-2} &   &   &   &   &   &   &   & \gategroup[8,steps=2,style={dashed,rounded corners=2pt,draw=purple!65,fill=purple!65,fill opacity=0.10,line width=0.7pt},background,label style={label position=below,anchor=north,yshift=-0.30cm,text=purple!65}]{\scriptsize $\overline{CX}_{AB},\overline{CX}_{AC}$ (relabel)} &   &   & \meter{} \\
\lstick{\scriptsize $5$} &   & \targ{} &   & \ctrl{-1} &   &   &   &   &   &   &   &   &   &   &   &   &   &   &   &   &   &   &   &   &   &   &   &   &   & \swap{2} &   & \meter{} \\
\lstick{\scriptsize $6$} &   &   &   &   &   & \targ{} &   &   &   &   & \ctrl{-3} &   &   &   &   &   &   &   &   &   &   & \ctrl{-4} &   &   &   &   & \gate[style={fill=gray!12,dashed,draw=gray!80}]{X} &   &   &   &   & \meter{} \\
\lstick{\scriptsize $7$} &   &   &   &   & \targ{} &   & \ctrl{4} &   &   &   &   &   &   &   &   &   &   &   &   &   &   &   &   &   &   &   & \gate[style={fill=gray!12,dashed,draw=gray!80}]{X} &   &   & \targX{} &   & \meter{} \\
\lstick{\scriptsize $8$} & \gate{H} & \ctrl{-3} & \ctrl{1} &   &   &   &   &   &   &   &   &   &   &   &   &   &   &   &   &   &   &   &   &   &   &   &   &   & \swap{2} &   &   & \meter{} \\
\lstick{\scriptsize $9$} &   &   & \targ{} &   &   &   &   &   &   & \ctrl{-6} &   &   &   &   &   &   &   &   &   &   &   &   &   &   &   &   &   &   &   &   &   & \meter{} \\
\lstick{\scriptsize $10$} &   &   &   & \gate{H} & \ctrl{-3} & \ctrl{-4} &   &   &   &   &   &   &   &   &   &   &   &   &   &   &   &   &   &   &   &   & \gate[style={fill=gray!12,dashed,draw=gray!80}]{X} &   & \targX{} &   &   & \meter{} \\
\lstick{\scriptsize $11$} &   &   &   &   &   &   & \targ{} &   &   &   &   & \ctrl{-8} &   &   &   &   &   &   &   &   &   &   &   &   &   &   & \gate[style={fill=gray!12,dashed,draw=gray!80}]{X} &   &   &   &   & \meter{}
\end{quantikz}
\end{adjustbox}
\caption{\textbf{Physical circuit for the logical GHZ-state preparation.} Qubits $0${\textendash}$3$ are the surface code, $4${\textendash}$11$ the colour code. The colour code is initialised in $\ket{\widebar{+}\,\widebar{0}\,\widebar{0}}_C$ and protected by a $Z$-parity flag on qubits $\{4,6,9,11\}$, measured on the flag ancilla $a_1$. The surface code is prepared in $\ket{\widebar{0}\,\widebar{0}}\suptiny{0}{0}{\mathrm{(SC)}}$ (a single Hadamard on qubit $0$ followed by three CNOTs) and then mapped to $\ket{\widebar{+}\,\widebar{+}}\suptiny{0}{0}{\mathrm{(SC)}}$ by the transversal $H^{\otimes4}$. A smooth merge measures $\Zbar{\!A}\suptiny{0}{0}{\mathrm{(CC)}}\!\otimes\!\Zbar{\!A}\suptiny{0}{0}{\mathrm{(SC)}} = Z_0 Z_2 Z_4 Z_6$ onto the merge ancilla $a_0$. The conditional $\Xbar{\!A}\suptiny{0}{0}{\mathrm{(CC)}}$ correction on the colour qubits $\{6,7,10,11\}$ (equal to $\Xbar{\!A}\suptiny{0}{0}{\mathrm{(CC)}}$ up to the stabiliser $S\suptiny{1}{0}{\mathrm{(CC)}}_0$) inside the merge block is not executed in hardware for this protocol: it is tracked in post-processing via a Pauli-frame update. The logical $\widebar{\mathrm{CNOT}}_{AB},\widebar{\mathrm{CNOT}}_{AC}$ are realised by relabelling ($5\!\leftrightarrow\!7$, $8\!\leftrightarrow\!10$) before the final readout.}
\label{fig:circuit_ghz}
\end{figure*}

%%%%%%%%%%%%%%%%%%%%%%%%%%%%%%%%%%%%%%%%%%%%%%%%%%%%%%%%%%%%%%%%%%%%%%%%%%%

\vspace*{-2mm}
\subsubsection{Depolarising-Noise Model}\label{appendix:noise_model}
\vspace*{-2mm}

{\noindent}We model circuit errors as depolarising errors, which reproduces well the experimentally observed infidelities despite its conceptual simplicity, which does not take the microscopic physical processes underlying noisy gates and operations in the ion trap into account explicitly. Noise is applied in simulations by randomly placing Pauli errors $E$ according to the experimental physical error rates after every single-qubit operation as well as two-qubit gates. The errors operators are
\begin{subequations}
\begin{align}
    E_1 &\in \{\sigma_k, \forall k \in \{1, 2, 3\}  \}\,, \\
    E_2 &\in \{\sigma_k \otimes \sigma_l, \forall k, l \in \{0, 1, 2, 3\}  \} \backslash \{I\otimes I\} \,,
\end{align}
\end{subequations}
where $\sigma_k = \{I, X, Y, Z\}$ with $k=0, 1, 2, 3$ are the Pauli matrices. The error channels for our depolarising noise
model read
\begin{subequations}
\begin{align}
    \mathcal{E}_1(\rho)&=(1-p_1)\rho + \frac{p_1}{3}\sum_{i=0}^{3}E_1^i\rho E_1^i\,,\\
    \mathcal{E}_2(\rho)&=(1-p_2)\rho + \frac{p_2}{15}\sum_{i=0}^{15}E_2^i\rho E_2^i\,,
\end{align}
\end{subequations}
so that any single-qubit error is applied uniformly to the ideal operation with equal probability $p_1/3$ and the single-qubit operation is executed ideally with probability $1 - p_1$; two-qubit errors are applied uniformly after the ideal two-qubit gates with equal probability $p_2/15$ and any two-qubit gate is executed ideally with probability $1-p_2$. In all simulations we used physical error rates
\begin{subequations}\label{Eq:gate_infidelity}
\begin{align}
    p_1&=0.005\\
    p_2&=0.015
\end{align}
\end{subequations}
for the corresponding operations specified in Ref.~\cite{PogorelovEtAl2021}.

%%%%%%%%%%%%%%%%%%%%%%%%%%%%%%%%%%%%%%%%%%%%%%%%%%%%%%%%%%%%%%%%%%%%%%%%%%%

\clearpage

\begin{figure*}[p]\centering
\begin{adjustbox}{max width=\textwidth,max totalheight=0.30\textheight}
\begin{quantikz}[row sep={0.55cm,between origins}, column sep=0.20cm]
\lstick{\scriptsize $0$} &   &   &   &   &   &   &   &   &   &   &   &   &   &   &   &   &   &   &   &   &   &   &   &   & \gate{H}\gategroup[4,steps=4,style={dashed,rounded corners=2pt,draw=blue!55!cyan,fill=blue!55!cyan,fill opacity=0.10,line width=0.7pt},background,label style={label position=above,anchor=south,yshift=0.10cm,text=blue!55!cyan}]{\scriptsize Surface $\ket{\widebar{+}\widebar{0}}$} & \ctrl{1} &   &   &   & \gategroup[15,steps=7,style={dashed,rounded corners=2pt,draw=green!55!black,fill=green!55!black,fill opacity=0.10,line width=0.7pt},background,label style={label position=above,anchor=south,yshift=0.10cm,text=green!55!black}]{\scriptsize Merge $\widebar{Z}_{A}\suptiny{0}{0}{\mathrm{CC}}\!\otimes\!\widebar{Z}_{A}\suptiny{0}{0}{\mathrm{SC}}$} &   & \ctrl{4} &   &   &   &   &   &   &   & \meter{}\gategroup[15,steps=1,style={dashed,rounded corners=2pt,draw=gray!60,fill=gray!60,fill opacity=0.10,line width=0.7pt},background,label style={label position=above,anchor=south,yshift=0.10cm,text=gray!60}]{\scriptsize Readout} \\
\lstick{\scriptsize $1$} &   &   &   &   &   &   &   &   &   &   &   &   &   &   &   &   &   &   &   &   &   &   &   &   &   & \targ{} &   &   &   &   &   &   &   &   &   &   &   &   &   & \meter{} \\
\lstick{\scriptsize $2$} &   &   &   &   &   &   &   &   &   &   &   &   &   &   &   &   &   &   &   &   &   &   &   &   &   &   & \gate{H} & \ctrl{1} &   &   &   &   & \ctrl{2} &   &   &   &   &   &   & \meter{} \\
\lstick{\scriptsize $3$} &   &   &   &   &   &   &   &   &   &   &   &   &   &   &   &   &   &   &   &   &   &   &   &   &   &   &   & \targ{} &   &   &   &   &   &   &   &   &   &   &   & \meter{} \\
\lstick{\scriptsize $a_0$} &   &   &   &   &   &   &   &   &   &   &   &   &   &   &   &   &   &   &   &   &   &   &   &   &   &   &   &   &   & \targ{} & \targ{} & \targ{} & \targ{} & \meter{} & \setwiretype{c} & \ctrl{8} &  \setwiretype{n} &   &   &   \\
\lstick{\scriptsize $a_1$} &   &   &   &   &   &   &   &   &   &   & \gate{H}\gategroup[10,steps=7,style={dashed,rounded corners=2pt,draw=teal!70,fill=teal!70,fill opacity=0.10,line width=0.7pt},background,label style={label position=below,anchor=north,yshift=-0.30cm,text=teal!70}]{\scriptsize Colour $X$ flag} & \ctrl{2} & \ctrl{7} & \ctrl{4} & \ctrl{9} & \gate{H} & \meter{} &   &   &   &   &   &   &   &   &   &   &   &   &   &   &   &   &   &   &   &   &   &   &   \\
\lstick{\scriptsize $a_2$} &   &   &   &   &   &   &   &   &   &   &   &   &   &   &   &   &   &   & \targ{}\gategroup[9,steps=5,style={dashed,rounded corners=2pt,draw=teal!70,fill=teal!70,fill opacity=0.10,line width=0.7pt},background,label style={label position=below,anchor=north,yshift=-0.30cm,text=teal!70}]{\scriptsize Colour $Z$ flag} & \targ{} & \targ{} & \targ{} & \meter{} &   &   &   &   &   &   &   &   &   &   &   &   &   &   &   &   &   \\
\lstick{\scriptsize $4$} & \gate{H}\gategroup[8,steps=9,style={dashed,rounded corners=2pt,draw=blue!70,fill=blue!70,fill opacity=0.10,line width=0.7pt},background,label style={label position=below,anchor=north,yshift=-0.30cm,text=blue!70}]{\scriptsize Colour $\ket{\widebar{+}\widebar{+}\widebar{+}}$} &   & \ctrl{2} & \ctrl{1} &   &   &   & \targ{} &   &   &   & \targ{} &   &   &   &   &   &   & \ctrl{-1} &   &   &   &   &   &   &   &   &   &   & \ctrl{-3} &   &   &   &   &   & \gate[style={fill=gray!12,dashed,draw=gray!80}]{X} &   & \gate{T}\gategroup[8,steps=1,style={dashed,rounded corners=2pt,draw=red!60!black,fill=red!60!black,fill opacity=0.10,line width=0.7pt},background,label style={label position=below,anchor=north,yshift=-0.30cm,text=red!60!black}]{\scriptsize Transversal $\overline{CCZ}$} &   & \meter{} \\
\lstick{\scriptsize $5$} &   &   &   & \targ{} &   &   &   &   & \targ{} &   &   &   &   &   &   &   &   &   &   &   &   &   &   &   &   &   &   &   &   &   &   &   &   &   &   & \gate[style={fill=gray!12,dashed,draw=gray!80}]{X} &   & \gate{T^{\dagger}} &   & \meter{} \\
\lstick{\scriptsize $6$} &   &   & \targ{} & \ctrl{1} & \targ{} &   &   &   &   &   &   &   &   & \targ{} &   &   &   &   &   &   & \ctrl{-3} &   &   &   &   &   &   &   &   &   & \ctrl{-5} &   &   &   &   &   &   & \gate{T^{\dagger}} &   & \meter{} \\
\lstick{\scriptsize $7$} &   &   &   & \targ{} &   & \targ{} &   &   &   &   &   &   &   &   &   &   &   &   &   &   &   &   &   &   &   &   &   &   &   &   &   &   &   &   &   &   &   & \gate{T} &   & \meter{} \\
\lstick{\scriptsize $8$} & \gate{H} & \ctrl{2} &   &   &   &   & \targ{} & \ctrl{-4} &   &   &   &   &   &   &   &   &   &   &   &   &   &   &   &   &   &   &   &   &   &   &   &   &   &   &   & \gate[style={fill=gray!12,dashed,draw=gray!80}]{X} &   & \gate{T^{\dagger}} &   & \meter{} \\
\lstick{\scriptsize $9$} & \gate{H} &   &   &   &   &   & \ctrl{-1} &   & \ctrl{-4} &   &   &   & \targ{} &   &   &   &   &   &   & \ctrl{-6} &   &   &   &   &   &   &   &   &   &   &   &   &   &   &   & \gate[style={fill=gray!12,dashed,draw=gray!80}]{X} &   & \gate{T} &   & \meter{} \\
\lstick{\scriptsize $10$} &   & \targ{} & \targ{} &   & \ctrl{-4} &   &   &   &   &   &   &   &   &   &   &   &   &   &   &   &   &   &   &   &   &   &   &   &   &   &   &   &   &   &   &   &   & \gate{T} &   & \meter{} \\
\lstick{\scriptsize $11$} & \gate{H} &   & \ctrl{-1} &   &   & \ctrl{-4} &   &   &   &   &   &   &   &   & \targ{} &   &   &   &   &   &   & \ctrl{-8} &   &   &   &   &   &   &   &   &   &   &   &   &   &   &   & \gate{T^{\dagger}} &   & \meter{}
\end{quantikz}
\end{adjustbox}
\caption{\textbf{Physical circuit for the logical $\widebar{\mathrm{CCZ}}$.} Qubits $0${\textendash}$3$ are the surface code, $4${\textendash}$11$ the colour code. The colour code is initialised in $\ket{\widebar{+}\widebar{+}\widebar{+}}\suptiny{0}{0}{\mathrm{(CC)}}$ with two flag checks: an $X$-parity flag (ancilla $a_1$) and a $Z$-parity flag (ancilla $a_2$), both coupled to qubits $\{4,6,9,11\}$. The surface code is prepared in $\ket{\widebar{+}\widebar{0}}\suptiny{0}{0}{\mathrm{(SC)}}$; no surface flag is used. The merge measures $\Zbar{\!A}\suptiny{0}{0}{\mathrm{(CC)}}\!\otimes\!\Zbar{\!A}\suptiny{0}{0}{\mathrm{(SC)}} = Z_0 Z_2 Z_4 Z_6$ onto the merge ancilla $a_0$. The conditional $\Xbar{\!A}\suptiny{0}{0}{\mathrm{(CC)}}$ correction on the colour qubits $\{4,5,8,9\}$ inside the merge block is applied for this protocol by post-selection on the merge outcome, since the transversal $T/T^{\dagger}$ layer prevents a Pauli-frame update. The transversal $\widebar{\mathrm{CCZ}}$ applies $T$ to qubits $\{4,7,9,10\}$ and $T^{\dagger}$ to $\{5,6,8,11\}$, followed by readout.}
\label{fig:circuit_ccz}
\end{figure*}

%%%%%%%%%%%%%%%%%%%%%%%%%%%%%%%%%%%%%%%%%%%%%%%%%%%%%%%%%%%%%%%%%%%%%%%%%%%

\begin{figure*}[p]\centering
\begin{adjustbox}{max width=\textwidth,max totalheight=0.30\textheight}
\begin{quantikz}[row sep={0.55cm,between origins}, column sep=0.20cm]
\lstick{\scriptsize $0$} &   &   &   &   &   &   &   &   &   &   &   &   &   &   &   &   &   &   &   &   &   &   & \gate{H}\gategroup[4,steps=4,style={dashed,rounded corners=2pt,draw=blue!55!cyan,fill=blue!55!cyan,fill opacity=0.10,line width=0.7pt},background,label style={label position=above,anchor=south,yshift=0.10cm,text=blue!55!cyan}]{\scriptsize Surface $\ket{\widebar{+}\widebar{0}}$} & \ctrl{1} &   &   &   & \gategroup[17,steps=7,style={dashed,rounded corners=2pt,draw=green!55!black,fill=green!55!black,fill opacity=0.10,line width=0.7pt},background,label style={label position=above,anchor=south,yshift=0.10cm,text=green!55!black}]{\scriptsize Merge $\widebar{Z}_{C}\suptiny{0}{0}{\mathrm{CC}}\!\otimes\!\widebar{Z}_{A}\suptiny{0}{0}{\mathrm{SC}}$} &   & \ctrl{5} &   &   &   & \gate[style={fill=gray!12,dashed,draw=gray!80}]{X} &   & \gategroup[17,steps=9,style={dashed,rounded corners=2pt,draw=orange!80!black,fill=orange!80!black,fill opacity=0.10,line width=0.7pt},background,label style={label position=below,anchor=north,yshift=-0.30cm,text=orange!80!black}]{\scriptsize Split $\widebar{X}_{C}\suptiny{0}{0}{\mathrm{CC}}$} &   &   &   &   &   &   &   & \gate[style={fill=gray!12,dashed,draw=gray!80}]{Z} &   & \gate{H}\gategroup[4,steps=1,style={dashed,rounded corners=2pt,draw=magenta!60,fill=magenta!60,fill opacity=0.10,line width=0.7pt},background,label style={label position=above,anchor=south,yshift=0.10cm,text=magenta!60}]{\scriptsize $H^{\otimes4}$} &   &   &   &   &   & \ctrl{4}\gategroup[5,steps=3,style={dashed,rounded corners=2pt,draw=blue!55!cyan,fill=blue!55!cyan,fill opacity=0.10,line width=0.7pt},background,label style={label position=above,anchor=south,yshift=0.10cm,text=blue!55!cyan}]{\scriptsize Readout $\widebar{X}\suptiny{0}{0}{\mathrm{SC}}$ (rot.)} &   &   \\
\lstick{\scriptsize $1$} &   &   &   &   &   &   &   &   &   &   &   &   &   &   &   &   &   &   &   &   &   &   &   & \targ{} &   &   &   &   &   &   &   &   &   & \gate[style={fill=gray!12,dashed,draw=gray!80}]{X} &   &   &   &   &   &   &   &   &   &   &   & \gate{H} &   &   &   &   &   &   & \ctrl{3} &   \\
\lstick{\scriptsize $2$} &   &   &   &   &   &   &   &   &   &   &   &   &   &   &   &   &   &   &   &   &   &   &   &   & \gate{H} & \ctrl{1} &   &   &   &   & \ctrl{3} &   &   &   &   &   &   &   &   &   &   &   &   & \gate[style={fill=gray!12,dashed,draw=gray!80}]{Z} &   & \gate{H} &   &   &   &   &   &   &   &   \\
\lstick{\scriptsize $3$} &   &   &   &   &   &   &   &   &   &   &   &   &   &   &   &   &   &   &   &   &   &   &   &   &   & \targ{} &   &   &   &   &   &   &   &   &   &   &   &   &   &   &   &   &   &   &   & \gate{H} &   &   &   &   &   &   &   &   \\
\lstick{\scriptsize $a_0$} &   &   &   &   &   &   &   &   &   &   &   &   &   &   &   &   &   &   &   &   &   &   &   &   &   &   &   &   &   &   &   &   &   &   &   &   &   &   &   &   &   &   &   &   &   &   &   &   &   &   &   & \targ{} & \targ{} & \meter{} \\
\lstick{\scriptsize $a_1$} &   &   &   &   &   &   &   &   &   &   &   &   &   &   &   &   &   &   &   &   &   &   &   &   &   &   &   & \targ{} & \targ{} & \targ{} & \targ{} & \meter{} & \setwiretype{c} & \ctrl{-5} &  \setwiretype{n} &   &   &   &   &   &   &   &   &   &   &   &   &   &   &   &   &   &   &   \\
\lstick{\scriptsize $a_2$} &   &   &   &   &   &   &   &   &   &   &   &   &   &   &   &   &   &   &   &   &   &   &   &   &   &   &   &   &   &   &   &   &   &   &   & \gate{H} & \ctrl{3} & \ctrl{6} & \ctrl{5} & \ctrl{4} & \gate{H} & \meter{} & \setwiretype{c} & \ctrl{-6} &  \setwiretype{n} &   &   &   &   &   &   &   &   &   \\
\lstick{\scriptsize $a_3$} &   &   &   &   &   &   &   &   &   &   &   &   &   &   &   &   &   &   &   &   &   &   &   &   &   &   &   &   &   &   &   &   &   &   &   &   &   &   &   &   &   &   &   &   &   &   &   & \targ{}\gategroup[10,steps=3,style={dashed,rounded corners=2pt,draw=gray!60,fill=gray!60,fill opacity=0.10,line width=0.7pt},background,label style={label position=below,anchor=north,yshift=-0.30cm,text=gray!60}]{\scriptsize Readout $\widebar{Z}_{B}\suptiny{0}{0}{\mathrm{CC}}$} & \targ{} & \meter{} &   &   &   &   \\
\lstick{\scriptsize $a_4$} &   &   &   &   &   &   &   &   &   &   &   & \targ{}\gategroup[9,steps=5,style={dashed,rounded corners=2pt,draw=teal!70,fill=teal!70,fill opacity=0.10,line width=0.7pt},background,label style={label position=below,anchor=north,yshift=-0.30cm,text=teal!70}]{\scriptsize Colour $Z$ flag} & \targ{} & \targ{} & \targ{} & \meter{} &   &   &   &   &   &   &   &   &   &   &   &   &   &   &   &   &   &   &   &   &   &   &   &   &   &   &   &   &   &   &   &   &   &   &   &   &   &   \\
\lstick{\scriptsize $4$} & \gategroup[8,steps=10,style={dashed,rounded corners=2pt,draw=blue!70,fill=blue!70,fill opacity=0.10,line width=0.7pt},background,label style={label position=below,anchor=north,yshift=-0.30cm,text=blue!70}]{\scriptsize Colour $\ket{\widebar{+}}_A,\ket{\widebar{+}_\theta}_B,\ket{\widebar{0}_{2\theta}}_C$} & \gate{R_x(2\theta)} &   & \ctrl{2} & \ctrl{1} &   &   &   & \targ{} &   &   & \ctrl{-1} &   &   &   &   &   & \gate{T}\gategroup[8,steps=1,style={dashed,rounded corners=2pt,draw=red!60!black,fill=red!60!black,fill opacity=0.10,line width=0.7pt},background,label style={label position=below,anchor=north,yshift=-0.30cm,text=red!60!black}]{\scriptsize Transversal $\overline{CCZ}$} &   & \gategroup[8,steps=2,style={dashed,rounded corners=2pt,draw=purple!65,fill=purple!65,fill opacity=0.10,line width=0.7pt},background,label style={label position=above,anchor=west,xshift=0.6cm,yshift=-0.80cm,fill=white,fill opacity=1,inner sep=1.2pt,text=purple!65}]{\scriptsize $\overline{CX}_{AB}$ (relabel)} &   &   &   &   &   &   &   & \ctrl{-4} &   &   &   &   &   &   &   &   & \targ{} &   &   &   &   &   &   &   &   &   &   &   &   &   &   &   &   &   \\
\lstick{\scriptsize $5$} &   &   &   &   & \targ{} &   &   &   &   & \targ{} &   &   &   &   &   &   &   & \gate{T^{\dagger}} &   & \swap{2} &   &   &   &   &   &   &   &   &   &   &   &   &   &   &   &   &   &   &   & \targ{} &   &   &   &   &   &   &   &   &   &   &   &   &   &   \\
\lstick{\scriptsize $6$} &   &   &   & \targ{} & \ctrl{1} & \targ{} &   &   &   &   &   &   &   & \ctrl{-3} &   &   &   & \gate{T^{\dagger}} &   &   &   &   &   &   &   &   &   &   &   &   &   &   &   &   &   &   &   &   & \targ{} &   &   &   &   &   &   &   &   &   &   &   &   &   &   &   \\
\lstick{\scriptsize $7$} &   &   &   &   & \targ{} &   & \targ{} &   &   &   &   &   &   &   &   &   &   & \gate{T} &   & \targX{} &   &   &   &   &   &   &   &   &   &   &   &   &   &   &   &   &   & \targ{} &   &   &   &   &   &   &   &   &   &   &   &   &   &   &   &   \\
\lstick{\scriptsize $8$} & \gate{H} & \gate{R_z(\theta)} & \ctrl{2} &   &   &   &   & \targ{} & \ctrl{-4} &   &   &   &   &   &   &   &   & \gate{T^{\dagger}} &   &   &   &   &   &   &   &   &   &   & \ctrl{-8} &   &   &   &   &   &   &   &   &   &   &   &   &   &   &   &   &   &   & \ctrl{-6} &   &   &   &   &   &   \\
\lstick{\scriptsize $9$} & \gate{H} &   &   &   &   &   &   & \ctrl{-1} &   & \ctrl{-4} &   &   & \ctrl{-6} &   &   &   &   & \gate{T} &   &   & \swap{2} &   &   &   &   &   &   &   &   &   &   &   &   &   &   &   &   &   &   &   &   &   &   &   &   &   &   &   & \ctrl{-7} &   &   &   &   &   \\
\lstick{\scriptsize $10$} &   &   & \targ{} & \targ{} &   & \ctrl{-4} &   &   &   &   &   &   &   &   &   &   &   & \gate{T} &   &   &   &   &   &   &   &   &   &   &   &   &   &   &   &   &   &   &   &   &   &   &   &   &   &   &   &   &   &   &   &   &   &   &   &   \\
\lstick{\scriptsize $11$} & \gate{H} &   &   & \ctrl{-1} &   &   & \ctrl{-4} &   &   &   &   &   &   &   & \ctrl{-8} &   &   & \gate{T^{\dagger}} &   &   & \targX{} &   &   &   &   &   &   &   &   &   &   &   &   &   &   &   &   &   &   &   &   &   &   &   &   &   &   &   &   &   &   &   &   &  
\end{quantikz}
\end{adjustbox}
\caption{\textbf{Physical circuit for the logical small-angle rotation.} Qubits $0${\textendash}$3$ are the surface code, $4${\textendash}$11$ the colour code. The colour code is initialised in $\ket{\widebar{+}}_A,\,\ket{\widebar{+}_\theta}_B,\,\ket{\widebar{0}_{2\theta}}_C$, where $\ket{\widebar{+}_\theta}_B=\widebar{R}_Z(\theta)\ket{\widebar{+}}_B$ and $\ket{\widebar{0}_{2\theta}}_C=\widebar{R}_X(2\theta)\ket{\widebar{0}}_C$ (realised at the physical level by $R_z(\theta)$ on qubit $8$ and $R_x(2\theta)$ on qubit $4$) and a single $Z$-parity flag on qubits $\{4,6,9,11\}$, measured on the flag ancilla $a_4$. The surface code is prepared in $\ket{\widebar{+}\widebar{0}}\suptiny{0}{0}{\mathrm{(SC)}}$. After the transversal $\widebar{\mathrm{CCZ}}$ and the relabelling $\widebar{\mathrm{CNOT}}_{AB}$ ($5\!\leftrightarrow\!7$, $9\!\leftrightarrow\!11$), the merge measures $\Zbar{\!C}\suptiny{0}{0}{\mathrm{(CC)}}\!\otimes\!\Zbar{\!A}\suptiny{0}{0}{\mathrm{(SC)}}$ and the split measures $\Xbar{\!C}\suptiny{0}{0}{\mathrm{(CC)}} = X_4 X_5 X_6 X_7$. The conditional corrections, $\Xbar{\!A}\suptiny{0}{0}{\mathrm{(SC)}}$ on surface qubits $\{0,1\}$ for the merge and $\Zbar{\!A}\suptiny{0}{0}{\mathrm{(SC)}}$ on surface qubits $\{0,2\}$ for the split (drawn inside their blocks), are not executed in hardware for this protocol: they are tracked in post-processing via a Pauli-frame update. A transversal $H^{\otimes4}$ then precedes the two logical readouts, $\Zbar{\!B}\suptiny{0}{0}{\mathrm{(CC)}} = Z_8 Z_9$ (ancilla $a_3$) and the surface observable $\widebar{X}\suptiny{0}{0}{\mathrm{(SC)}}$ on qubits $\{0,2\}$ (ancilla $a_0$).}
\label{fig:circuit_rotation}
\end{figure*}

%%%%%%%%%%%%%%%%%%%%%%%%%%%%%%%%%%%%%%%%%%%%%%%%%%%%%%%%%%%%%%%%%%%%%%%%%%%

\end{document}